%% file: template.tex
\documentclass{article}

\usepackage{arxiv}

\usepackage[utf8]{inputenc} 
\usepackage[T1]{fontenc}    
\usepackage{hyperref}       
\usepackage{url}            
\usepackage{booktabs}       
\usepackage{amsfonts}       
\usepackage{nicefrac}       
\usepackage{microtype}      
\usepackage{graphicx}
\usepackage[sort]{natbib}
\setcitestyle{comma,numbers,open={[},close={]}}
\usepackage{doi}

\title{Neural network-based multiscale modeling of finite strain magneto-elasticity with relaxed convexity criteria}

\author{
	Karl A. Kalina\\
	Chair of Computational and\\
	Experimental Solid Mechanics\\
	TU Dresden,
	01062 Dresden, Germany \\
	\And
	Philipp Gebhart\\
	Chair of Mechanics of Multifunctional Structures\\
	TU Dresden,
	01062 Dresden, Germany \\
	\And
	J\"{o}rg Brummund\\
	Chair of Computational and\\
	Experimental Solid Mechanics\\
	TU Dresden,
	01062 Dresden, Germany \\
	\And
	Lennart Linden\\
	Chair of Computational and\\
	Experimental Solid Mechanics\\
	TU Dresden,
	01062 Dresden, Germany \\
	\And
	WaiChing Sun\\
	Department of Civil Engineering\\
	and Engineering Mechanics\\
	Columbia University, 
	NY 10027, New York, United States \\
	\And
	Markus K\"{a}stner\thanks{Corresponding author, email: \texttt{markus.kaestner@tu-dresden.de}.} \\
	Chair of Computational and\\
	Experimental Solid Mechanics\\
	TU Dresden, 
	01062 Dresden, Germany \\
}

\renewcommand{\shorttitle}{Neural network-based multiscale modeling of finite strain magneto-elasticity}

\hypersetup{
pdftitle={Preprint_KalinaEtAl_2023},
pdfsubject={},
pdfauthor={Kalina},
pdfkeywords={First keyword, Second keyword, More},
}

\input{StyleSetup}

\theoremstyle{definition} 
\newtheorem{rmk}{Remark}
\usepackage[font=small]{caption}

\begin{document}
\maketitle

\begin{abstract}
We present a framework for the multiscale modeling of finite strain magneto-elasticity based on physics-augmented neural networks (NNs). By using a set of problem specific invariants as input, an energy functional as the output and by adding several non-trainable expressions to the overall total energy density functional, the model fulfills multiple physical principles by construction, e.g., thermodynamic consistency, material symmetry and a stress-free and non-magnetized unloaded configuration. Three NN-based models with varying requirements in terms of an extended polyconvexity condition and the growth condition of the magneto-elastic potential are presented. First, polyconvexity, which is a global concept, is enforced via input convex neural networks (ICNNs).
Afterwards, we formulate a relaxed local version of the polyconvexity and fulfill it in a weak sense by adding a tailored loss term. As an alternative, a loss term to enforce the weaker requirement of strong ellipticity locally is proposed, which can be favorable to obtain a better trade-off between compatibility with data and physical constraints. Databases for training of the models are generated via computational homogenization for both compressible and quasi-incompressible magneto-active polymers (MAPs). Thereby, to reduce the computational cost, 2D statistical volume elements and an invariant-based sampling technique for the pre-selection of relevant states are used. All models are calibrated by using the database, whereby interpolation and extrapolation are considered separately. Furthermore, the performance of the NN models is compared to a conventional model from the literature. 
The numerical study suggests that the proposed physics-augmented NN approach is advantageous over the conventional model for MAPs. Thereby, the two more flexible NN models in combination with the weakly enforced local polyconvexity lead to good results, whereas the model based only on ICNNs has proven to be too restrictive.
\end{abstract}

\section*{Graphical abstract}
\includegraphics{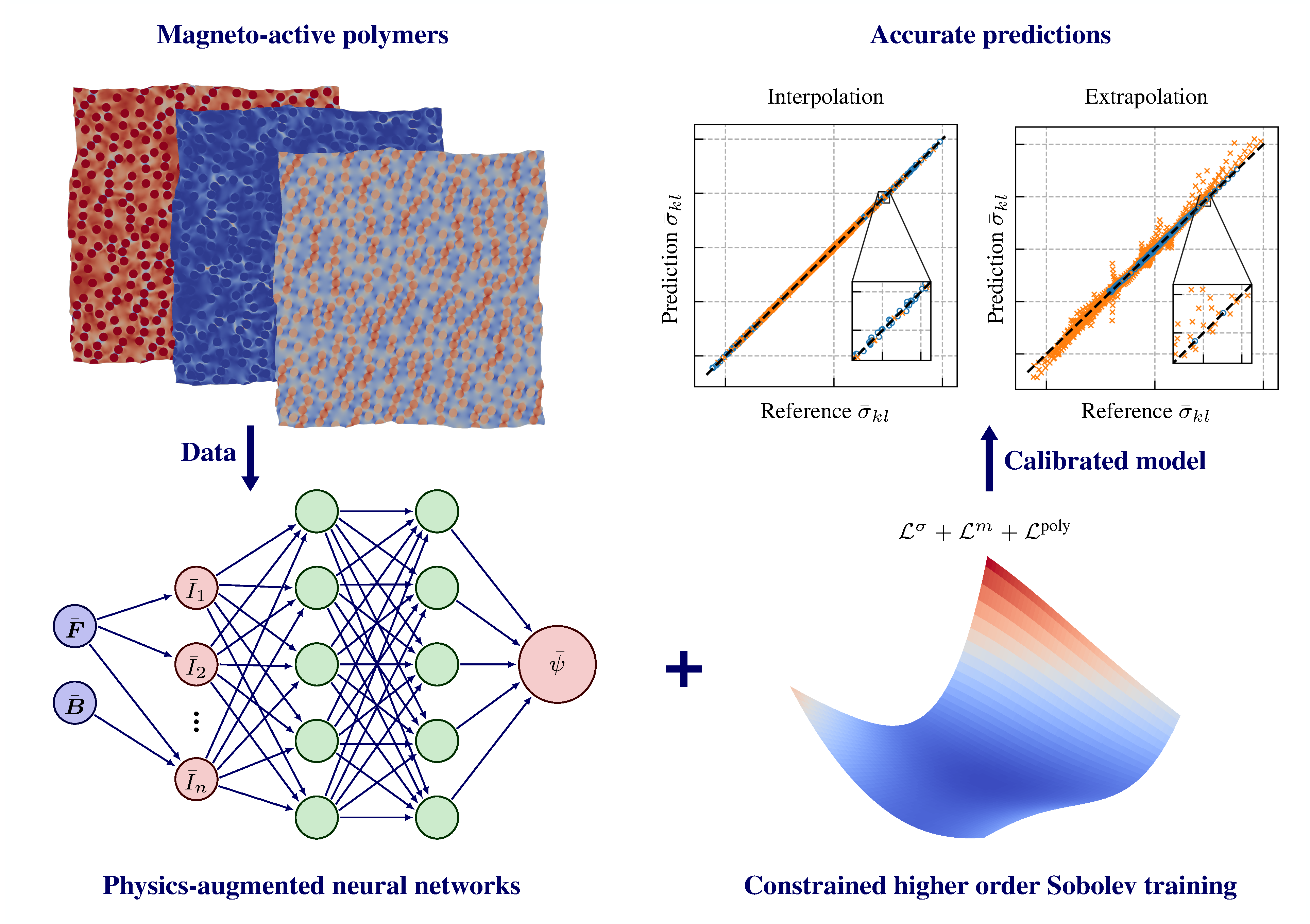}


\section{Introduction}
\label{sec:int}

Due to the possibility of targeted control by means of external stimuli, smart materials are of increasing interest in industry and research. \emph{Magneto-active polymers (MAPs)}, which exhibit a significant change in mechanical properties when exposed to external magnetic fields, form a subgroup of this material class. Due to their special properties, they are attractive options for several engineering applications such
as actuators and sensors \cite{Tian2011,Volkova2017}, grippers \cite{Becker2019}, valves \cite{Bose2012}, tunable isolators and vibration absorbers \cite{Li2013,Deng2006}, microfluidic transport systems \cite{Behrooz2016} or medical
robots \cite{Kim2019,Hu2018}.

\subsection{Magneto-active polymers}\label{subsec:MAPs}

MAPs consist of a soft polymer matrix filled with magnetizable particles that can be arranged in an unstructured distribution \cite{Tian2011a} or, if external
fields are applied during crosslinking of the polymer, in a chain-like \cite{Danas2012,Hiptmair2015} or even more
complex structure \cite{Martin2003}. Possible matrix materials include elastomers and  gels, where MAPs based on them are also referred to as \emph{magnetorheological elastomers (MREs)} and \emph{ferrogels}, respectively \cite{Bastola2020}. Typically, MREs are filled with micron-sized magnetizable multidomain particles \cite{Danas2012,Hiptmair2015,Linke2016}, whereas single-domain nanoparticles are more common for ferrogels \cite{Weeber2018}. However, it is also possible to embed micron-sized particles in ferrogels \cite{Puljiz2018}. If \emph{magnetically soft} filler particles as carbonyl iron \cite{Danas2012,Li2013} -- which show no significant hystereses -- are used, magnetically induced stresses and deformations of the composite are reversible. However, due to possible viscoelastic effects of the matrix, a rate dependency can occur \cite{Hiptmair2015,Moreno2021}. If instead \emph{magnetically hard} fillers as NdFeB \cite{Linke2016,Kalina2017} -- which exhibit significant hystereses -- are incorporated into the matrix, the system's behavior is irreversible. Also mixtures of magnetically hard and soft particles are possible \cite{Linke2016,Moreno-Mateos2022}. 
Under the influence of magnetic fields, the microstructure of MAPs changes due to magnetic interactions and the resulting restructuring of the embedded particles \cite{Schumann2017}. For the composite, this results in three main magneto-mechanical coupling effects \cite{Kalina2021a}: magnetically induced deformation \emph{(magnetostrictive effect)} \cite{Ginder2002,Danas2012,Diguet2010,Stepanov2008}, magnetically induced stiffness change \emph{(magnetorheological effect)} \cite{Hiptmair2015,Lokander2003} and \emph{field-induced plasticity} \cite{Melenev2011,Stepanov2008}, where the latter only occurs with extremely soft matrices \cite{Stepanov2008}. For more details on the synthesis and experimental characterization of MAPs, the reader is referred to the review article Bastola~and~Hossain~\cite{Bastola2020}.

In general, modeling strategies for MAPs can be divided into \emph{microscopic} and \emph{macroscopic} approaches, where the former explicitly resolve the heterogeneous microstructure, while the latter consider the composite an effective medium. 
\emph{Particle interaction models} \cite{Cremer2015,Fischer2020,Puljiz2018,Romeis2017} represent the first class of microscopic approaches. They are based on energy minimization and describe the particles as magnetic dipoles. The matrix can be modeled as an elastic continuum or by springs between the particles \cite{Menzel2019}. Particle interaction models are numerically very efficient but, due to the dipole approximation, are only sufficiently accurate for systems with large relative particle distances. A more accurate description of particle interactions for small relative distances between inclusions is possible by using a multipole expansion \cite{Biller2014}.
The second class of microscopic models for MAPs is based on a \emph{continuum theory} as shown by \cite{PonteCastaneda2011,Keip2016,Danas2017,Metsch2019,Kalina2017}, among others. Here, both the local mechanical and magnetic fields within the composite system are explicitly resolved. Accordingly, these approaches are not limited to the modeling of MAPs with low particle volume fractions. Furthermore, they have the benefit of being easily adaptable to a wide range of non-dissipative and \emph{dissipative} component materials, e.g., for viscoelasticity or magnetically hard behavior \cite{Kalina2017,Rambausek2022,Moreno-Mateos2022}. A major drawback of continuum approaches is the computationally intensive numerical solution of the underlying nonlinear magneto-mechanical boundary value problem (BVP), which is usually done by using the \emph{finite element method (FEM)}. Therefore, \emph{homogenization schemes} as presented in \cite{Chatzigeorgiou2014} are typically used to predict the effective behavior of realistic magneto-active composites from \emph{representative volume elements (RVEs)}, where both analytical \cite{PonteCastaneda2011,Lefevre2020} and numerical solution techniques \cite{Metsch2019} can be applied. A classical tool for capturing not only the microscopic structure but also the macroscopic shape of the samples is the FE${}^2$ method \cite{Keip2016,Keip2017,Zabihyan2020}, which is also known as \emph{coupled multiscale scheme}. As shown in \cite{Rudykh2013,Goshkoderia2017,Polukhov2020,Galipeau2013}, microscopic continuum approaches in combination with computational homogenization approaches also offer the possibility to analyze the \emph{stability} of MAPs at multiple scales under different loading conditions.

\emph{Macroscopic models}, in contrast, do not resolve the microstructure explicitly. Instead, a constitutive law that replicates the response of MAPs as effective media is used to replace the homogenization of RVEs. This allows for the representation of real samples and components with reasonable computational cost \cite{Psarra2017,Moreno-Mateos2023}. The first class of these models are motivated from \emph{experimental observations}, where models for isotropic \cite{Bustamante2011} and transversely isotropic magneto-elasticity \cite{Dorfmann2004a,Bustamante2010,Salas2015} or for magneto-viscoelasticity \cite{Saxena2013,Haldar2016} are available. Of course, all these models are \emph{purely phenomenological}, i.e., a calibration by using experimental data has to be done \cite{Danas2012,Haldar2016}.
However, it is important to note that at least inhomogeneous mechanical fields occur independently from the macroscopic specimen geometry, which is due to the nature of the coupling of magnetic and mechanical fields \cite{Keip2017,Gebhart2022,Kalina2020}. In addition, the magnetization is also non-homogeneous if no ellipsoidal samples are used \cite{Bodelot2018}. The parameterized macroscopic model thus contains the influence of the sample geometry, unless an expensive inverse parametrization is performed. In other words, the model represents the behavior of a MAP-sample, rather than the constitutive behavior of the pure composite material.
An alternative are \emph{microscopically guided macroscopic models} calibrated with data obtained from analytical or computational \emph{homogenization} techniques. This approach eliminates the difficulties of the fitting to experiments described above, as it allows for the identification of parameters independent of the macroscopic specimen geometry. In the recent years, several works applying this technique for MAPs have been published, e.g., \cite{Mukherjee2019a,Mukherjee2021,Mukherjee2022,Kalina2020,Kalina2020a,Gebhart2022,Gebhart2022a,Garcia-Gonzalez2021}, among others. To simulate real components with low computational cost while implicitly accounting for microscopic effects, these models can also be applied in \emph{decoupled multiscale schemes} \cite{Lefevre2020,Kalina2020a,Gebhart2022a} as introduced by Terada~et~al.~\cite{Terada2013} for hyperelasticity. A clear disadvantage of macroscopic material models for MAPs is that their formulation can be extremely complicated and high accuracy may not be achievable for general multiaxial loading cases.

\subsection{Constitutive modeling with neural networks}

Due to the time-consuming formulation of conventional constitutive models for complex materials and the limited functional relationships inherent in most of these expressions, alternatives based on \emph{machine learning} methods are becoming increasingly popular, with \emph{neural networks (NNs)} being the most widely used. The following brief overview is mainly limited to finite strain hyperelasticity.

NNs were first introduced into constitutive modeling by Ghaboussi~et~al.~\cite{Ghaboussi1991} at the beginning of the 1990s. At this early stage, however, most approaches were developed without a reasonable physical background, which is why they are also referred to as \emph{black-box models}. Thus, these networks can replicate data well but cannot guarantee a physically meaningful behavior for unseen data, which often leads to \emph{poor extrapolations}. To address this weak point, a rather new development in NN-based constitutive modeling and generally in scientific machine learning is to include fundamental underlying physics, which is labeled as \emph{physics-informed} \cite{Raissi2019,Henkes2022}, \emph{mechanics-informed} \cite{Asad2022}, \emph{physics-augmented} \cite{Klein2022,Linden2023}, \emph{physics-constrained} \cite{Kalina2023}, or
\emph{thermodynamics-based} \cite{Masi2021}. This can be done in strong sense, e.g., by using adapted network architectures, or in weak sense, e.g., by changing the loss term of the training, cf. \cite{Rosenkranz2023,Weber2021}, and enables an improvement of the extrapolation capability and the usage of sparse training data \cite{Linden2023,Masi2021,Fuhg2023}.

Several works that model elasticity along these lines exist, e.g., \cite{Shen2004,Liang2008} approximate the elastic potential by using a \emph{feedforward neural network (FNN)} with three deformation-type invariants as input, which leads to the fulfillment of several constitutive requirements
by construction, e.g., \emph{thermodynamic consistency}, \emph{objectivity}, or \emph{material symmetry}. However, training
of these models was done by directly using the elastic potential instead of the stress. Meanwhile, approaches based on architectures that use \emph{invariants} as input and the \emph{hyperelastic potential} as output are a very well established approach, e.g., \cite{Linden2023,Klein2021,Kalina2022a,Thakolkaran2022,Linka2021,Fuhg2022b}, among others. A special training is applied which allows the calibration of the network directly by tuples of stress and strain. For this purpose, the derivative of the energy with respect to the deformation is included into the loss, which is also called \emph{Sobolev training} \cite{Czarnecki2017,Vlassis2020}. Other models directly formulate NN-based potentials
in terms of the components of strain or deformation tensors \cite{Asad2022,Vlassis2020,Fernandez2020a}, which gives more flexibility in the case of anisotropy.
In addition, several approaches build \emph{polyconvex} NNs \cite{Klein2021,Tac2022a,Chen2022,Linka2023,Tac2023a}, which is favorable in finite element (FE) simulations and guarantees strong ellipticity \cite{Ebbing2010,Schroder2010}. Different techniques are applied, whereby the most widespread for incorporating this condition is the use of \emph{input convex neural networks (ICNNs)} according to Amos~et~al.~\cite{Amos2017}. Finally, very recently, in Linden~et~al.~\cite{Linden2023} an ICNN-based approach fulfilling all common conditions of hyperelasticity, namely thermodynamic consistency, symmetry of the stress tensor, objectivity, material symmetry, polyconvexity, growth condition, as well as zero energy and stress in the undeformed configuration by construction. 

Besides the NN approaches developed for hyperelasticity, there are also several works that provide constitutive models in the same line for other field problems. E.g., in \cite{Lu2019}, an application of NNs as surrogate models describing the anisotropic \emph{electrical} response of graphene/polymer nanocomposites is shown, while in \cite{Cardelli2016,Quondam-Antonio2023} hystereses of \emph{ferromagnetic materials} are modeled by NNs. However, these models do not account for crucial principles as the second law of thermodynamics or material symmetry.
Regarding \emph{coupled problems}, only a few works are provided so far: Adiabatic thermally expanded hyperelasticity is modeled with FNNs in \cite{Zlatic2023}, where approaches with differing physical background are compared. The coupled finite strain electro-elastic behavior of electro-active polymers is modeled with ICNNs in \cite{Klein2022} for the linear electric regime.

\subsection{Objectives and contributions of this work}

The literature overview on NNs given above shows that for purely mechanical problems, and in particular for elasticity, numerous very sophisticated approaches exist that combine modern machine learning methods with a reasonable physical basis. In contrast, comparable approaches for coupled problems are rare. To the best of our knowledge, only the work of Klein~et~al.~\cite{Klein2022}, which deals with the modeling of electro-elasticity, exists. However, the application of such NN-based models to MAPs is very sensible, since they exhibit an extremely complex material response that cannot always be represented with the highest quality using the currently available conventional macroscopic models. Thus, within this contribution, we present several NN-based macroscopic models for \emph{isotropic MAPs}, which are characterized by a consequent enrichment with physical knowledge. By using a set of problem specific \emph{invariants} as input, an \emph{energy functional} as the output and by adding several non-trainable expressions to the overall total potential, the model fulfills multiple \emph{physical principles} by construction. These are thermodynamic consistency, compatibility with the balance of angular momentum, objectivity, material symmetry, as well as zero energy, stress-free and non-magnetized unloaded configuration.
Following \cite{Klein2022,Linden2023}, we denote the developed models as \emph{physics-augmented neural networks (PANNs)}. Three PANN models with different levels of rigor with respect to \emph{polyconvexity} and \emph{growth condition} are presented. Thereby, we follow an extended polyconvexity theory according to \v{S}ilhav\'y~\cite{Silhavy2018,Silhavy2019}. 
First, polyconvexity, which is a global concept and thus must hold for all states of the domain of definition, is enforced via input convex neural networks (ICNNs), i.e., by construction.
Afterwards, we formulate a \emph{relaxed local version of the polyconvexity} and fulfill it in a \emph{weak sense} by adding a tailored
loss term. Furthermore, as an alternative, a loss term to enforce the weaker requirement of strong ellipticity locally is
proposed, which can be favorable to obtain a better trade-off between compatibility with data and physical constraints.
The models are calibrated by using a comprehensive database generated with a computational homogenization approach, where \emph{compressible} and \emph{quasi-incompressible MAPs} are considered. Furthermore, the performance of the PANNs is compared to the conventional model by Gebhart~and~Wallmersperger~\cite{Gebhart2022a} which serves as a benchmark.
To reduce the computational cost for the data generation, relevant states are identified in advance by an \emph{invariant-based sampling} method and 2D \emph{statistical volume elements (SVEs)} are used for the simulations. 

The organization of the paper is as follows: In Sect.~\ref{sec:magneto-mechanics}, the
underlying equations of finite strain magneto-mechanics, basic principles of magneto-hyperelasticity and a scale transition scheme are
given. After this, microscale constitutive models for particles and matrix are formulated in Sect.~\ref{sec:micromodels}, which is followed by the summary of a macroscopic model from the literature in Sect.~\ref{sec:Geb}. The proposed NN-based models and details on the training are presented in Sect.~\ref{sec:macro_NN}. The introduced models are exemplarily
calibrated to data generated with a computational homogenization approach in Sect.~\ref{sec:examples}. After a discussion of the results, the paper is closed by concluding
remarks and an outlook to necessary future work in Sect.~\ref{sec:conc}.

\paragraph{Notation}
Within this work, tensors of rank one and two are given by boldface italic letters, i.e., $\ve A, \ve B \in \Ln_1$ or $\te C, \te D \in \Ln_2$, where $\Ln_n$ denotes the space of tensors with rank $n\in \N$ with $\N$ being the set of natural numbers without zero.
Tensors with rank three and four are marked by boldface upright and blackboard symbols, i.e., $\ttte E \in \Ln_3$
and $\tttte F \in \Ln_4$.
Single and double contractions of two tensors are given by $\ve C \cdot \te D = C_{kl} D_{li} \ve e_k \otimes \ve e_i$ and $\te C:\te D=C_{kl}D_{lk}$, respectively. Therein, $\ve e_k\in \Ln_1$ and $\otimes$ denote a Cartesian basis vector and the dyadic product, where the Einstein summation convention is used. The cross product of two rank one tensors is given by $\ve A \times \ve B = e_{ijk} A_j B_k \ve e_i$, with $e_{ijk}$ being the antisymmetric Levi-Civita symbol.
Transpose and inverse of a second order tensor $\te C$ are given by $\te C^T$ and $\te C^{-1}$, respectively.
Additionally, $\tr \te C$, $\det \te C$, $\cof \te C := \det(\te C) \te C^{-T}$, $\sym \te C$ and $\skw \te C$ are used to indicate trace, determinant, cofactor as well as symmetric and skew-symmetric parts, respectively.
The space of unit vectors is given by $\Vn:=\{\ve n \in \Ln_1 \, | \, \ve n\cdot \ve n = 1\}$.
The set \mbox{$\Sym:=\left\{\te A \in \Ln_2 \, |\, \te A = \te A^T\right\}$} signifies the space of symmetric second order tensors.
Furthermore, the orthogonal group and special orthogonal group in the Euclidean vector space $\R^3$ are given by $\Othree:=\left\{\te A \in \Ln_2\,|\,\te A^T \cdot \te A = \one\, , \det \te A = \pm 1 \right\}$ and $\SO:=\left\{\te A \in \Ln_2\,|\,\te A^T \cdot \te A = \one,\,\det \te A = 1\right\}$, respectively, while $\GL:=\left\{\te A \in \Ln_2\,|\,\det \te A > 0\right\}$ is the set of invertible second order tensors with positive determinant. Therein, $\one:=\delta_{ij}\ve e_i \otimes \ve e_j\in \Ln_2$ is the second order identity tensor, where $\delta_{ij}$ denotes the Kronecker delta. Norms of rank one and two tensors are given by $|\ve A| := \sqrt{A_iA_i}$ and $\|\te C\| := \sqrt{C_{ij}C_{ij}}$, respectively.

Nabla operators with respect to current configuration $\B$ and reference configuration $\B_0$ are given by $\nabla$ and $\nablaX$, respectively. Using the definitions given above, the divergence of a second order tensor follows to $\nabla \cdot \te C = C_{kl,k} \ve e_l$ or $\nabla \cdot \te C^T = C_{kl,l} \ve e_k$.
Furthermore, the jump of tensor quantities across material surfaces of discontinuity $\Sj$ and $\Sj_0$, with unit normal vectors $\ve n \in \Vn$ and $\ve N \in \Vn$ pointing from subdomains $\B^-$ to $\B^+$ and $\B_0^-$ to $\B_0^+$, is denoted by \mbox{$\jump{(\cdot)}:=(\cdot)^+ - (\cdot)^-$}.
For reasons of readability, the arguments of functions are usually omitted within this work. However, energy functions are given with their arguments to show the dependencies, except when derivatives are written. Furthermore, in the following the symbol of a function is identical with the symbol of the function value itself.

\section{Finite strain magneto-mechanics}\label{sec:magneto-mechanics}

\begin{figure}
	\includegraphics{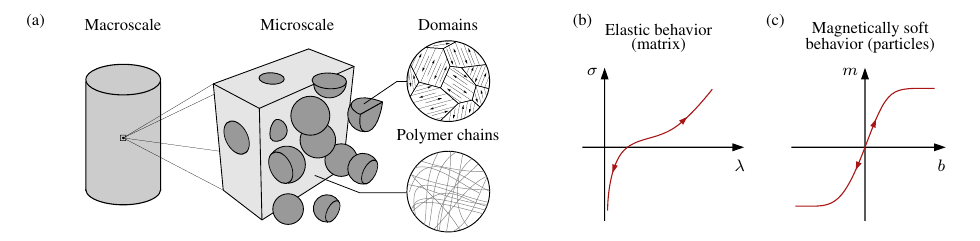}
	\caption{Magneto-active polymers: (a) structure of the different scales, as well as (b) and (c) typical constitutive behavior of the polymer matrix and the magnetically soft filler particles.}
	\label{fig:MAPs}
\end{figure}

In this section, we introduce the basic principles and equations of \emph{coupled magneto-mechanical BVPs}, including kinematics, Maxwell and balance equations, as well as general magneto-elastic constitutive relations and requirements on it. All of these
equations are valid on the microscopic as well as the macroscopic scales of the considered MAPs, see Fig.~\ref{fig:MAPs}(a). Additionally, to link the two scales, we summarize an appropriate homogenization scheme according to Chatzigeorgiou~et~al.~\cite{Chatzigeorgiou2014}.

\subsection{Kinematics}
Let us consider the motion of a body whose reference configuration at time $t_0 \in \R_{\ge 0}$ is given by \mbox{$\mathcal{B}_0 \subset \R^3$}. The current configuration of the body at time \mbox{$t\in \mathcal T:=\{\tau\in\R \,|\,\tau \ge t_0\}$} is defined as \mbox{$\mathcal{B} \subset \R^3$}. To describe the motion of this body, a smooth bijective mapping $\ve \varphi: \mathcal{B}_0 \times \mathcal T \to \mathcal{B}$, which links material points $\ve X\in\mathcal{B}_0$ to $\ve x=\ve \varphi\left(\ve X, t\right) \in \mathcal{B}$, is introduced.
Displacement $\ve u \in \Ln_1$ and velocity $\ve v \in \Ln_1$ of each material point are given by $\ve u(\ve X, t) := \ve \varphi(\ve X, t) - \ve X$ and $\ve v := \dot{\ve u}$, with $\dot{(\cdot)}$ being the material time derivative. The deformation gradient and its determinant are defined as $\te F := (\nablaX \ve \varphi)^T \in \GL$ and \mbox{$J:=\det \te F \in \R_{>0}$}. Thus, the compatibility condition $\nablaX\times\te F^T=\zero$ with $\jump{\te F} \times \ve N = \zero$ on the material surface of discontinuity $\Sj_0$ is automatically fulfilled. 
Finally, we define the right Cauchy-Green deformation tensor \mbox{$\te C:=\te F^T \cdot \te F \in \Sym$} as a deformation measure which is free of rigid body motions and positive definite. For more details on kinematics we refer the reader to the textbooks of Haupt~\cite{Haupt2000} or Holzapfel~\cite{Holzapfel2000}.

\subsection{Maxwell and balance equations}\label{subsec:MaxBal}
In the stationary case and for vanishing current densities, the \emph{Maxwell equations} with corresponding jump conditions are given by
\begin{linenomath*}
	\begin{alignat}{5}
		\nabla \cdot \ve b   &= 0 &&\;\text{with}\; &\jump{\ve b} \cdot \ve n  &= 0 &&\;\text{on }\Sj &&\; \text{and} \label{eq:GaussEuler}\\
		\nabla \times \ve h &= \zero  &&\;\text{with}\; &\jump{\ve h} \times \ve n &= \zero &&\;\text{on }\Sj &&\; , \label{eq:AmpereEuler}
	\end{alignat}
\end{linenomath*}
see \cite{deGroot1972,Kalina2020}.
Therein, $\ve b \in \Ln_1$, $\ve h \in \Ln_1$, and $\ve n \in \Vn$ are the magnetic induction, the magnetic field, and a unit normal vector on the surface of discontinuity $\mathcal S$.
The quantities $\ve b$ and $\ve h$ are linked by the permeability of free space $\mu_0 = 4\pi\cdot \SI{d-7}{\tesla\meter\ampere}^{-1}$ and the magnetization $\ve m \in \Ln_1$  via the equation 
\begin{linenomath*}
	\begin{equation}
		\ve b = \mu_0 (\ve h + \ve m) \; . \label{eq:linking}
	\end{equation}
\end{linenomath*}

As a result of the present magnetic fields, \emph{ponderomotive force}, \emph{moment} and \emph{power densities} occur in addition to purely mechanical problems and have bo be taken into account in the corresponding \emph{balance principles}.
The specific form of these densities is not uniquely defined and depends on the chosen
field-matter interaction theory \cite{deGroot1972,Pao1978,Hutter2006}, where the
statistical model by de~Groot~and~Suttorp~\cite{deGroot1972} is chosen here.
We note again that only \emph{quasi-stationary} processes according to Eqs.~\eqref{eq:GaussEuler} and \eqref{eq:AmpereEuler}, i.e., those with sufficiently slow change of magnetic fields, non-polarizable matter as well as areas without free charge and current densities, are considered. Thus, the ponderomotive force, moment and power densities per current volume are $\ve f^\text{pon} := \ve m \cdot \nabla \ve b$, $\ve c^\text{pon}:=\ve m\times \ve b$, and $p^\text{pon}:= \ve f^\text{pon} \cdot \ve v - \ve m \cdot \dot{\ve b}$.
In the following, a \emph{ponderomotive stress tensor} $\te \sigma^\text{pon}\in\Ln_2$ with $\nabla \cdot (\te \sigma^\text{pon})^T = \ve f^\text{pon}$ and $\skw(\te \sigma^\text{pon}) = -\frac{1}{2} \ttte e \cdot \ve c^\text{pon}$ is used, where $\ttte e:= e_{ijk} \ve e_i \otimes \ve e_j \otimes \ve e_k \in \Ln_3$. The \emph{non-symmetric ponderomotive stress tensor} thus is given by
\begin{linenomath*}
	\begin{equation}
		\te \sigma^\text{pon} = \te \sigma^\text{Max} + (\ve b \cdot \ve m) \one - \ve m \otimes \ve b \in \Ln_2 \text{ with } \te \sigma^\text{Max}:= \frac{1}{\mu_0} \left(\ve b \otimes \ve b - \frac{1}{2}|\ve b|^2 \one \right) \in \Sym \; , \label{eq:pon}
	\end{equation}
\end{linenomath*}
see \cite{deGroot1972,Kalina2020}. Note that the symmetric Maxwell stress tensor $\te \sigma^\text{Max}$ is also present in the vacuum.

Accounting for conservation of mass given by $\varrho_0 = J \varrho$ with $\varrho_0$ and $\varrho$ denoting the mass densities of reference and
current configuration and using the \emph{total stress tensor} $\te \sigma^\text{tot}:=\te \sigma + \te \sigma^\text{pon} \in \Sym$, the balances of linear and angular momentum  are given by
\begin{linenomath*}
	\begin{align}
		\nabla \cdot \te \sigma^\text{tot} &= \zero \; \text{with} \; \jump{\te \sigma^\text{tot}} \cdot \ve n = \zero \; \text{on} \; \mathcal S \; \text{and} \label{eq:linmom} \\
		\skw(\te \sigma + \ve b \otimes \ve m) &= \skw(\te \sigma^\text{tot}) = \zero  \; . \label{eq:angmom}
	\end{align}
\end{linenomath*}
Thus, it becomes apparent that the \emph{mechanical stress tensor} $\te \sigma \in \Ln_2$ itself is \emph{non-symmetric} in the general case. This only applies to the total stress tensor $\te \sigma^\text{tot} \in \Sym$. The \emph{Clausius-Duhem ineqaulity}, which follows as a consequence of the second law of thermodynamics and by accounting for the previously introduced Maxwell and balance equations as well as the balances of energy and entropy, is given by
\begin{linenomath*}
	\begin{equation}
		- J^{-1} \dot W + \left(\te \sigma + \frac{1}{2\mu_0} |\ve b|^2 \one \right) : (\dot{\te F} \cdot \te F^{-1})^T + \ve h \cdot \dot{\ve b} \ge 0 \label{eq:CDU}
	\end{equation}
\end{linenomath*}
if thermal
effects are neglected. In the equation above, $W$ is a scalar valued \emph{total free Helmholtz energy density} with respect to $\B_0$. 

By using the pull-back operations $\ve B:= \ve b \cdot \cof \te F$, $\ve H := \ve h \cdot \te F$ and $\ve M := \ve m \cdot \te F$, $\te P^\text{tot} := \tg \upsigma^\text{tot} \cdot \cof \te F$  as well as $\ve N = \ve n \cdot \te F / |\ve n \cdot \te F|$, we get the \emph{Lagrangian forms} of Eqs.~\eqref{eq:GaussEuler}--\eqref{eq:linking} as well as \eqref{eq:linmom}--\eqref{eq:CDU}:
\begin{linenomath*}
	\begin{alignat}{5}
		\nablaX \cdot \ve B   &= 0 &&\;\text{with}\; &\jump{\ve B} \cdot \ve N  &= 0 &&\;\text{on }\Sj_0 &&\; \text{,} \label{eq:GaussLag}\\
		\nablaX \times \ve H &= \zero  &&\;\text{with}\; &\jump{\ve H} \times \ve N &= \zero &&\;\text{on }\Sj_0 &&\; , \label{eq:AmpereLag} \\
		\mu_0 J \te C^{-1} \cdot (\ve H + \ve M) &= \ve B  &&\;, & & && &&  \label{eq:linkingLag}\\
		\nablaX \cdot (\te P^\text{tot})^T &= \zero  &&\;\text{with}\; &\jump{\te P^\text{tot}} \cdot \ve N &= \zero &&\;\text{on }\Sj_0 &&\; , \label{eq:linmomLag}
		\\
		\skw(\te P^\text{tot} \cdot \te F^T) &= \zero  &&\;\text{and} & & && &&  \label{eq:angmomLag}\\
		- \dot{W} + \te P^\text{tot} : \dot{\te F}^T + \ve H \cdot \dot{\ve B} &\ge 0 &&\;\text{,} & & && &&  \label{eq:CDULag}
	\end{alignat}
\end{linenomath*}
see \cite{Kalina2020}. The quantities $\ve B$, $\ve H$, $\ve M$ as well as $\te P^\text{tot}$ are denoted as Lagrangian magnetic induction, Lagrangian magnetic field, Lagrangian magnetization as well as first Piola-Kirchhoff stress tensor. 

To solve the Maxwell equations \eqref{eq:GaussEuler}, \eqref{eq:AmpereEuler} or \eqref{eq:GaussLag}, \eqref{eq:AmpereLag}, it is favorable to use a magnetic potential. In this work, the \emph{magnetic vector potential} $\ve A: \mathcal{B}_0 \times \mathcal T \to \Ln_1, (\ve X, t) \mapsto \ve A(\ve X,t)$ with $\ve B =: \nablaX \times \ve A$ and $\jump{\ve A} \times \ve N = \zero$ on $\Sj_0$ is used. Thus, Eq.~\eqref{eq:GaussLag} is fulfilled automatically. For reasons of uniqueness, the Coulomb gauge, i.e., $\nablaX \cdot \ve A  = 0$ in $\mathcal B_0$ and $\jump{\ve A} \cdot \ve N = 0$ on $\mathcal S_0$, is additionally used. Note that the Coulomb gauge is automatically fulfilled in the 2D case, cf. \ref{app:2D}.

\subsection{Magneto-elasticity}\label{subsec:magnetoElasticity}
\subsubsection{Constitutive equations of magneto-elasticity and general requirements}
In this work, we restrict ourselves to the consideration of reversible magneto-mechanical processes, i.e., we describe elastic and magnetically soft materials \cite{Jiles2016}, which we call \emph{magneto-elastic} in the following. The aim of constitutive modeling for this class is to relate deformation gradient and magnetic induction to total stress and magnetic field induced at a material point. In \emph{magneto-hyperelasticity}, the mapping between these quantities is not defined directly, but via the potential
\begin{equation}
	\psi: \GL \times \Ln_1  \to \R, \; (\te F, \ve B) \mapsto \psi(\te F, \ve B) \; ,
\end{equation}
which corresponds to the already introduced total Helmholtz free energy density $W$ \cite{Dorfmann2004a}.

By evaluating Ineq.~\eqref{eq:CDULag} according to the concept of Coleman~and~Noll~\cite{Coleman1963}, the Lagrangian magnetic field $\ve H$ and the total first Piola-Kirchhoff stress tensor $\te P^\text{tot}$ follow from the constitutive relations
\begin{equation}
	\te P^\text{tot} = \diffp{\psi}{\te F} \text{ and } \ve H = \diffp{\psi}{\ve B} \; , \label{eq:constFB}
\end{equation}
which imply energy conservation and path-independency. The model is thus a priori \emph{thermodynamically consistent}, i.e., in accordance with the second law \cite{Kalina2020,Linden2023}.
Additional mathematical and physical considerations are usually made for hyperelastic potentials \cite{Haupt2000,Holzapfel2000}. We will briefly summarize and adapt them for magneto-hyperelasticity below, where we follow the recent work Linden~et~al.~\cite{Linden2023} on hyperelasticity. 

In order to assure compliance with the balance of angular momentum \eqref{eq:angmomLag}, $\psi(\te F,\ve B)$ has to be formulated such that the condition
\begin{equation}
	\skw\left(\diffp{\psi}{\te F} \cdot \te F^T\right) = \zero
\end{equation}
holds, where this requirement leads to the \emph{symmetry of the total Cauchy stress tensor} $\te \sigma^\text{tot}$.
According to the principle of \emph{material objectivity} in active interpretation, the behavior of the material must not change when it is subjected to an arbitrary rigid body motion. In the context of magneto-hyperelasticity this means 
\begin{equation}
	\psi(\te F, \ve B) = \psi(\te Q \cdot \te F, \ve B) 
	\; \forall \te F \in \GL, \; \ve B \in \Ln_1, \; \te Q \in \SO \; .
	\label{eq:objectivity}
\end{equation}
Furthermore, the potential should also account for the underlying anisotropy class of the material. This principle is termed \emph{material symmetry} and is given by
\begin{equation}
	\psi(\te F, \ve B) = \psi(\te F \cdot \te Q^T, \ve B \cdot \te Q^T) 
	\; \forall \te F \in \GL, \; \ve B \in \Ln_1, \; \te Q \in \G \subseteq \Othree \; .
	\label{eq:symmetry}
\end{equation}
In hyperelasticity, various coercivity conditions can also be taken into account, with the \emph{volumetric growth condition} being the most prevalent one \cite{Holzapfel2000,Linden2023}. This condition accounts for the finding that a material cannot be extended to an infinite volume or compacted to a volume of zero \cite{Holzapfel2000,Klein2021} and thus the energy has to increase towards infinity. In the same manner, a \emph{magnetic growth condition} can be formulated, i.e., the energy should also tend to infinity if $|\ve B|$ or $|\ve b|$ are close to infinity, which of course is not of particular relevance due to technical limitations. The described conditions are given by
\begin{align}
	\psi(\te F, \ve B) \rightarrow \infty \;
	\text{as}\; \big(J \rightarrow 0^+ \;\lor\; J\rightarrow\infty\big) \quad 
	\text{and} \quad
	\psi(\te F, \ve B) \rightarrow \infty \; \text{as}\; |\ve B| \to \infty \; .
	\label{eq:growthConditions}
\end{align}
The \emph{magnetic saturation behavior}, i.e.,
\begin{equation}
	\lim\limits_{|\ve b|\to\infty}|\ve m| = J^{-1} m_\text{s} \; \text{with} \; \ve m = \frac{1}{\mu_0} \ve b - \diffp{\psi}{\ve B} \cdot \te F^{-1}\; ,
	\label{eq:saturation}
\end{equation}
is another important property which should be represented by the potential $\psi(\te F, \ve B)$, where $m_\text{s}$ is the material's saturation magnetization for $\te F = \one$.
It results from the fact that a typical ferromagnetic bulk material consist of a polycrystalline substructure with a stochastic distribution of magnetic domains, see Fig.~\ref{fig:MAPs}. Since the magnitude of the domains' magnetic moments does not change \cite{Jiles2016,Keip2019}, the overall magnetization reaches a limit if all magnetic moments point into the direction of the applied field $\ve b$. The saturation magnetization of composite materials like MAPs results from the volume fraction of the embedded particles and their characteristic saturation value.
Besides the conditions described above, the \emph{undeformed} and/or \emph{magnetic induction-free} configuration should have zero energy, zero total stress and zero magnetic field, i.e.,  
\begin{align} \label{eq:norm}
	\psi(\te F = \one, \ve B = \zero)=0 \; \text{,} \; \te P^\text{tot}(\te F = \one, \ve B= \zero) = \zero \; \text{and} \; \ve H(\te F, \ve B = \zero) = \zero \; ,
\end{align}
should hold. Finally, it is expected that the stored energy always increases, i.e., it is \emph{non-negative}, if a deformation $\te F$ is applied or the material is exposed to a magnetic loading by $\ve B$, thus
\begin{align}
	\psi(\te F, \ve B) \ge 0 \; \forall \, \te F \in \GL, \, \ve B \in \Ln_1 \; . 
	\label{eq:posEn}
\end{align}

The introduced constitutive equations are given in the $\te F$-$\ve B$ setting, which is a pure energy formulation. The solution of a variational principle thus leads to a minimization problem. If further convexity criteria are fulfilled, the existence of a minimizer is guaranteed \cite{Schroder2010,Ebbing2010}, see Sects.~\ref{subsec:ell} and \ref{subsec:poly}. 
Similarly, the formulation in an $\te F$-$\ve H$ setting is also very common for the modeling of MAPs, see \ref{app:FH}.

To fulfill numerous of the introduced principles by construction, it is favorable to formulate the potential in terms of \emph{invariants} from the right Cauchy-Green deformation tensor $\te C$, the Lagrangian magnetic induction $\ve B$ and, for anisotropy, a set of structural tensors, i.e., $\psi(\I)$ with $\I:=(I_1,I_2,\ldots,I_m)\in\R^m$. For the \emph{isotropic} case, a complete and irreducible set%
\footnote{Following Ebbing~\cite{Ebbing2010} and Linden~et~al.\cite{Linden2023}, \emph{complete} means that any invariant quantity, e.g., the free energy, can be expressed as a function of the elements of $\I$ and \emph{irreducible} means that no element $I_\alpha$ of $\I$ can be expressed as a polynomial in the remaining elements of the set $\I$.}
is given by
\begin{align}
	I_1 := \tr \te C \; , \; I_2 := \tr(\cof \te C) \; , \; I_3 := \det \te C \; ,\; I_4 := |\ve B|^2
	\; , \; I_5 := \te C :(\ve B \otimes \ve B) \; \text{and} \; 
	I_6 := \te C^2 :(\ve B \otimes \ve B) \; ,
	\label{eq:invariants}
\end{align}
see \cite{Dorfmann2004a,Gebhart2022}. Thus, the constitutive relations according to Eq.~\eqref{eq:constFB} follow then to
\begin{align}
	\te P^\text{tot} = \sum_{\alpha=1}^6 \diffp{\psi}{I_\alpha}\diffp{I_\alpha}{\te F}
	\text{ and }
	\ve H = \sum_{\alpha=1}^6 \diffp{\psi}{I_\alpha}\diffp{I_\alpha}{\te B} \; .
	\label{eq:constInv}
\end{align}
With that, \emph{thermodynamic consistency}, \emph{symmetry of the total Cauchy stress}, \emph{material objectivity} and \emph{material symmetry} are automatically fulfilled \cite{Linden2023}.

Besides the conditions described above, there exist further principles in order to ensure physically reasonable behavior, where \emph{strong ellipticity} and \emph{polyconvexity} should be underlined here. We will summarize these concepts in the following.

\subsubsection{Rank-one convexity and strong ellipticity}\label{subsec:ell}

Another important condition for constitutive models is the \emph{strong ellipticity}, which ensures traveling waves with non-negative real-valued wave speeds \cite{Ebbing2010,Schroder2010,Marsden1984}. For reasons of brevity, we will only use the term \emph{ellipticity} in the following.
Global ellipticity of twice differentiable and smooth energy functions as considered in this work is equivalent to \emph{rank-one convexity}, where \emph{global} means that a condition or a property holds for the entire definition domain. According to common terminology, we call a condition $C$ that is satisfied not only with $C\ge 0$ but with $C>0$ as \emph{strict}.

\paragraph{Global ellipticity}
In this work, we follow an extension of the concept of ellipticity to magneto-elasticity according to Destrade~and~Ogden~\cite{Destrade2011} which is also used in \cite{Polukhov2020,Rudykh2013,Galipeau2013}.
Accordingly, \emph{(strict) global ellipticity} of a magneto-mechanical system, i.e., for all possible states, requires 
\begin{align}
	\forall \te F \in\GL, \ve B\in\Ln_1: \; \forall \ve a \in \Vn, \; (\ve N,\ve V) \in \mathcal E : \;  \ve a \cdot \te{\mathit\Gamma}(\ve N,\ve V) \cdot \ve a
	= \varrho_0 c^2 \,
	(>) \ge 0  \; , \label{eq:ellipticity}
\end{align}
where $\mathcal E:= \{\ve N, \ve V\in\Vn \, |\, \ve N \cdot \ve V = 0\}$, $c\in\R$ denotes the wave speed and $\te{\mathit\Gamma}(\ve N,\ve V) \in \Sym$ is the \emph{generalized acoustic tensor} which is defined as 
\begin{align}
	\te{\mathit\Gamma}(\ve N, \ve V) := \te Q(\ve N) - \frac{[\te R(\ve N) \cdot \ve V]\otimes [\te R(\ve N) \cdot \ve V]}{\te K :(\ve V \otimes \ve V)} \; , (\ve N,\ve V) \in \mathcal E \; ,\label{eq:acousticTensor}
\end{align}
with $\te Q(\ve N) = A_{iJkL} N_J N_L \ve e_i \otimes \ve e_k$, $\te R(\ve N) = G_{iJK} N_J \ve e_i \otimes \ve e_K$. The introduced tangent moduli are given by
\begin{equation}
	\tttte A := \diffp{{}^2\psi}{\te F \partial \te F} \in \Ln_4 \; , \te K := \diffp{{}^2 \psi}{\ve B \partial \ve B} \in \Sym \text{ and }
	\ttte G := \diffp{{}^2 \psi}{\te F \partial \ve B} \in \Ln_3 \; .
\end{equation}
The stated condition~\eqref{eq:ellipticity} and the acoustic tensor~\eqref{eq:acousticTensor} are briefly derived in \ref{app:ell}.
To ensure Ineq.~\eqref{eq:ellipticity}, positive (semi-)definiteness of $\te{\mathit\Gamma}$ is required for all admissible states and all possible directions $\ve N$ and $\ve V$ ensuring $\ve N\cdot \ve V = 0$, which follows from the incremental form of Gauss's law \eqref{eq:GaussLag}, see \ref{app:ell}.

\paragraph{Local ellipticity}
A weaker requirement is to ensure \emph{ellipticity locally}, i.e., only in the direct neighborhood of an $\te F$-$\ve B$ state or within an \emph{elliptic domain} $\mathscr{E\!l\!l} \subset \GL\times \Ln_1$, cf. \cite{Ghiba2015}. The local version of Ineq.~\eqref{eq:ellipticity} for a state $(\te F,\ve B) \in \GL \times \Ln_1$ thus reads
\begin{align}
	\forall \ve a \in \Vn, \; (\ve N,\ve V) \in \mathcal E : \;  \ve a \cdot \te{\mathit\Gamma}(\ve N,\ve V) \cdot \ve a
	= \varrho_0 c^2 \,
	(>) \ge 0  \; . \label{eq:ellipticityLocal}
\end{align}
To check whether the above local condition for a state is satisfied, the positive (semi-)definiteness of $\te{\mathit\Gamma}$ usually has to be checked numerically.
This can be done by using \emph{Sylvester’s criterion}, i.e., for positive semi-definiteness the minors of $\te{\mathit\Gamma}$ have to fulfill
\begin{align}
	\gamma_1^i:=\Gamma_{i(i)} \ge 0 \,  i \in\{1,2,3\} \; \wedge \;
	\gamma_2^{ij}:=\Gamma_{i(i)}\Gamma_{j(j)} - \Gamma_{ij}\Gamma_{(i)(j)} \ge 0
	, \;  j \ne i \in\{1,2,3\} \; \wedge \; \gamma_3:=\det \te{\mathit\Gamma} \ge 0  
	\label{eq:minors}
\end{align}
for all $(\ve N,\ve V)\in\mathcal E$, see \cite{Vlassis2021c}. In the conditions~\eqref{eq:minors}, it is not summed over bracketed indices. For positive definiteness, it is sufficient to check the main minors, which have to be greater to zero in that case \cite{Schroder2010}:
\begin{align}
	\left(\gamma_1 = \gamma_1^1 > 0 \; \wedge \; 
	\gamma_2 = \gamma_2^{12} > 0 \; \wedge \;
	\gamma_3 >0\right) \; \forall (\ve{N},\ve{V})\in\mathcal E \; .
	\label{eq:ell_cond}
\end{align}
Thus, a model is locally elliptic for a given $\te F$-$\ve B$-state if condition~\eqref{eq:minors} holds or strictly local elliptic if condition~\eqref{eq:ell_cond} holds. A 2D version of the statement~\eqref{eq:ell_cond} is given in \ref{app:2D}.

\subsubsection{Polyconvexity}\label{subsec:poly}

Since it is complicated to formulate potentials such that they fulfill the \emph{global ellipticity} / \emph{rank-one convexity} condition by construction, the concept of \emph{polyconvexity}, originally introduced by Ball~\cite{Ball1976,Ball1977}, is quite common in constitutive modeling of elastic materials, cf. \cite{Linden2023,Klein2021,Ebbing2010,Schroder2010,Schroder2003}. 

\paragraph{Polyconvexity as a global statement}
Polyconvexity is a global property and is sufficient for quasiconvexity, which in turn guarantees rank-one convexity and thus finally guarantees that the Euler equations of the associated functional are elliptic \cite{Schroder2010,Ebbing2010}.
In addition, polyconvexity guarantees sequential weak lower semicontinuity (s.w.l.s.). Together with the coercivity it is sufficient for the existence of minimizers \cite{Ebbing2010,Schroder2010}.
The converse of the implications mentioned above is not true.

An extension of the polyconvexity theory to \emph{magneto-elasticity} has been proposed by Ortigosa~and~Gil~\cite{Ortigosa2016a} and also by \v{S}ilhav\'y~\cite{Silhavy2018,Silhavy2019}. 
Accordingly, the potential $\psi(\te F, \ve B)$ is polyconvex if and only if there exists a
function such that
\begin{equation}
	\psi(\te F, \ve B) = \mathcal P(\gt I) \text{ with } \gt I =  \left(\te F, \cof \te F, \det \te F, \te F \cdot \ve B, \ve B\right) \in \mathcal R := \Ln_2 \times \Ln_2 \times \R \times \Ln_1 \times \Ln_1 \; ,
\end{equation}
where $\mathcal P(\gt I)$ is convex with respect to the set $\gt I$ for all possible states $\gt I\in \mathcal R$.
In order to ensure that the potential is polyconvex, we thus have to guarantee that the \emph{Hessian} $\gt H$ is positive semi-definite, i.e., 
\begin{equation}  
	\Delta \gt I \bigcdot \gt H  \bigcdot \Delta \gt I \ge 0 \; \forall \Delta \gt I 
	\text{ with }
	\gt H := \diffp{{}^2 \mathcal P}{\gt I\partial \gt I}  \label{eq:polyconvex}
\end{equation}
holds for all states $\gt I \in \mathcal R$, where $\bigcdot$ is a generalized tensor product, e.g., it gives $\gt I \bigcdot \gt I = \|\te F\|^2 + \|\cof \te F\|^2 + \det^2 \te F + |\te F\cdot \ve B|^2 + |\ve B|^2$.

To proof for the polyconvexity of a given energy expression depending on \emph{invariants} $I_\alpha$, i.e., $\psi(I_1,I_2,\ldots,I_m)$, we have to describe these quantities in terms of the set $\gt I$, i.e., $I_\alpha(\ve{\mathfrak I})$\footnote{The invariants given in Eq.~\eqref{eq:invariants} can be represented as follows by the polyconvex set $\gt I$:
	\begin{align*}
		I_1 = \|\te F\|^2 \; , \; I_2 = \|\cof \te F\|^2 \; , \; I_3 = J^2 \; , \; I_4 = |\ve B|^2 \; , \; I_5 = |\te F \cdot \ve B|^2 \; \text{and} \; I_6 = |\te F^T \cdot (\te F \cdot \ve B)|^2 \; .
\end{align*}}, which allows us to calculate
\begin{equation}
	\diffp{{}^2 \mathcal P}{\ve{\mathfrak I}\partial \ve{\mathfrak I}} = \sum_{\alpha, \beta}\diffp{{}^2\mathcal \psi}{I_\alpha\partial I_\beta}\diffp{I_\alpha}{\ve{\mathfrak I}} \otimes \diffp{I_\beta}{\ve{\mathfrak I}} + 
	\sum_{\alpha}\diffp{\mathcal \psi}{I_\alpha}\diffp{{}^2I_\alpha}{\ve{\mathfrak I}\partial \ve{\mathfrak I}} \; .
	\label{eq:compHessian}
\end{equation}
Note that the set of invariants must be chosen such that none of its elements $I_\alpha$ violates the polyconvexity condition since a model has to satisfy it for all possible states $\gt I \in \mathcal R$, cf. \cite{Linden2023,Klein2021,Klein2022}. Here, the invariant $I_6$ given in Eq.~\eqref{eq:invariants} is not polyconvex \cite{Silhavy2019}.

\begin{rmk}
	Note the following for multiscale problems. There, polyconvex microscopic energies do not necessarily guarantee macroscopic quasiconvexity and thus rank-one convexity, see \cite{Polukhov2020,Abeyaratne1984}. The terms microscopic and macroscopic are defined in more detail in the next subsection. Even a loss of ellipticity is possible for certain areas $(\te F, \ve B)$, see \cite{Polukhov2020, Rudykh2013}.
\end{rmk}

\paragraph{Concept of local polyconvexity}
Now, we formulate a \emph{relaxed local version of the polyconvexity} introduced above. Accordingly, we say $\psi(\te F,\ve B)$ is locally polyconvex in the neighborhood of a state $(\te F, \ve B)\in\GL\times\Ln_1$ if the following holds:
\begin{align}
	\exists \, r_{F}, r_B > 0: \; \forall \|\Delta \te F\| < r_F, |\Delta \ve B| < r_B: \; 
	&\psi(\ve{\mathcal S}) = 
	\mathcal P (\gt I(\ve{\mathcal S})) \text{ with } \ve{\mathcal S}:=(\te F + \Delta \te F, \ve B + \Delta \ve B) \; ,\nonumber \\
	&\Delta \gt I \bigcdot \diffp{{}^2 \mathcal P}{\ve{\mathfrak I}\partial \ve{\mathfrak I}} \Big|_{\gt I(\ve{\mathcal S})}   \bigcdot \Delta \gt I \ge 0 \; \forall \Delta \gt I \; . \label{eq:localPoly}
\end{align}
If the functional $\mathcal P(\gt I)$ is continuously differentiable twice, such a region always exists since we allow arbitrarily small values for $r_F$ and $r_B$. Condition \eqref{eq:localPoly} is less restrictive then polyconvexity in its original form, which is a global statement. This relaxed requirement is nevertheless very useful, since a positive semi-definite Hessian in a state and its environment implies \emph{local ellipticity} as defined in Ineq.~\eqref{eq:ellipticityLocal} within this domain, i.e., for $(\te F+\Delta \te F, \ve B + \Delta \ve B) \in \mathscr{E\!l\!l} \subset \GL\times \Ln_1$. A proof of this is given in \ref{app:poly}.

To test whether the local condition~\eqref{eq:localPoly} for a state is satisfied, the positive semi-definiteness of $\gt H$ has to be checked, which is usually done numerically. To this end, one have to consider the \emph{eigenvalues} $\lambda^{\mathfrak H}_\alpha$ of the \emph{Hessian} $\gt H$:
\begin{equation}  
	\Delta \gt I \bigcdot \gt H  \bigcdot \Delta \gt I \ge 0 \; \forall \Delta \gt I 
	\; \Leftrightarrow \;
	\lambda^{\mathfrak H}_\alpha \ge 0 \; \forall \alpha \in \{1,2,\ldots,25\} \; . \label{eq:cond_loc_poly}
\end{equation}
It is thus also possible to determine a domain $\mathscr{P\!o\!l\!y}\subset \GL \times \Ln_1$ in which a model is locally polyconvex.

\subsection{Scale transition scheme}\label{subsec:hom}
Now, when considering MAPs, we want to distinguish between two different scales, the \emph{microscale} and the \emph{macroscale}. The former is characterized by a heterogeneous structure consisting of a polymeric matrix $\B_0^\text{m}\subset \R^3$ and ferromagnetic particles $\B_0^\text{p}\subset \R^3$ of characteristic length $\ell\in \R_{>0}$, where both subphases are considered as homogeneous.\footnote{This assumption is quite reasonable for MAPs. The characteristic chain length of the typically used polymers is in the nanometer range and is thus significantly smaller than the diameter of the embedded particles. These, in turn, are multidomain particles, which can thus be understood as a continuum in a good approximation, see also Fig.~\ref{fig:MAPs}(a).}
At the macroscale, we consider a body $\bar \B_0 \subset \R^3$, which is assumed as homogeneous, with characteristic length $\bar \ell \in \R_{>0}$. For the introduced lengths, the relation known as \emph{scale separation} $\bar{\ell} \gg \ell$ apply \cite{Schroder2014,Kalina2023}. To denote macroscopic quantities, we will mark them with $\bar{(\cdot)}$ in the following. At this point it should be noted again, that all equations introduced so far are valid on both scales. 

To connect microscopic and macroscopic quantities, a \emph{magneto-mechanical homogenization} scheme is applied, cf. the works \cite{Chatzigeorgiou2014,Polukhov2020,Gebhart2022,Kalina2020,Danas2017,Keip2016,Keip2017,Moreno-Mateos2022}, among others. The basic principles will be summarized in the following.
Each macroscopic point $\bve{X} \in \bar\B_0$ gets assigned properties resulting from the behavior of the microscale.	For this purpose, an \emph{RVE} with domain $\B_0^\text{RVE} \subset \R^3$ and volume $V^\text{RVE}$ of the material, which have to be large enough to capture the statistics of the microstructure\footnote{As an alternative it is also possible to use several smaller cells, so called statistical volume elements (SVEs), which build an RVE by calculating the statistical mean from the respective volume averages. This strategy will be applied later, cf. Sect.~\ref{subsec:SVE}.}, is considered to be in the vicinity of $\bve X$, cf. Fig.~\ref{fig:MAPs}(a). The effective macroscopic quantities $\bte F$, $\bte P^\text{tot}$, $\bve B$, and $\bve H$ are then identified by \emph{averaging} the field distributions within the RVE, respectively:
\begin{align}
	\bte F := \frac{1}{V^\text{RVE}}\int\limits_{\B_0^\text{RVE}} \te F \, \dx V \; , \;
	\bte P^\text{tot} := \frac{1}{V^\text{RVE}}\int\limits_{\B_0^\text{RVE}} \te P^\text{tot} \, \dx V \; , \;
	\bve B := \frac{1}{V^\text{RVE}}\int\limits_{\B_0^\text{RVE}} \ve B \, \dx V \; \text{and} \; 
	\bve H := \frac{1}{V^\text{RVE}}\int\limits_{\B_0^\text{RVE}} \ve H \, \dx V \; .
	\label{eq:average}
\end{align}
Other relevant effective tensor-valued quantities, e.g., $\bte C$, $\bte \sigma$, or $\bve m$, must be calculated from these averaged quantities using their defining equations, since it would lead to inconsistencies if these were also determined by volume averaging \cite{Schroder2014}.

\begin{figure}
	\begin{center}
		\includegraphics{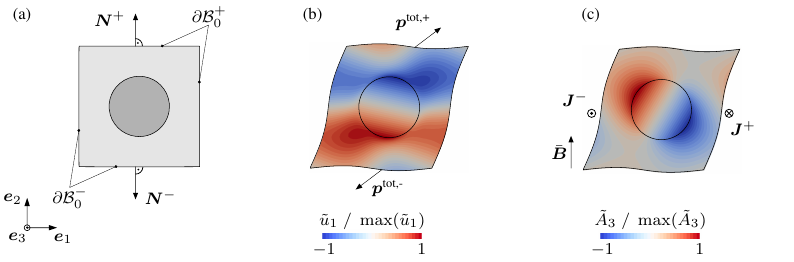}
	\end{center}
	\caption{Periodic boundary conditions on a cubic unit cell: (a) initial geometry
		with positive and negative boundary, and (b), (c) deformed geometry with fluctuation
		fields of displacement and vector potential as well as schematic representation of the
		antiperiodic fluxes $\ve p^\text{tot}$ and $\ve J$. The picture is adapted from Kalina~\cite{Kalina2021a}.}
	\label{fig:peri}
\end{figure}

Boundary conditions (BCs) for the microscopic BVP could be derived from the equivalence of  macroscopic and averaged microscopic energies, which is also known as the \emph{Hill-Mandel condition} \cite{Hill1963}, and is given by 
\begin{equation}
	\frac{1}{V^\text{RVE}}\int\limits_{\B_0^\text{RVE}} \left(\te P^\text{tot} : \dot{\te F}^T + \ve H \cdot \dot{\ve B}\right) \, \dx V = \bte P^\text{tot} : \dot{\bte F}^T + \bve H \cdot \dot{\bve B}
	\label{eq:HillMandel}
\end{equation}
in finite strain magneto-mechanics \cite{Chatzigeorgiou2014}. For the magneto-hyperelastic behavior considered here, Eq.~\eqref{eq:HillMandel} expresses the equality of the rates of the macroscopic and the averaged microscopic potentials, so that $\bar \psi$ can be calculated by volume averaging \cite{Kalina2023}. The Hill-Mandel condition is satisfied if, for example, \emph{periodic} fluctuations of displacement $\tilde{\ve u}$ and vector potential $\,\tilde{\!\ve A}$ are chosen in combination with \emph{antiperiodic} fluxes $\ve p^\text{tot}:=\te P^\text{tot} \cdot \ve N$ and $\ve J := \ve H \times \ve N$:
\begin{align}
	\ve u \in \mathcal U(\bte F) &:= \left\{\ve u \in \Ln_1 \, | \, \ve u = (\bte F - \one)\cdot \ve X + \tilde{\ve u} \text{ with } \tilde{\ve u}^+ = \tilde{\ve u}^-\right\} \; , \label{eq:peri1}\\
	\ve A \in \mathcal A(\bte B) &:= \left\{\ve A \in \Ln_1 \, | \, \ve A = \frac{1}{2} \bve B \times \ve X + \, \tilde{\!\ve A} \text{ with } \,\tilde{\!\ve A}^+ = \,\tilde{\!\ve A}^-\right\} \; , \label{eq:periA3D}\\
	\ve p^\text{tot} \in \mathcal P^\text{tot} &:= \left\{\ve p^\text{tot} \in \Ln_1 \, | \, (\ve p^\text{tot})^+ = - (\ve p^\text{tot})^-\right\} \; \text{and}\\
	\ve J \in \mathcal J &:= \left\{\ve J \in \Ln_1 \, | \, \ve J^+ = - \ve J^-\right\} \; . \label{eq:peri4}
\end{align}
In the equations above, $(\cdot)^+$ and $(\cdot)^-$ mark opposing sites of the RVE and $\ve N \in \Vn$ is a  unit normal vector on the RVE surface $\partial\B_0^{-/+}$, see Fig.~\ref{fig:peri}(a).\footnote{The set $\mathcal A(\bte B)$ differs in the 2D case, see \ref{app:2D}.} The periodic fields $\tilde{\ve u}$ and $\tilde{\!\ve A}$ are exemplarily shown for a cubic unit cell under combined magneto-mechanical loading in Fig.~\ref{fig:peri}(b),(c).
Note that there exist further BCs to fulfill the Hill-Mandel equation, cf. Zabihyan~et~al.~\cite{Zabihyan2018}. However, within this work we only use the periodic BCs defined by Eqs.~\eqref{eq:peri1}--\eqref{eq:peri4}. Within FE simulations, we realize these constraints via the concept of \emph{master nodes}, see Haasemann~et~al.~\cite{Haasemann2006}. This also allows us to compute the \emph{macroscopic tangent moduli}
\begin{align}
	\btttte A = \diffp{\bte P^\text{tot}}{\bte F} \; , \; \bte K = \diffp{\bve H}{\bve B} \; \text{and}
	\; \bttte G = \diffp{\bte P^\text{tot}}{\bve B} \label{eq:tangents}
\end{align}
corresponding to the RVE response in a straightforward manner. Thus, by using the computational homogenization approach, we get the mapping
\begin{align}
	\mathcal H: \GL \times \Ln_1 \to \R \times \Ln_2 \times \Ln_1 \times \Ln_4 \times \Ln_2 \times \Ln_3, \; (\bte F, \bve B) \mapsto (\bar \psi, \bte P^\text{tot}, \bve H, \btttte A, \bte K, \bttte G) \; .
\end{align}
All FE simulations shown in this work have been done by using an in-house code based on Matlab.

\section{Microscale constitutive models}\label{sec:micromodels}

In the following, we define constitutive models to describe the typical behavior of MAP components, i.e., elastic \emph{polymer matrix} materials and ferromagnetic \emph{metal particles}, see Fig.~\ref{fig:MAPs}(b) and (c). As mentioned before, we consider both constituents as approximately homogeneous continua, respectively. 

\subsection{Matrix}
The stochastic arrangement of the polymer chains within the matrix results in an \emph{isotropic} behavior of this phase at the microscale level. In addition, many polymers can be described as elastic in a very good approximation \cite{Kalina2020a}. From a magnetic point of view, they approximately behave like vacuum, i.e., the permeability is given by $\mu = \mu_0$. Thus, the energy of the polymer matrix divides into an \emph{elastic part} and a \emph{vacuum part}:
\begin{align}
	\psi^\text{m}(\te F, \ve B) := \psi^\text{el}(\te F) + \psi^\text{vac}(\te F, \ve B) \; \text{with} \;
	\psi^\text{vac}(\te F, \ve B) := \frac{1}{2\mu_0} J^{-1} |\te F \cdot \ve B|^2 \; .
	\label{eq:matrix}
\end{align}
The derivatives of the vacuum energy with respect to $\te F$ and $\ve B$ give
\begin{equation}
	\diffp{\psi^\text{vac}}{\te F} \cdot \cof \te F^{-1} = \te \sigma^\text{Max} \text{ and }
	\diffp{\psi^\text{vac}}{\ve B} \cdot \te F^{-1} = \frac{1}{\mu_0} \ve b \; . \label{eq:vacenergy}
\end{equation}
Note that the introduced vacuum energy is independent of the symmetry group of the considered material. In addition, according to \v{S}ilhav\'y~\cite{Silhavy2019}, the vacuum invariant $I_\text{vac}:=J^{-1} |\te F \cdot \ve B|^2$ is polyconvex with respect to $\gt I$ and thus also the vacuum energy term $\psi^\text{vac}(\te F, \ve B)$.

For \emph{compressible matrix materials} like saturated gels \cite{Gebhart2022,Gebhart2022a}, we choose the two-parametric neo-Hookean model 
\begin{equation}
	\psi^\text{el}(I_1,I_3) := \frac{1}{2} \left[G (I_1 - \ln I_3 -3) + \frac{\lambda}{2}(I_3 - \ln I_3 - 1) \right]  \text{ with } G,\lambda > 0 \; , \; G = \frac{E}{2(1+\nu)} \;,\; \lambda = \frac{E \nu}{(1+\nu)(1-2\nu)}
	\label{eq:comp}
\end{equation}
according to Ciarlet~\cite{Ciarlet1988}. In the equation above $G$ and $\lambda$ denote Lam\'{e} parameters and $E$ and $\nu$ are Young's modulus and Poisson's ratio, respectively. The elastic energy part given in Eq.~\eqref{eq:comp} is also polyconvex \cite{Linden2023}.

For \emph{quasi-incompressible matrix} materials, e.g., silicone elastomers \cite{Kalina2020a,Danas2012} which are most common for MREs, we choose a two-parametric neo-Hookean model based on the \emph{Flory split} $\te F = \te F^\text{vol} \cdot \te F^\text{iso}$ with $\te F^\text{iso} = J^{-1/3} \te F$, $\det \te F^\text{iso} \equiv 1$. The elastic energy is thus given by 
\begin{equation}
	\psi^\text{el}(I_1^\text{iso},I_3) := G \left(I_1^\text{iso}-3\right) + \frac{K}{4}\left(I_3 - \ln I_3 -1 \right)
	\text{ with } G,K > 0 \; , \; G = \frac{E}{2(1+\nu)} \;,\; K = \frac{2}{3}G\frac{1+\nu}{1-2\nu} \; ,
	\label{eq:incomp}
\end{equation}
where the introduced elastic constant $K$ is the compression modulus. The  isochoric invariant $I_1^\text{iso}$ is defined as \mbox{$I_1^\text{iso}:= J^{-2/3} \|\te F\|^2$}, which is also polyconvex \cite{Hartmann2003a}. This also holds for the model given in Eq.~\eqref{eq:incomp}.

Finally, we observe that the models for the compressible and quasi-incompressible matrix materials defined by Eqs.~\eqref{eq:matrix} and \eqref{eq:comp} as well as Eqs.~\eqref{eq:matrix} and \eqref{eq:incomp} are both polyconvex with respect to $\gt I$ in total, respectively.

\subsection{Particles}
Usually, \emph{carbonyl iron particles} are used for magnetically soft MAPs \cite{Kalina2020a,Danas2012}. These particles also exhibit an approximately isotropic behavior, which is due to the stochastic arrangement of the underlying crystal and domain structure. They show ferromagnetic properties and are significantly stiffer compared to the surrounding matrix. Thus,  magneto-mechanical coupling effects in the particles are negligible and a split of the energy 
\begin{align}
	\psi^\text{p}(\te F, \ve B) := \psi^\text{el}(\te F) + \psi^\text{mag}(\te F, \ve B) + \psi^\text{vac}(\te F, \ve B)
\end{align}
into \emph{elastic}, \emph{magnetic} and \emph{vacuum} contributions is possible, see also \cite{Gebhart2022,Gebhart2022a,Polukhov2020,Kalina2020,Danas2017}. From Eqs.~\eqref{eq:linking}, \eqref{eq:constFB} and \eqref{eq:vacenergy} we directly find that the introduced magnetic energy part is associated with the magnetization as follows:
\begin{align}
	\ve m = -\diffp{\psi^\text{mag}}{\ve B} \cdot \te F^{-1} \; .
	\label{eq:magnetization}
\end{align}
To model the characteristic saturation behavior, we choose a \emph{Langevin-type} model \cite{Gebhart2022,Polukhov2020,Lefevre2020,Danas2017} defined by
\begin{align}
	\psi^\text{mag}(I_\text{vac}) := \frac{\mu_0 m_\text{s}}{3 \chi} \left[\ln\left(\frac{3\chi}{\mu_0 m_\text{s}} \sqrt{I_\text{vac}}\right) - 
	\ln\left[\sinh\left(\frac{3\chi}{\mu_0 m_\text{s}} \sqrt{I_\text{vac}}\right)\right)\right] \; ,
	\label{eq:magenergy}
\end{align}
where $m_\text{s}>0$ and $0\le\chi\le 1$ denote the particles' saturation magnetization and initial susceptibility, respectively. The second parameter is linked to the initial permeability by $\mu=\mu_0/(1-\chi)$. The elastic energy contribution $\psi^\text{el}(\te F)$ is again described by the neo-Hookean model given in Eq.~\eqref{eq:comp}. The chosen parameters for the microscale constitutive models are given in Tab.~\ref{tab:parameters_micro}.\footnote{For reasons of numerical stability, the Young's modulus of the particles is not set to the true value of $\SI{2.1e6}{\mega\pascal}$. To model the particles as approximately rigid, it is sufficient to choose $E$ four orders of magnitude larger compared to the value of the matrix phase.} In the following, we denote the composite materials resulting from the combination of one of the matrix materials with the particles as \emph{compressible} and \emph{quasi-incompressible MAPs}, respectively. 

\begin{table}
	\begin{center}
		\caption{Chosen parameters for the constitutive models of the compressible and quasi-incompressible matrix as well as the magnetizable particles.}
		\label{tab:parameters_micro}
		\begin{footnotesize}
			\begin{tabular}{lllll}
				Parameter & Symbol & Compressible matrix & Quasi-incompressible matrix & Particles\\
				\hline\hline
				Young's modulus & $E$ & $\SI{100}{\kilo\pascal}$ & $\SI{100}{\kilo\pascal}$ & $\SI{1000}{\mega\pascal}$ \\
				Poison's ratio & $\nu$ & $0.4$ & $0.49$ & $0.3$ \\
				Saturation magnetization & $m_\text{s}$ & $-$ & $-$ & $\SI{1000}{\kilo\ampere\meter}^{-1}$ \\
				Initial susceptibility & $\chi$ & $-$ & $-$ & $0.9$ \\
			\end{tabular}
		\end{footnotesize}
	\end{center}
\end{table}

Note that the magnetic energy term \eqref{eq:magenergy} itself is not polyconvex with respect to $\gt I$, which also becomes apparent from Eq.~\eqref{eq:magnetization}. However, the total energy density $\psi(\te F,\ve B)$ may still be polyconvex, but this cannot be shown analytically because of the nonlinear functional dependencies. Thus, alternatively, \emph{local polyconvexity} according to Eq.~\eqref{eq:localPoly}, which implies \emph{local ellipticity}, can be checked numerically for the model. Fortunately, the test is only relevant for deformations close to $\te U \approx \one$ with $\te F = \te R \cdot \te U$, $\te R\in \SO$, $\te U\in\Sym$, since the strain in particles is close to zero due to their very high stiffness in relation to the matrix phase. The expected rigid body rotations are also moderate. In addition, only \emph{2D plane strain} simulations are done later. The numerical test has shown that the model is locally polyconvex for the domain $(\te F(\phi_3,\Phi_3,\lambda,J), \ve B(B,\theta)) \in \mathscr{P\!o\!l\!y}$ specified by
$\phi_3\in [-\pi/18,\pi/18]$, $\Phi_3\in [0,\pi]$, $\lambda\in [0.999,1.001]$, $J\in [0.999,1.001]$, $B\in [0,4]\,\text{T}$, $\theta\in [0,2\pi]$ with
\begin{align}
	\te F = \te R_{x_3}(\phi_3) \cdot \te U \text{ with } \te U := \te R_{X_3}(\Phi_3) \cdot \diag(\lambda, \lambda^{-1} J, 1) \cdot \te R_{X_3}^T(\Phi_3) \text{ and } \ve B = B (\cos \theta \ve \, \ve e_1 + \sin \theta \ve \, \ve e_2) \; .
\end{align}
Therein, $\te R_{x_3}(\phi_3)\in\SO$ denotes a tensor for rotation around the $x_3$ axis. For the test, the 2D version of the local polyconvexity has been used, cf. \ref{app:poly_ell_2D}.

\begin{rmk}
	In order to construct a polyconvex particle model for which this property can be analytically proven, the split into a magnetic and a vacuum part would have to be abandoned.
	In the linear magnetic regime this is straightforward, see \cite{Javili2013}. However, for the nonlinear case this is disadvantageous, since the saturation behavior according to Eq.~\eqref{eq:saturation} cannot be easily modeled in this way. The authors are also unaware of such a model.
\end{rmk}

\section{Conventional macroscale constitutive model for comparison}\label{sec:Geb}
The macroscopic description of MAPs as a homogeneous continuum allows real structures to be simulated under complex loading conditions at a reasonable computational cost. Therefore, macroscopic models are of high relevance. As explained in Sect.~\ref{subsec:MAPs}, \emph{microscopically guided} approaches \cite{Kalina2020,Kalina2020a,Gebhart2022,Gebhart2022a,Mukherjee2019a,Lefevre2020}, that serve as surrogate models for the complex homogenized material behavior, are preferable. In this section we thus summarize the model developed by Gebhart~and~Wallmersperger~\cite{Gebhart2022a}. It is later used to evaluate the performance of the NN-based models given in Sect.~\ref{sec:macro_NN}.

\subsection{Model definition}
To account for effects arising from the particle interactions on
the microscale, the structure of the macroscale model is more complex compared to the energies given in Sect.~\ref{sec:micromodels}. Thus, an additive split of the total energy density in \emph{elastic, coupling, magnetic} and \emph{vacuum} contributions is assumed:
\begin{align}
	\bar \psi^\text{Geb}(\bte F,\bve B) := \bar \psi^\text{el}(\bte F) + \bar \psi^\text{coup}(\bte F, \bve B)
	+ \bar \psi^\text{mag}(\bte F, \bve B) + \bar \psi^\text{vac}(\bte F, \bve B) \; .
	\label{eq:GebhartModel}
\end{align}
The individual energy expressions according to \cite{Gebhart2022a} are defined in the following.

The elastic part $\bar \psi^\text{el}(\bte F)$, which is related to the elastic ground stress, comprises contributions from the composite's purely elastic deformation. It is described by the four-parameter \emph{Mooney-Rivlin model}
\begin{align}
	\bar \psi^\text{el}(\bar I_1, \bar I_2, \bar J) := \frac{G_1}{2}\left(\bar I_1 -3\right)
	+ \frac{G_2}{2}\left(\bar I_2 -3\right) 
	- d \ln \bar J + \frac{K}{\beta^2} \left(\beta \ln \bar J + \bar J^{-\beta} -1 \right) \; ,
	\label{eq:Rivlin}
\end{align}
where $G_1,G_2, K, \beta > 0$ to ensure polyconvexity and $d:=G_1 + 2 G_2$.

The coupling part $\bar \psi^\text{coup}(\bte F, \bve B)$ is linked to magnetically induced mechanical stress contributions. Thus, it models phenomena originating from magneto-mechanical particle interactions on the microscale. To account for \emph{saturation effects}, the following empirically identified expression is chosen:
\begin{align}
	\bar \psi^\text{coup}(\bar I_6,\bar J) := \frac{\kappa_1}{2} \ln\left(1+\bar J^{-4}\kappa_2^{-2}\bar I_6\right) + \frac{\kappa_3}{2} \ln\left(1+\bar J^{-2}\kappa_4^{-2}\bar I_6\right) \; . \label{eq:couplingGeb}
\end{align}
In the equation above, $\kappa_1$ and $\kappa_3$ characterize
the coupling effect's intensity, while it's
saturation is determined by the
material parameters $\kappa_2$ and $\kappa_4$. The suggested ansatz is able to rebuild the typical S-shaped behavior of the magnetically induced part of the mechanical stress tensor $\bte \sigma$. It results from the saturation behavior of the magnetizable particles, see \cite{Kalina2020a,Gebhart2022a}. As already stated in Sect.~\ref{subsec:poly}, the invariant $\bar I_6$ is not polyconvex with respect to $\bar{\gt I}$ and thus also the mixed invariants $\bar J^{-4}\bar I_6$ as well as  $\bar J^{-2}\bar I_6$ do not fulfill this global condition. However, one can find a necessary criterion for the \emph{relaxed local polyconvexity} of the overall model $\bar \psi^\text{Geb}(\bte F, \bve B)$. Thus it has to hold $\kappa_1\kappa_2^{-2} + \kappa_3\kappa_4^{-2} \ge 0$ for the parameters of the coupling contribution given in Eq.~\eqref{eq:couplingGeb}, see \cite{Gebhart2022a}.

Finally, the part $\bar \psi^\text{mag}(\bte F, \bve B)$ only includes \emph{purely magnetic contributions} and
is associated with the composite's magnetization $\bve m$. It is given by
\begin{align}
	\bar \psi^\text{mag}(\bar I_\text{mag}) &:=\alpha_1\left[\ln\left(\alpha_2\sinh\left(\alpha_3^{-1} \sqrt{\bar I_\text{mag}}\right)\right] - \ln\left[\sinh\left(\alpha_2\alpha_3^{-1} \sqrt{\bar I_\text{mag}}\right)\right)\right] \nonumber \\
	&-\frac{\kappa_1}{2} \ln\left(1+ \kappa_2^{-2}\bar I_\text{mag}\right) - \frac{\kappa_3}{2} \ln\left(1+\kappa_4^{-2}\bar I_\text{mag}\right) \; , \label{eq:magGeb}
\end{align}
where $\bar I_\text{mag}:=|\bar J^{-1} \bte F \cdot \bve B|^2 = |\bve b|^2$ is an invariant related to the Eulerian magnetic induction. This
invariant has the special property to only affect the magnetization and not the mechanical
Cauchy stress tensor $\bte \sigma$, see Gebhart~and~Wallmerperger~\cite{Gebhart2022}. This is favorable for the parametrization process of the model later on.
The first term in Eq.~\eqref{eq:magGeb}, a three parametric \emph{Brillouin-type} model \cite{Lefevre2020}, accounts for the typical magnetic saturation behavior. The two additional terms are added for the compensation of perturbation effects on $\bve m$ resulting from the coupling part defined in Eq.~\eqref{eq:couplingGeb}.
According to \v{S}ilhav\'y~\cite{Silhavy2019}, the invariant $\bar I_\text{mag}$ is not polyconvex and thus also the magnetic energy term is not. Again, a necessary condition for \emph{relaxed local polyconvexity} of $\bar \psi^\text{Geb}(\bte F,\bve B)$ with respect to $\bar{\gt I}$ is given by \mbox{$1/3 \alpha_1\alpha_3^{-2}(1-\alpha_2^2)+\mu_0^{-1}\ge 0$}.

The vacuum term is equivalent to the microscale models introduced above and is given by $\bar \psi^\text{vac}(\bar I_\text{vac}) := \frac{1}{2\mu_0} \bar I_\text{vac}$ with $\bar I_\text{vac} = \bar J^{-1} |\bte F \cdot \bve B|^2$, which is polyconvex \cite{Silhavy2019}.

\subsection{Model calibration}\label{subsec:modelCal}
To determine the model parameters, Gebhart~and~Wallmersperger~\cite{Gebhart2022,Gebhart2022a} have developed a \emph{sequential calibration strategy} tailored to the structure of the model \eqref{eq:GebhartModel}. We will use a slightly adapted version of it in this work. The procedure enables to calibrate the model based on data originating from RVE simulations, whereby two datasets are needed. The first set $\mathcal D^\text{mech}:=\{{}^1\mathcal T^\text{mech}, {}^2\mathcal T^\text{mech},\ldots,{}^k\mathcal T^\text{mech}\}$ with ${}^i \mathcal T^\text{mech}:=({}^i \bte F, {}^i\bte \sigma^\text{el})^\text{RVE} \in \GL \times \Sym$ contains only purely mechanical states, i.e., $\bve B \equiv \zero$, while the second set $\mathcal D^\text{coup}:=\{{}^1\mathcal T^\text{coup}, {}^2\mathcal T^\text{coup},\ldots,{}^k\mathcal T^\text{coup}\}$ with ${}^i \mathcal T^\text{coup}:=({}^i \bte F, {}^i \bve B, {}^i\bte \sigma, {}^i\bve m)^\text{RVE} \in \GL \times \Ln_1 \times \Ln_2 \times \Ln_1$ comprises coupled magneto-mechanical states. 
In order to investigate the generalizability of a model, it must be evaluated against data not used for calibration. Therefore, according to common practice in machine learning \cite{Linden2023,Klein2021,Vlassis2021c}, we divide the overall datasets $\mathcal D^\square$ into \emph{calibration} and \emph{test} sets, respectively, i.e., 
\begin{align}
	\mathcal D^\square = \mathcal D^\square_\text{cal} \cup \mathcal D^\square_\text{test} \text{ and }\varnothing = \mathcal D^\square_\text{cal} \cap \mathcal D^\square_\text{test} \; .
	\label{eq:cal_test}
\end{align}
After calibrating the model on a dataset $\mathcal D^\square_\text{cal}$, its predictions for $\mathcal D^\square_\text{test}$ can then be checked. To ensure good generalization of the model so that the magneto-mechanical response can be predicted for arbitrary states $(\bte F, \bve B)\in\GL\times\Ln_1$, the calibrated model should also be able to achieve good predictions for the test data set.

The calibration procedure is divided into three steps: Purely elastic data from $\mathcal D^\text{mech}_\text{cal}$ are used to calibrate the elastic energy contribution (i). Afterwards, using $\mathcal D^\text{coup}_\text{cal}$, the
magnetically induced part of the mechanical Cauchy stress tensor is utilized to parametrize the coupling energy term (ii). Finally, the spatial magnetization is fitted by calibrating the magnetic energy term (iii).
The complete procedure is given in Tab.~\ref{tab:fitting_Geb}.
\begin{rmk}
	\label{rmk:total_stress}
	It should be noted that it is essential to accurately represent the \emph{mechanical stress tensor} $\bte \sigma$, since this is linked to typical macroscopic effects such as the magnetorheological or magnetostrictive effect \cite{Kalina2020,Kalina2020a,Gebhart2022,Gebhart2022a}. It is not sufficient to accurately predict only the total stress tensor $\bte \sigma^\text{tot}$. This is also evident from the fact that $\bte \sigma^\text{tot}$ does not show the characteristic S-shaped course, since the Maxwell stresses $\bte \sigma^\text{Max}$, which show a quadratic dependence on $\bve b$, dominate. Moreover, as a rule, the total stress tensor is approximately two orders of magnitude larger than the mechanical stress. If a macroscopic model is able to predict $\bte \sigma$ and $\bve m$ accurately, the prediction for $\bve \sigma^\text{tot}$ is automatically given with equivalent quality, which follows from the structure of $\bte \sigma^\text{pon}$, see Eq.~\eqref{eq:pon}.
	
	This particular characteristic can also be explained by the macroscopic experimental setup for MAPs. Usually, a sample is placed between two magnetic poles \cite{Danas2012,Bodelot2018,Schumann2017,Psarra2017}, i.e. the background magnetic field is generated far away from it. Because of this, the \emph{Maxwell stresses} in the surrounding air, often referred to as free space, are not zero \cite{Danas2017}. Thus, it becomes clear once again that the real constitutive behavior is reflected in the mechanical stress.
	It should also be noted that in a macroscopic simulation, the surrounding free space must be included to capture the jumps in the magnetic and mechanical quantities \cite{Kalina2020a,Lefevre2020,Moreno-Mateos2023,Psarra2017,Rambausek2023}.
\end{rmk}

\begin{table}
	\begin{center}
		\caption{Sequential procedure for the calibration of the macroscopic model \eqref{eq:GebhartModel} according to Gebhart~and~Wallmersperger~\cite{Gebhart2022a}. The separate fitting of $\bte \sigma$ and $\bve m$ is possible since the invariant $\bar I_\text{mag}$ has no influence on $\bte \sigma$, see \cite{Gebhart2022,Gebhart2022a}.}
		\label{tab:fitting_Geb}
		\begin{footnotesize}
			\noindent\rule[0.5ex]{\textwidth}{0.1pt}
			\begin{itemize}
				\item[(i)] Determine parameters $\hat\p^\text{el}:=(G_1,G_2,K,\beta)$ of the elastic part $\psi^\text{el}(\bar I_1,\bar I_2, \bar J)$ from the purely mechanical dataset $\mathcal D^\text{mech}_\text{cal}$ by
				\begin{align*}
					\hat\p^\text{el} = \underset{\p^\text{el}\in\mathcal C^\text{el}}{\arg\min} \sum_{i=1}^{|\mathcal D^\text{mech}_\text{cal}|} \left\| \bte \sigma^\text{el}({}^i\bte F, \p^\text{el}) - {}^i\bte\sigma^\text{el} \right\|^2 \; \text{with} \;
					\mathcal C^\text{el} := \left\{G_1,G_2,K,\beta \in \R_{>0} \right\} \; .
				\end{align*}
				\item[(ii)] Find parameters $\hat\p^\text{coup}:=(\kappa_1,\kappa_2,\kappa_3,\kappa_4)$ of $\bar \psi^\text{coup}(\bar I_6,\bar J)$ from the dataset $\mathcal D^\text{coup}_\text{cal}$ and fix $\hat \p^\text{el}$ from step (i):
				\begin{align*}
					\hat\p^\text{coup} = \underset{\p^\text{coup}\in\mathcal C^\text{coup}}{\arg\min} \sum_{i=1}^{|\mathcal D^\text{coup}_\text{cal}|} \left\| \bte \sigma({}^i\bte F, {}^i\bve B, \p^\text{coup}, \hat\p^\text{el}) - {}^i\bte\sigma \right\|^2 \; \text{with} \;
					\mathcal C^\text{coup} := \left\{\kappa_1,\kappa_2,\kappa_3,\kappa_4 \in \R \, | \, \kappa_1\kappa_2^{-2} + \kappa_3\kappa_4^{-2} \ge 0 \right\} \; .
				\end{align*}
				\item[(iii)] Identify the remaining parameters $\hat\p^\text{mag}:=(\alpha_1,\alpha_2,\alpha_3)$ of $\bar \psi^\text{mag}(\bar I_\text{mag})$ from the dataset $\mathcal D^\text{coup}_\text{cal}$ and fix $\hat \p^\text{coup}$ from step (ii):
				\begin{align*}
					\hat\p^\text{mag} = \underset{\p^\text{mag}\in\mathcal C^\text{mag}}{\arg\min} \sum_{i=1}^{|\mathcal D^\text{coup}_\text{cal}|} \left| \bve m({}^i\bte F, {}^i\bve B, \p^\text{mag}, \hat\p^\text{coup}) - {}^i\bve m \right|^2 \; \text{with} \;
					\mathcal C^\text{mag} := \left\{\alpha_1,\alpha_2,\alpha_3 \in \R \, | \, 1/3 \alpha_1\alpha_3^{-2}(1-\alpha_2^2)+\mu_0^{-1}\ge 0 \right\} \; .
				\end{align*}
			\end{itemize}
			\noindent\rule[0ex]{\textwidth}{0.1pt}
		\end{footnotesize}
	\end{center}
\end{table}

\section{Physics-augmented neural network-based macroscale constitutive models}\label{sec:macro_NN}

Although the model summarized in Sect.~\ref{sec:Geb} is, to the best of the authors' knowledge, one of the most accurate macroscopic models for MAPs\footnote{This has been shown in the comparative study given in \cite{Gebhart2022a}.}, especially for the compressible case, the approximation quality for complex load cases is limited, see also the investigation in Sect.~\ref{sec:examples}. Thus, instead, an alternative approach which builds on NNs seems obvious. Thereby, approaches that combine the approximation capability of NNs with a rigorous physical formulation have proven to be particularly powerful \cite{Linden2023,Klein2022,Kalina2023,Vlassis2021c,Kalina2022a}. As mentioned in the beginning, we denote this type of networks as \emph{physics-augmented neural networks (PANNs)} herein. In this section, we formulate three PANN models which are differently rigorous with regard to the incorporated physics. Thereby, \emph{feedforward neural networks (FNNs), input convex neural networks (ICNNs)} according to Amos~et~al.~\cite{Amos2017} as well as \emph{positive neural networks (PNNs)} are used. In addition a tailored \emph{internal normalization} technique is applied.  Details on the specif architectures and the internal normalization technique are given in \ref{app:NNs}.

\subsection{Polyconvex model}

\subsubsection{Formulation of the model}
First, we formulate a \emph{polyconvex} energy expression which thus has to fulfill condition \eqref{eq:polyconvex} by construction. This is done by using ICNNs with a set of polyconvex invariants as inputs, see also \cite{Linden2023,Klein2021,Klein2022}. Thus, convex and non-decreasing activation functions and non-negative weights are used, cf. \ref{app:NNs}. 
Since $\bar I_6$ is not polyconvex, we have to limit the set to $\bI^\text{poly} := (\bar I_1,\bar I_2,\bar I_3, \bar I_4, \bar I_5)$. However, according to Linden~et~al.~\cite{Linden2023}, it may be necessary to incorporate further invariants $\bar I_\gamma^*(\bI^\text{poly}), \gamma \in \N_{\le A}$ with $A\in \N_{\ge 0}$ into the argument list, e.g., in order to increase
the approximation quality or to fulfill additional physical conditions. First, since the activation function acting on $\bar J$ must only be convex and not necessarily non-decreasing \cite{Klein2021}, the additional invariant $\bar I_1^* := -2\bar J$ is used. This is essential to represent
negative stresses \cite{Klein2021}. Furthermore, to improve the model flexibility, we use the polyconvex invariant $\bar I_2^* = \bar I_\text{vac} = \bar J^{-1} | \bte F \cdot \bve B|^2$. Thus, we end up with 
$\bI^{\text{poly},*} := (\bar I_1,\bar I_2,\bar I_3, \bar I_4, \bar I_5, \bar I_1^*, \bar I_2^*)$. 

Without loss of generality, we use an ansatz with a split of the potential in an \emph{elastic part}, describing the purely elastic behavior of the composite, and a part which captures contributions from \emph{coupling, magnetic} and \emph{vacuum parts} according to Eq.~\eqref{eq:GebhartModel}. The model is given by
\begin{align}
	\bar \psi^\text{I}(\bI^{\text{poly},*},\bar J) := \bar \psi^\text{el}(\bar I_1,\bar I_2,\bar I_3,\bar I_1^*,\bar J) + \psi^\text{cmv}(\bI^{\text{poly},*},\bar J) \; . \label{eq:modelI}
\end{align}
Both parts are represented by \emph{ICNNs} and additional terms, depending on $\bar J$, used to satisfy various physical conditions, see Sect.~\ref{subsec:magnetoElasticity}. To fulfill all common conditions of finite strain hyperelasticity by construction, except for the positivity of the energy, we follow the approach recently suggested by Linden~et~al.~\cite{Linden2023}. Thus, the \emph{elastic part} is given by
\begin{align}
	\bar \psi^\text{el}(\bar I_1,\bar I_2,\bar I_3,\bar I_1^*,\bar J) := \bar \psi_\text{ICNN}^\text{el}(\bar I_1,\bar I_2,\bar I_3,\bar I_1^*)
	+ \bar \psi^\text{el}_\text{str}(\bar J) + \bar \psi^\text{el}_\text{en} + \bar \psi^\text{el}_\text{gro}(\bar J) \; ,
	\label{eq:NNel}
\end{align} 
where 
\begin{align}
	\bar \psi^\text{el}_\text{str}(\bar J) &:= - \mathfrak n^\text{el}(\bar J -1) \text{ with }
	\mathfrak n^\text{el} := 2 \left(\diffp{\bar \psi^\text{el}_\text{ICNN}}{\bar I_1} + 2\diffp{\bar \psi^\text{el}_\text{ICNN}}{\bar I_2} + \diffp{\bar \psi^\text{el}_\text{ICNN}}{\bar I_3} - \diffp{\bar \psi^\text{el}_\text{ICNN}}{\bar I_1^*}\right)\Bigg|_{\bte F = \one} \in \R \; , \label{eq:normStress}\\
	\bar \psi^\text{el}_\text{en} &:= \bar \psi^\text{el}_\text{ICNN}(\bar I_1,\bar I_2,\bar I_3,\bar I_1^*) \big|_{\bte F = \one} \; , \label{eq:normEnergy}\\
	\bar \psi^\text{el}_\text{gro}(\bar J) &:= \lambda_\text{gro} \left( \bar J + \bar J^{-1} - 2\right)^2 \; . \label{eq:growth}
\end{align}
The parameter $\lambda_\text{gro}$ has to be chosen such that the energy grows fast enough during compression, e.g., a value of around \num{1e-2} or \num{1e-3} the material's initial stiffness has shown to be reasonable.  
By using invariants, we \emph{a priori} enforce \emph{symmetry} of the elastic Cauchy stress, \emph{objectivity} and \emph{isotropy}. Due to the use of the ICNN, the model is in addition \emph{polyconvex} and thus \emph{globally elliptic} by construction. 
With the additional terms given in Eqs.~\eqref{eq:normStress}--\eqref{eq:growth}, the elastic model furthermore automatically ensures a \emph{stress- and energy-free undeformed configuration} as well as the \emph{volumetric growth condition} to be fulfilled a priori, while the polyconvexity is not violated, see \cite{Linden2023} for more details.
In the same way, we define the potential
\begin{align}
	\bar \psi^\text{cmv}(\bI^{\text{poly},*},\bar J) := \bar \psi^\text{cmv}_\text{ICNN}(\bI^{\text{poly},*})
	+ \bar \psi^\text{cmv}_\text{str}(\bar J) + \bar \psi^\text{cmv}_\text{en} \; ,
\end{align}
with
\begin{align}
	\bar\psi^\text{cmv}_\text{str}(\bar J) &:= - \mathfrak n^\text{cmv}(\bar J -1) \; ,\;
	\mathfrak n^\text{cmv} := 2 \left(\diffp{\bar \psi^\text{cmv}_\text{ICNN}}{\bar I_1} + 2\diffp{\bar \psi^\text{cmv}_\text{ICNN}}{\bar I_2} + \diffp{\bar \psi^\text{cmv}_\text{ICNN}}{\bar I_3} - \diffp{\bar \psi^\text{cmv}_\text{ICNN}}{\bar I_1^*}\right)\Bigg|_{\bte F = \one, \bve B = \zero} \in \R \; , \label{eq:norm_stress} \\
	\bar \psi^\text{cmv}_\text{en} &:= \bar \psi^\text{cmv}_\text{ICNN}(\bI^{\text{poly},*}) \big|_{\bte F = \one, \bve B = \zero} \; .
\end{align} 
Note that it is not necessary to add derivatives with respect to $\bar I_4$, $\bar I_5$ and $\bar I_2^*$ to Eq.~\eqref{eq:norm_stress}, since these invariants depend on $\bte F,\bve B$ and a vanishing total stress, i.e., $\bte P^\text{tot}=\zero$, is only requested for $\bte F = \one, \bve B= \zero$. The condition $\bve H(\bte F, \bve B=\zero) = \zero$ for the magnetic field is automatically fulfilled since $\partial_{\bve B} \bar I_4|_{\bve B= \zero} =  \partial_{\bve B} \bar I_5|_{\bve B= \zero} = \partial_{\bve B} \bar I_2^*|_{\bve B= \zero} = \zero$. 

The proposed model $\bar \psi^\text{I}(\bI^{\text{poly},*},\bar J)$ thus a priori fulfills \emph{thermodynamic consistency}, \emph{symmetry of the total Cauchy stress}, \emph{objectivity}, \emph{material symmetry}, \emph{growth conditions}, \emph{polyconvexity} and thus \emph{ellipticity} as well as \emph{zero energy}, \emph{stress} and \emph{magnetic field in the unloaded state}. For the best of the authors knowledge, the \emph{magnetic saturation behavior}, cf. Eq.~\eqref{eq:saturation}, cannot be incorporated directly. Also, the chosen architecture do not enforce condition \eqref{eq:posEn} by construction, see the discussion in Linden~et~al.~\cite{Linden2023}.

\begin{figure}
	\includegraphics{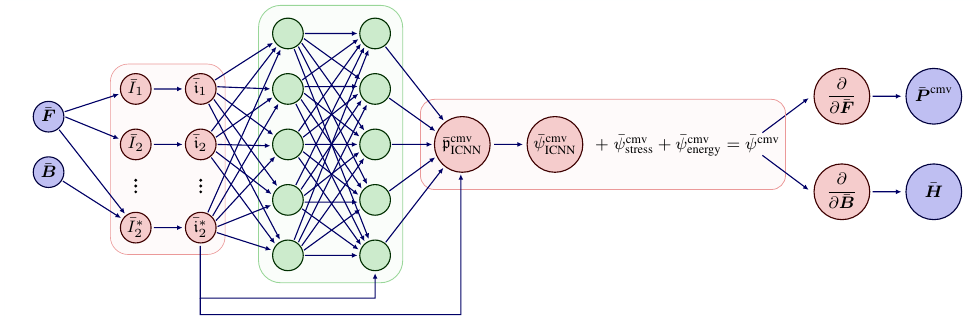}
	\caption{Illustration of the part $\bar \psi^\text{cmv}(\bI^{\text{poly},*},\bar J)$ of the polyconvex NN~model~\eqref{eq:modelI}. From the tensor-valued inputs $\bte F$ and $\bve B$, the polyconvex invariant set $\bI^{\text{poly},*}$ is calculated. In a non-trainable normalization layer, $\bI^{\text{poly},*}$ is transformed to the set $\bfI^{\text{poly},*}$. This set is passed through the ICNN which gives the normalized potential $\bar{\mathfrak p^\text{cmv}_\text{ICNN}}$ and from this, via another non-trainable layer, the true potential $\bar \psi^\text{cmv}_\text{ICNN}$. After adding terms to enforce stress and energy to be zero in the unloaded state, the stress and magnetic field are obtained from partial derivation.}
	\label{fig:modelI}
\end{figure}

The overall structure of the part $\bar \psi^\text{cmv}(\bI^{\text{poly},*},\bar J)$ of the polyconvex model is shown in Fig.~\ref{fig:modelI}. 
As depicted there, a standard ICNN architecture is combined with two internal non-trainable normalization layers for in- and output. This allows to restrict the weights to a range favorable for optimization without having to normalize the data. In particular, this offers some advantages when using invariants, since each invariant is typically in a different range and is computed from $\bte F$ and $\bve B$ during training and prediction process. In addition, it simplifies to incorporate trained models into FE codes. For more details see \ref{app:NNs}.

\subsubsection{Loss terms for training}
The \emph{weights} and \emph{biases} of the two ICNNs are summarized in the parameter sets $\w^\text{el}$ and $\w^\text{cmv}$. To determine them, a sequential training strategy tailored to the structure of the model \eqref{eq:modelI} is applied. Again, as already introduced in Sect.~\ref{subsec:modelCal}, the two datasets  $\mathcal D^\text{mech}:=\{{}^1\mathcal T^\text{mech}, {}^2\mathcal T^\text{mech},\ldots,{}^k\mathcal T^\text{mech}\}$ with ${}^i \mathcal T^\text{mech}:=({}^i \bte F, {}^i\bte \sigma^\text{el})^\text{RVE} \in \GL \times \Sym$ and $\mathcal D^\text{coup}:=\{{}^1\mathcal T^\text{coup}, {}^2\mathcal T^\text{coup},\ldots,{}^k\mathcal T^\text{coup}\}$ with ${}^i \mathcal T^\text{coup}:=({}^i \bte F, {}^i \bve B, {}^i\bte \sigma, {}^i\bve m)^\text{RVE} \in \GL \times \Ln_1 \times \Ln_2 \times \Ln_1$ are used, where a division into calibration and test sets is done, respectively, see Eq.~\eqref{eq:cal_test}. 

In the first step of the calibration procedure, the purely elastic data from $\mathcal D^\text{mech}_\text{cal}$ are used to calibrate the elastic energy contribution by 
\begin{align}
	\hat \w^\text{el} &= \underset{\w^\text{el}\in\ICNN}{\arg\min} \; \mathcal L^{\sigma^\text{el}} \; \text{with} \label{eq:cal_el}\\
	\mathcal L^{\sigma^\text{el}} &:= \frac{1}{|\mathcal D^\text{mech}_\text{cal}|} \frac{1}{n_\sigma^\text{el}}\sum_{i=1}^{|\mathcal D^\text{mech}_\text{cal}|}
	\left\|{}^i \bte{\sigma}^\text{el} - \bte{\sigma}^\text{el}({}^i\bte F, \w^\text{el})\right\|^2
	\; \text{with} \; n_{\sigma^\text{el}} := \max\left(\|{}^1\bte\sigma^\text{el}\|^2, \|{}^2\bte\sigma^\text{el}\|^2, \ldots, \|{}^{|\mathcal D^\text{mech}|}\bte\sigma^\text{el}\|^2\right) \; . \label{eq:loss_el}
\end{align}
The used set $\ICNN$ is introduced in \ref{app:NNs}, Eq.~\eqref{eq:setICNN}.
After calibration of $\bar \psi^\text{el}(\bar I_1,\bar I_2,\bar I_3,\bar I_1^*,\bar J)$, a training of $\bar \psi^\text{cmv}(\bI^{\text{poly},*},\bar J)$ by using $\mathcal D^\text{coup}_\text{cal}$ follows. It is done with \emph{loss functions} for mechanical stress tensor 
\begin{align}
	\mathcal L^\sigma &:= \frac{1}{|\mathcal D^\text{coup}_\text{cal}|} \frac{1}{n_\sigma}\sum_{i=1}^{|\mathcal D^\text{coup}_\text{cal}|}
	\left\|{}^i \bte{\sigma} - \bte{\sigma}({}^i\bte F, {}^i\bve B, \hat \w^\text{el}, \w^\text{cmv})\right\|^2
	\; ,\;   n_\sigma := \max\left(\|{}^1\bte\sigma\|^2, \|{}^2\bte\sigma\|^2, \ldots, \|{}^{|\mathcal D^\text{coup}_\text{cal}|}\bte\sigma\|^2\right)  \label{eq:loss_sigma}
\end{align}
and magnetization
\begin{align}
	\mathcal L^m &:= \frac{1}{|\mathcal D^\text{coup}_\text{cal}|} \frac{1}{n_m}\sum_{i=1}^{|\mathcal D^\text{coup}_\text{cal}|}
	\left|{}^i \bve{m} - \bve{m}({}^i\bte F, {}^i\bve B, \w^\text{cmv})\right|^2
	\; , \; n_m := \max(|{}^1\bve m|^2, |{}^2\bve m|^2, \ldots, |{}^{|\mathcal D^\text{coup}_\text{cal}|}\bve m|^2) \; . \label{eq:loss_m}
\end{align}
The training is thus given by
\begin{align}
	\hat \w^\text{cmv} &= \underset{\w^\text{cmv}\in\ICNN}{\arg\min} \; \mathcal L^\sigma + \mathcal L^m \; .
\end{align}
At this point it should be noted again, that it is not sufficient to just fit the total stress $\bte \sigma^\text{tot}$ instead of $\bte \sigma$, see Remark~\ref{rmk:total_stress}.

Later on we will see that it is too restrictive to describe the macroscopic behavior of MAPs with a polyconvex model. Although $\bte \sigma^\text{tot}$ and $\bve m$ can be learned with sufficient quality, an accurate reproduction of the mechanical stress tensor $\bte \sigma = \bte \sigma^\text{tot} - \bte \sigma^\text{pon}$ has shown to be not possible. Thus, two further approaches with weaker requirements are introduced in the following.   

\subsection{Models with weakly enforced local polyconvexity}
Since polyconvexity is a global statement, cf. the definition in Eq.~\eqref{eq:polyconvex}, a model that should have this property must satisfy it for all possible states. Alternatively, however, it is possible to consider the \emph{relaxed local polyconvexity} condition~\eqref{eq:localPoly}. 
This is not necessarily true globally, and it is therefore useful to enforce it for the calibration dataset in a \emph{weak sense} via a penalty term.
This allows a less restrictive model formulation. In the following, two NN models with weakly enforced local polyconvexity are introduced. 

\subsubsection{Formulation of a model without split of the non-elastic energy contribution}
For this model, again, we use an ansatz with a split of the potential in an elastic part and a part which captures the remaining magneto-mechanical contributions. Thereby, the elastic part is still formulated as a polyconvex energy functional. Only for the second energy part the requirements are relaxed, which allows us to use the complete and irreducible isotropic set given by  $\bI := (\bar I_1,\bar I_2,\bar I_3, \bar I_4, \bar I_5, \bar I_6)$ for it. The model is thus given by
\begin{align}
	\bar \psi^\text{II}(\bI, \bar I_1^*, \bar J) := \bar \psi^\text{el}(\bar I_1,\bar I_2,\bar I_3,\bar I_1^*,\bar J) + \psi^\text{cmv}(\bI,\bar J) \; . \label{eq:modelII}
\end{align} 
Therein, the elastic part is defined according to Eqs.~\eqref{eq:NNel}--\eqref{eq:growth}. To satisfy the \emph{growth conditions} \eqref{eq:growthConditions} also for $\bar \psi^\text{II}(\bI, \bar I_1^*, \bar J)$  by construction, a PNN is used to model the second part. Thus, we define
\begin{align}
	\bar \psi^\text{cmv}(\bI,\bar J) := \bar \psi^\text{cmv}_\text{PNN}(\bI)
	+ \bar \psi^\text{cmv}_\text{str}(\bar J) + \bar \psi^\text{cmv}_\text{en} \; , \; \bar \psi^\text{cmv}_\text{PNN}(\bI) \ge 0 \; \forall \bI \in \R^6\; ,
\end{align}
with
\begin{align}
	\bar \psi^\text{cmv}_\text{str}(\bar J) &:= - \mathfrak n^\text{cmv}(\bar J -1) \; ,\;
	\mathfrak n^\text{cmv} := 2 \left(\diffp{\bar \psi^\text{cmv}_\text{PNN}}{\bar I_1} + 2\diffp{\bar \psi^\text{cmv}_\text{PNN}}{\bar I_2} + \diffp{\bar \psi^\text{cmv}_\text{PNN}}{\bar I_3}\right)\Bigg|_{\bte F = \one, \bve B=\zero} \in \R \; , \label{eq:strCMV}\\
	\bar \psi^\text{cmv}_\text{en} &:= \bar \psi^\text{cmv}_\text{PNN}(\bI) \big|_{\bte F = \one, \bve B=\zero} \; . \label{eq:enCMV}
\end{align} 

\subsubsection{Formulation of a model with split into coupling, magnetic and vacuum contributions}
Finally, we formulate a third NN-based model that is closer in structure to a conventional model according to Sect.~\ref{sec:Geb}. Thus, similar to Eq.~\eqref{eq:GebhartModel}, we split the energy into \emph{elastic, coupling, magnetic} and \emph{vacuum} parts. Consequently, besides, $\bar I_1^*:=-2\bar J$ for the elastic part, we use $\bar I_\text{mag}$ and $\bar I_\text{vac}$ for magnetic and vacuum part, respectively. The model is given by 
\begin{align}
	\bar \psi^\text{III}(\bI, \bar I_1^*, \bar J) := \bar \psi^\text{el}(\bar I_1,\bar I_2,\bar I_3,\bar I_1^*,\bar J) + \bar \psi^\text{coup}(\bI,\bar J)
	+ \bar \psi^\text{mag}(\bar I_\text{mag}) + \bar \psi^\text{vac}(\bar I_\text{vac}) \; ,
	\label{eq:modelIII}
\end{align}
where the elastic part is defined according to Eqs.~\eqref{eq:NNel} -- \eqref{eq:growth}. For coupling and magnetic parts, we define 
\begin{align}
	\bar \psi^\text{coup}(\bI,\bar J) := \bar \psi^\text{coup}_\text{FNN}(\bI)
	+ \bar \psi^\text{coup}_\text{str}(\bar J) + \bar \psi^\text{coup}_\text{en} \; \text{and} \; 
	\bar \psi^\text{mag}(\bar I_\text{mag}) := \bar \psi^\text{mag}_\text{FNN}(\bar I_\text{mag})
	+ \bar \psi^\text{mag}_\text{en} \; ,
\end{align}
where the terms $\bar \psi^\text{coup}_\text{str}(\bar J)$, $\bar \psi^\text{coup}_\text{en}$ and $\bar \psi^\text{mag}_\text{en}$ for the enforcement of zero energy and stress in the unloaded state are defined equivalently to Eqs.~\eqref{eq:strCMV} and \eqref{eq:enCMV}. The \emph{non-trainable vacuum part} is given by $\bar \psi^\text{vac}(\bar I_\text{vac}) := \frac{1}{2\mu_0} \bar I_\text{vac}$. Note that FNNs are chosen for coupling and magnetic parts, since these energies are expected to be negative. This becomes apparent from
\begin{align}
	\bve m = -\left(\diffp{\bar\psi^\text{coup}}{\bve B} 
	+ \diffp{\bar\psi^\text{mag}}{\bve B} \right) \cdot \bte F^{-1} \; .
\end{align}
Thus, the \emph{growth conditions} \eqref{eq:growthConditions} can no longer be guaranteed to be fulfilled.

\subsubsection{Training procedure and loss terms}
The training of both NN~models with the relaxed local polyconvexity criterion is done in three steps. First, the purely elastic data from $\mathcal D^\text{mech}_\text{cal}$ are used to calibrate the elastic energy contribution as given in Eq.~\eqref{eq:cal_el}.
Afterwards, using $\mathcal D^\text{coup}_\text{cal}$, a training based on the loss terms \eqref{eq:loss_sigma} and \eqref{eq:loss_m} for $\bte \sigma$ and $\bve m$ is executed to determine the remaining parameters, i.e., $\w^\text{cmv}$ or $\w^\text{coup}$ and $\w^\text{mag}$, respectively. Finally, in a third step, the pre-trained energy terms are \emph{re-calibrated} in such a way that the \emph{local polyconvexity} is enforced in a \emph{weak sense}. 
To this end, an additional loss term, which penalizes the violation of condition \eqref{eq:localPoly}, is added to the overall loss. Since local polyconvexity requires a \emph{positive semi-definite Hessian} $\bar{\gt H}$, negative eigenvalues $\bar \lambda^{\mathfrak H}_\alpha$ should give contributions to this new loss function. This can be considered as follows:
\begin{equation}
	\mathcal L^\text{poly} := \frac{1}{|\mathcal D^\text{coup}_\text{cal}|}\frac{1}{n_{\lambda^\mathfrak{H}}} \sum_{i=1}^{|\mathcal D^\text{coup}_\text{cal}|}\sum_{\alpha=1}^{25}\relu\left(-\bar\lambda^{\mathfrak H}_\alpha({}^i\bte F, {}^i\bve B, \w)\right)  \; \text{with} \; n_{\lambda^\mathfrak{H}} := \max(|{}^1\bar\lambda_1^\mathfrak{H}|,|{}^1\bar\lambda_2^\mathfrak{H}|, \ldots, |{}^{|\mathcal D^\text{coup}_\text{cal}|}\bar\lambda_{25}^\mathfrak{H}|) \; .
	\label{eq:loss_poly}
\end{equation}
In the equation above, $\w$ is a vector comprising the parameters of all NNs in the model.
Note that, in contrast to the loss terms $\mathcal L^\sigma$ and $\mathcal L^m$, the normalization values $n_{\lambda^\mathfrak{H}}$ change in each training step since the Hessian depends on the weights of the NN.
The overall loss for the third and final step is thus given by
\begin{align}
	\mathcal L := w_\sigma \mathcal L^\sigma + w_m \mathcal L^m + w_\text{poly} \mathcal L^\text{poly} \; ,
\end{align}
where the introduced non-trainable weights $w_\sigma$, $w_m$ and $w_\text{poly}$ have to be chosen. The specific values are given in the examples in Sect.~\ref{sec:examples}. The additional training step is also denoted as \emph{post-training} or \emph{post-calibration} in the following. Since the second derivatives of the potential with respect to the set $\bar{\gt I}$ are necessary to compute the Hessian $\bar{\gt H}$, it is a \emph{higher order Sobolev training} \cite{Vlassis2021}. Note that the parameters $\w^\text{el}$ of the ICNN-based model capturing the elastic ground stress are fixed, since this part is polyconvex and thus also locally polyconvex.

Alternatively, it is also possible to directly enforce \emph{local strict ellipticity}\footnote{Note that it is not of practical relevance to check of enforce for the local ellipticity, which requires Ineq.~\eqref{eq:minors}, instead of the strictly local ellipticity. This is due to the fact that, as a rule, no minors with values of exactly zero occur during numerical evaluation. It is therefore preferable to evaluate the less computationally intensive relationship~\eqref{eq:ell_cond}.} in a weak sense during the training. This requires to penalize the violation of condition~\eqref{eq:ell_cond}. The loss is thus given by  
\begin{align}
	\mathcal L^\text{ell} &:= \frac{1}{|\mathcal D^\text{coup}_\text{cal}|}\sum_{i=1}^{|\mathcal D^\text{coup}_\text{cal}|} \Bigg[ \frac{1}{n_{\gamma_1}} \relu\left(-{}^i\gamma_1^\text{min} \right)
	+ \frac{1}{n_{\gamma_2}} \relu\left(-\gamma_2^\text{min} \right) 
	+ \frac{1}{n_{\gamma_3}} \relu\left(-\gamma_3^\text{min} \right)
	\Bigg]\; \text{with} \; \label{eq:ell}\\
	{}^i\gamma_\alpha^\text{min}&:=\underset{(\ve N, \ve V) \in \mathcal E}{\inf}  \gamma_\alpha({}^i\bte F, {}^i\bve B, \w,\ve N, \ve V), \; \alpha \in\{1,2,3\} \; ,	\label{eq:opt_ell}
\end{align}
where $\mathcal E:= \{\ve N\in\Vn, \ve V\in\Vn \, |\, \ve N\cdot \ve V = 0\}$ and $n_{\gamma_1}:=\max(|{}^1\gamma_1^\text{min}|,|{}^2\gamma_1^\text{min}|,\ldots,|{}^{|\mathcal D^\text{coup}_\text{cal}|}\gamma_1^\text{min}|)$, $n_{\gamma_2}$, $n_{\gamma_3}$ analogue. During training with the loss given in Eq.~\eqref{eq:ell}, the additional optimization problems \eqref{eq:opt_ell} have to be solved for all states $({}^i\bte F, {}^i\bve B)$ included into the calibration set $\mathcal D^\text{coup}_\text{cal}$.
To solve these internal optimization problems, the \emph{unit vectors} $\ve N$, $\ve V$ are \emph{sampled} in $s$ steps. For the 2D problems considered later on, due to the orthogonality, both unit vector can be described in terms of only one angle, which makes this quite doable, e.g., with $s=180$ steps. The simplified loss terms for the 2D case are given in \ref{app:2D}. The complete loss is then given by 
\begin{align}
	\mathcal L := w_\sigma \mathcal L^\sigma + w_m \mathcal L^m + w_\text{ell} \mathcal L^\text{ell} \; .
\end{align}

\subsection{Comparison of the different models}

\begin{table}
	\begin{center}
		\caption{Comparison of the different macroscopic models for MAPs. Here, \cmark, \cbmark\, and \xmark\, mean that the considered condition is a priori fulfilled, weakly enforced by claiming necessary but not sufficient parameter restrictions or by adding penalty terms in the loss and not fulfilled, respectively.}
		\begin{footnotesize}
			\begin{tabular}{lcccc}
				\label{tab:comp}
				\emph{Condition} & \emph{Conventional model}~\eqref{eq:GebhartModel} & \emph{NN model I}~\eqref{eq:modelI} & \emph{NN model II}~\eqref{eq:modelII} & \emph{NN model III}~\eqref{eq:modelIII}\\
				\hline
				\hline
				Thermodynamic consistency & \cmark &\cmark & \cmark & \cmark \\
				Symmetric total Cauchy stress & \cmark & \cmark & \cmark & \cmark \\
				Objectivity & \cmark & \cmark & \cmark & \cmark \\
				Material symmetry & \cmark & \cmark & \cmark & \cmark \\
				Growth conditions & \xmark & \cmark & \cmark & \xmark \\
				Magnetic saturation limit & \xmark & \xmark & \xmark & \xmark \\
				Zero energy in unloaded state  & \cmark & \cmark & \cmark & \cmark \\
				Stress-free unloaded state & \cmark & \cmark & \cmark & \cmark \\
				Non-magnetized unloaded state & \cmark & \cmark & \cmark & \cmark \\
				Non-negative energy & \xmark & \xmark & \xmark & \xmark \\
				Polyconvexity & \xmark & \cmark & \xmark & \xmark \\
				Local polyconvexity & \cbmark & \cmark & \cbmark & \cbmark \\
				Global ellipticity & \xmark & \cmark & \xmark & \xmark \\
				Local ellipticity & \cbmark & \cmark & \cbmark & \cbmark 
			\end{tabular}
		\end{footnotesize}
	\end{center}
\end{table}

After the introduction of the conventional model \eqref{eq:GebhartModel} and the three NN-based models \eqref{eq:modelI}, \eqref{eq:modelII} and \eqref{eq:modelIII}, a short comparison regarding the fulfillment of the \emph{physical conditions} introduced in Sect.~\ref{subsec:magnetoElasticity} is done. Thereby, in principle, we distinguish between three different levels for the fulfillment of a condition: 
\begin{itemize}
	\item[\cmark] A condition is \emph{a priori fulfilled} by a model for all possible loading states $(\bte F, \bve B) \in\GL\times \Ln_1$, i.e., by construction,
	\item[\cbmark] it is \emph{weakly enforced}, i.e., by claiming necessary but not sufficient conditions or penalizing the violation with additional loss terms, and
	\item[\xmark] the fulfillment \emph{cannot be claimed} during calibration or it is not fulfilled.
\end{itemize}
In this way, a comparison of the four macroscopic models is given in Tab.~\ref{tab:comp}. At this point we will note again that it is of course always preferable to fulfill a condition globally and thus a priori if possible. However, if this limits the approximation ability of the model too much and the data can no longer be reproduced with sufficient quality, a relaxation of the requirement is necessary.

\section{Numerical examples}\label{sec:examples}
In order to demonstrate and compare the achievable prediction quality of the conventional model by Gebhart and Wallmersperger~\cite{Gebhart2022a} and the developed NN-based macroscopic models introduced in the previous section, we will calibrate them by using data generated with a \emph{computational homogenization} approach according to Sect.~\ref{subsec:hom} in this section. Due to the enormous numerical cost and the partially limited feasibility of 3D RVE simulations of MAPs with random microstructure, we restrict ourselves to \emph{2D plane strain} problems in the following, see \ref{app:2D} for details on the changes that go along with this simplification.

\subsection{Data generation procedure}\label{subsec:data}
We generate synthetic data for two different types of materials, MAPs with a \emph{compressible} gel-like matrix described by Eq.~\eqref{eq:comp} and MAPs with a \emph{quasi-incompressible} elastomer matrix described by Eq.~\eqref{eq:incomp}. For both, the magnetization behavior of the embedded particles is defined by Eq.~\eqref{eq:magenergy}. The chosen parameters are given in Tab.~\ref{tab:parameters_micro}. In the following, a study on the representativeness of microstructure cells is given and an invariant-based sampling approach is applied to pre-select relevant states.

\subsubsection{Study on representativeness of microstructure cells}\label{subsec:SVE}

To determine the effective material response of MAPs with stochastic particle distribution, it is necessary to select an RVE with a sufficient number of particles or to combine several smaller cells, so called \emph{statistical volume elements (SVEs)} \cite{Aldakheel2023a,Rassloff2021}, which build an RVE by calculating the statistical mean from the respective volume averages. The second strategy will be applied here. Thus, Eq.~\eqref{eq:average} has to be replaced by
\begin{align}
	\bte F := \langle \te F \rangle_n \; , \;
	\bte P^\text{tot} := \langle \te P^\text{tot} \rangle_n \; ,\; 
	\bve B :=\langle \ve B \rangle_n \; \text{and} \; 
	\bve H :=\langle \ve H \rangle_n  \text{ with }
	\langle (\cdot) \rangle_n := \frac{1}{n} \sum_{\alpha=1}^n \frac{1}{V^\text{SVE}_\alpha} \int\limits_{\mathcal B_{0,\alpha}^\text{SVE}} (\cdot) \, \text{d}V \; ,
	\label{eq:SVE}
\end{align}
where $\langle (\cdot) \rangle_n$ is an averaging operator over $n$ SVE cells with domain $\B^\text{SVE}_{0,\alpha}\subset\R^3$ and volume $V^\text{SVE}_\alpha\in\R_{>0}$. Also, the tangent terms according to Eq.~\eqref{eq:tangents} are calculated for each SVE and are then averaged.

Regarding the necessary number of particles or cells to be sufficient for representativeness, various values have been proposed in the literature for two-dimensional RVEs/SVEs \cite{Danas2017,Zabihyan2018,Kalina2020a}. Differences can be attributed to influences such as the chosen BCs, minimum particle distance, volume fraction or the dispersity of the particle diameters. Furthermore, it is decisive which physical quantity is considered to decide whether a cell or an ensemble of several cells is representative or not. According to \cite{Kalina2021a}, the mechanical part of the stress tensor, i.e., $\bte \sigma = \bte \sigma^\text{tot}-\bte \sigma^\text{pon}$, is by far more sensitive than $\bte \sigma^\text{tot}$, $\bve h$ or $\bve m$ and  must therefore be used for the evaluation, see also Remark~\ref{rmk:total_stress} for more details on the importance of $\bte \sigma$ compared to $\bte \sigma^\text{tot}$.

\begin{figure}
	\includegraphics{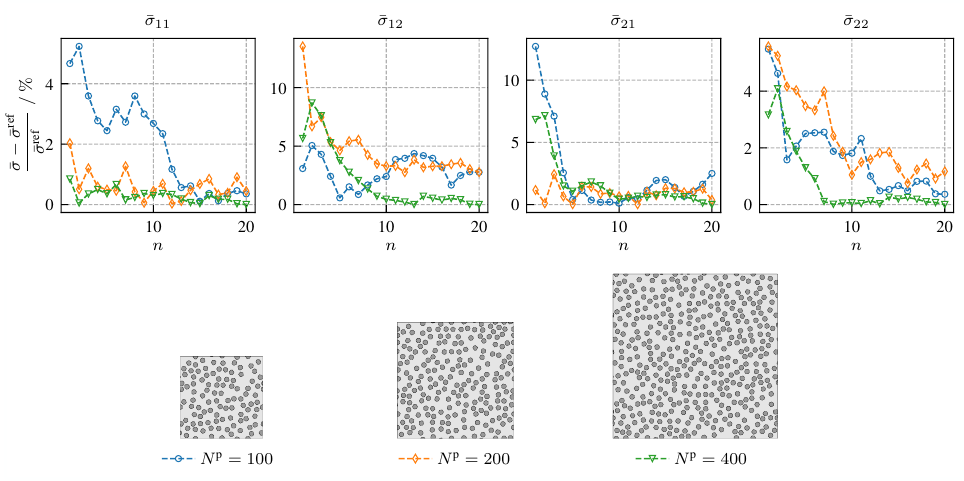}
	\caption{Study on the representativeness of statistical-periodic SVEs with $\phi=\SI{30}{\percent}$ particle volume fraction and minimal distance $r^\text{min}=0.15\,d$ for $\bte F = \one$ and $\bve B = \frac{\sqrt{2}}{2}(\ve e_1 + \ve e_2) \, \SI{2.5}{\tesla}$. Shown is the relative deviation of the predicted value for $\bte \sigma$ depending on the number of cells $n$ to the value $\bte \sigma^\text{ref}$ for $n=20$ cells with $N^\text{p}=400$ particles.}
	\label{fig:SVEs}
\end{figure}

In the following, the analysis will be performed for \emph{monodisperse, statistical-periodic microstructures} with a minimum inter-particle distance of $r^\text{min} := 0.15\,d$, where $d$ denotes the particle diameter.
Here, the placement of the particles and the subsequent creation of the periodic mesh has been done with the Python tool \emph{gmshModel}.\footnote{The Python tool gmshModel is freely available under the link \url{https://gmshmodel.readthedocs.io/en/latest/}. The generation of stochastic particle distributions in this tool is based on the Random Sequential Adsorption (RSA) algorithm.} Exemplarily, the compressible case is considered now.
To evaluate the statistical representativeness of an SVE ensemble, the predicted effective mechanical stress $\bte \sigma(\bte F, \bve B, \langle \bte P^\text{tot}\rangle_n,\langle \bve H\rangle_n)$ of $n\in\{1,2,\ldots,20\}$ cells with $\phi=\SI{30}{\percent}$, which have $N^\text{p} = \{100, 200, 400\}$ inclusions, were analyzed. The predictions for the loading $\bte F=\one$ and $\bve B = \frac{\sqrt{2}}{2}(\ve e_1 + \ve e_2) \, \SI{2.5}{\tesla}$, more precisely the deviations to the best mean value for $N^\text{p}=400$ and $n=20$, are plotted in Fig.~\ref{fig:SVEs}.\footnote{The choice $\bte F=\one$ is predestinated for the analysis of the representativeness, since the effective elastic ground stress is zero and only the magnetically induced actuation stresses occur.}
As shown there, for all investigated classes, i.e., $N^\text{p} = \{100, 200, 400\}$, a significant deviation from the reference is observable for all if $n$ is small. The SVEs with $100$ and $200$ particles show a noticeable deviation of around $\SI{2.5}{\percent}$ to the reference even with 20 cells for $\bar \sigma_{12}$. In contrast, for the SVEs with $400$ particles, no significant deviation occurs for $n\ge 10$. Thus, we use $n=10$ SVEs with $N^\text{p}=400$ particles each in the following.

It should be noted that such a strongly pronounced sensitivity of the effective material response on the cell size does not occur in the purely mechanical case. However, for consistency, also the purely elastic simulations are done with the same number of SVEs.

\subsubsection{Data sampling}
Now, after the definition of a reasonable number of SVEs to guarantee representativeness, we introduce a technique for \emph{sampling} the $\bte F$-$\bve B$ space in a predefined range. To reduce the number of SVE simulations later on, we have to reduce the number of sampling points to a minimum by taking into account the underlying physics of the problem. More precisely, we do the sampling in the space of invariants \cite{Kalina2022a,Kalina2023}, which is done in several steps and is described in the following.

Due to the principle of \emph{material frame invariance} and the restriction to \emph{isotropy} of the considered energy density functions, see Eqs.~\eqref{eq:objectivity} and \eqref{eq:symmetry}, it holds
\begin{equation}
	\bar \psi(\bte F, \bve B) = \bar \psi(\te Q_1 \cdot \bte F \cdot \te Q_2^T, \bve B \cdot \te Q_2^T) \; \forall \bte F \in \GL, \bve B \in \Ln_1, \te Q_1 \in \SO, \te Q_2 \in \Othree \; .
\end{equation}
Thus, from the material frame invariance we find that $\bar \psi(\bte F, \bve B)$ only depends on the right stretch tensor $\bte U$ and the Lagrangian magnetic induction $\bve B$ with
$\bte F = \bte R \cdot \bte U$, $\bte R \in \SO$. For sampling we thus choose $\bte R^\text{samp}:=\one$. In addition, due to isotropy and the restriction on 2D plane strain problems, we find that the following expressions are sufficient to sample all relevant $\bte F$-$\bve B$ states:
\begin{align}
	\bte F^\text{samp} = \bte U^\text{samp} \text{ with } \bte U^\text{samp} = \diag(\bar \lambda_1, \bar \lambda_1^{-1}\bar J, 1) \text{ and } \bve B^\text{samp} = \bar B (\cos \theta \ve \, \ve e_1 + \sin \theta \ve \, \ve e_2) \; .
\end{align}
Thus, altogether, we end up with a four parameter space, consisting of $(\bar \lambda_1, \bar J, \theta, \bar B) \in \R_{>0} \times \R_{>0} \times \R \times \R_{>0}$ to be sampled initially, which is done by \emph{Latin hypercube sampling (LHS)} within a predefined range. This gives a number of $k\in\N$ states. For each state a loading path with $l\in\N$ increments is generated afterwards, which gives a total of $k\cdot l$ states, i.e., ${}^a\bte F$, ${}^a\bte B$ with $a\in\{1,2,\ldots,k\cdot l\}$.

Since the intrinsic constitutive behavior of the material lives
in the \emph{space of invariants}, it is reasonable to transform into this space now, i.e., $({}^a\bte F, {}^a\bve B) \mapsto ({}^a\bar I_1, {}^a\bar I_2, {}^a\bar I_3, {}^a\bar I_4, {}^a\bar I_5, {}^a\bar I_6)$. This makes it possible to decide whether a state is truly unique from the point of view of the material or not. Furthermore, since the definition ranges of the single invariants differ, a normalization of each invariant according to ${}^a\bar I_\alpha \mapsto {}^a\bar{\mathfrak i}_\alpha \in [-1,1]$ is done for all states. Now, to reduce the number of $\bte F$-$\bve B$ tuples, a \emph{filtering} is done. The aim of this process is to find a set $\mathcal F$ including only states that are unique from the material's point of view. To start the process, all states of the first loading path are added to this set, i.e., $\mathcal F:=\{{}^1\bfI^\text{un}, {}^2\bfI^\text{un},\ldots,{}^l\bfI^\text{un}\}$, where ${}^b\bfI^\text{un}$ is a generalized vector including the six \emph{normalized invariants} of a unique state. Then, to decide whether another state ${}^a\bfI$ is inimitably, the relative Euclidean distance
\begin{align}
	{}^{ab}\epsilon := \frac{|{}^a\bfI- {}^b\bfI^\text{un}|}{\relu(|{}^b\bfI^\text{un}|-\delta) + \delta} \; \text{with} \; \delta := \frac{1}{3} \max(|{}^1\bfI|,|{}^2\bfI|,\ldots,|{}^{kl}\bfI|) \; \text{and} \;
	{}^b\bfI^\text{un} \in \mathcal F
\end{align}
of the state to all states currently included in $\mathcal F$ is calculated. Thereby, $\delta$ is a heuristic parameter to prevent small values to be overrepresented. If ${}^{ab}\epsilon\ge \epsilon_\text{tol}$ even for one $b \in\{1,2,\ldots,|\mathcal F|\}$ it is unique and has to be added to $\mathcal F$. To facilitates the application of a state within the computational homogenization later on, the full loading path belonging to a new \emph{unique state} is added to the set $\mathcal F$. After the filtering process is done, deformation gradients and Lagrangian magnetic induction fields belonging to the identified set of unique states are stored.

\begin{figure}
	\begin{center}
		\includegraphics{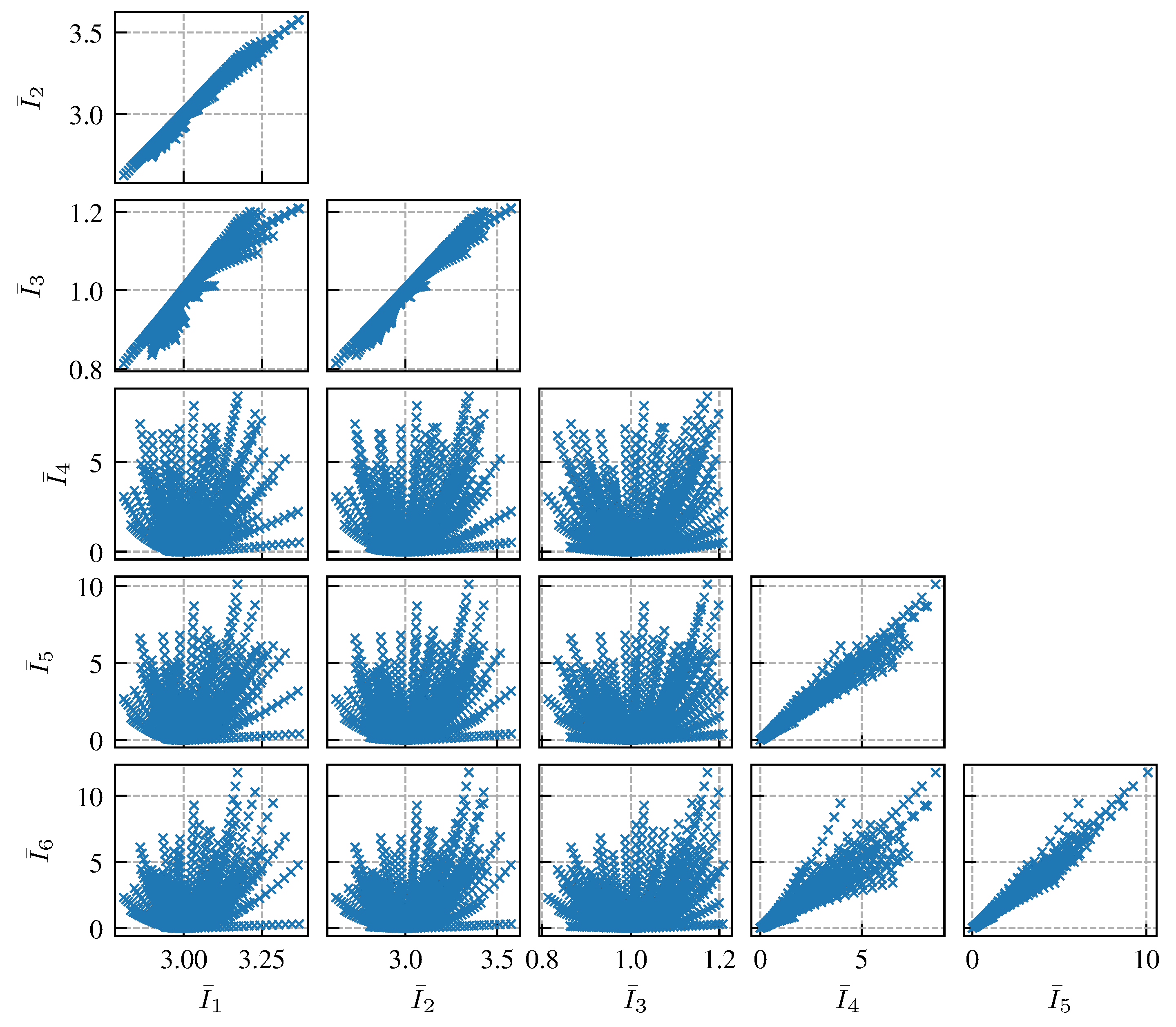}
	\end{center}
	\caption{Sampling of the invariant space for isotropic magneto-elasticity for plane strain and the compressible case. Shown is the coupling dataset $\mathcal D^\text{coup}$ with a number of 1,500 tuples in sectional planes, cf. Tab.~\ref{tab:sampling}.}
	\label{fig:Icomp}
\end{figure}

As mentioned in Sects.~\ref{sec:Geb} and \ref{sec:macro_NN}, purely mechanical and coupled magneto-mechanical loadings are needed for the calibration of the introduced models. Thus, two sets with unique states are generated for \emph{compressible} and \emph{quasi-incompressible MAPs}, respectively. The chosen sampling and sorting parameters as well as the resulting number of states found with the proposed data sampling and filtering strategy are given in Tab.~\ref{tab:sampling}. The identified states are exemplarily shown within sectional planes for the compressible case in Fig.~\ref{fig:Icomp}. 

\begin{table}
	\begin{center}
		\caption{Chosen parameters for the data sampling and filtering. $k$, $l$ and $\epsilon_\text{tol}$ are the number of LHC sampling points, the increment number as well as the tolerance for filtering. Note that the filtered states also contain the respective load path associated with a unique state.}
		\label{tab:sampling}
		\begin{footnotesize}
			\begin{tabular}{llllllllll}
				Type & $k$ & $l$ &  $\epsilon_\text{tol}$ & Range $\bar \lambda_1$ & Range $\bar J$ & Range $\theta$ & Range $\bar B$ & Filtered states\\
				\hline\hline			
				Compressible (mech) & 250 & 25 & $\SI{15}{\percent}$ & $[0.85,1.15]$ & $[0.9,1.1]$ & $-$ & $-$ & 500\\
				Compressible (coup) & 250 & 25 & $\SI{15}{\percent}$ & $[0.85,1.15]$ & $[0.9,1.1]$ & $[0,\pi]$ & $[0.2,2.5]$ & 1,500\\
				Quasi-incompressible (mech) & 250 & 25 & $\SI{15}{\percent}$ & $[0.85,1.15]$ & $[0.99,1.01]$ & $-$ & $-$ & 750\\
				Quasi-incompressible (coup) & 250 & 25 & $\SI{15}{\percent}$ & $[0.85,1.15]$ & $[0.99,1.01]$ & $[0,\pi]$ & $[0.2,2.5]$ & 2,225
			\end{tabular}
		\end{footnotesize}
	\end{center}
\end{table}

\begin{figure}
	\begin{center}
		\includegraphics{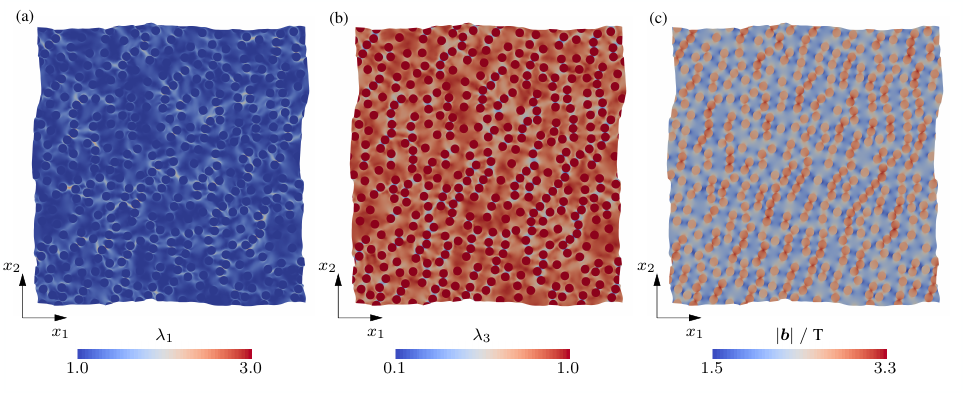}
	\end{center}
	\caption{Local fields in a deformed SVE with compressible matrix material for the loading specified by $\bte F=0.9372\, \ve e_1\otimes\ve e_1 + 0.9666\, \ve e_2 \otimes \ve e_2 + \ve e_3 \otimes \ve e_3$ and $\bve B=(0.7046\,\ve e_1+2.0774\,\ve e_2)\,\text{T}$: (a) and (b) principal stretches $\lambda_1$ and $\lambda_3$ with $\lambda_1\ge\lambda_2\ge\lambda_3$ and $\lambda_2 \equiv 1$, as well as (c) norm of the magnetic induction  $\ve b$.}
	\label{fig:SVE_local}
\end{figure}

Furthermore, microscopic field distributions in an SVE are given in Fig.~\ref{fig:SVE_local}, where the principal stretches $\lambda_1$ and $\lambda_3$ as well as the norm of the Eulerian magnetic induction $|\ve b|$ are shown. As can be seen, the deformations that occur are much more pronounced than the average value $\bte F:=\langle \te F\rangle_n$. In the tensile range, stretching occurs up to maximum values of $\lambda_1\approx 3$. In the compression range, isolated regions with $\lambda_3\approx 0.1$ can be found. In addition to the prescribed macroscopic deformation gradient $\bte F$, the local deformation fields result from the magnetic interactions of the particles. According to \cite{Biller2014,Puljiz2018}, with correspondingly low stiffness of the surrounding matrix, it is quite possible for them to move almost into contact. In Fig.~\ref{fig:SVE_local}(c), the magnetically induced chain formation can also be seen particularly well. This structuring process leads to the fact that $\bve b$ and $\bve m$ are not aligned due to the induced anisotropy and the effective stress $\bte \sigma$ gets unsymmetrical, cf.the balance of angular momentum given in Eq.~\eqref{eq:angmom}.

For all sampled states, the resulting effective material response is calculated by computational homogenization using the SVE approach, i.e., $(\bte F, \bve B) \mapsto (\bar \psi, \bte P^\text{tot}, \bve H, \btttte A, \bttte G, \bte K)$, to get the datasets $\mathcal D^\text{mech}$ and $\mathcal D^\text{coup}$. Note that some SVE simulations have been diverged before the last increment was reached, which is due to the chosen comparatively low Young's modulus of the matrix. In that case, the corresponding states have been removed from the dataset, even though only one of ten SVE simulations diverged. Overall, this concerns 126 states for the compressible and 3 for the quasi-incompressible MAP, respectively.

The evaluation of the \emph{local strict ellipticity} condition \eqref{eq:ell_cond} for all generated data has shown that all states satisfy it for both, the compressible and the quasi-incompressible MAP. Macroscopic material instabilities usually only occur with more pronounced deformations, cf. the study for periodic MAPs given in \cite{Polukhov2020}. In addition, it is to be expected that the considered stochastic microstructures have an enlarged stable region compared to the periodic cells.

\subsection{Interpolation behavior of the macroscopic models}
After the description of the data generation process, the obtained sets $\mathcal D^\text{mech}$ and $\mathcal D^\text{coup}$ are used for the calibration of the macroscopic models. As mentioned before, we divide the overall datasets into calibration and test sets, respectively, see Eq.~\eqref{eq:cal_test}. A ratio of $70/30$ for \emph{calibration/test data} is chosen within this first study. Thus, we mainly check the interpolation ability of the models, using the test data to analyze the generalization in addition. 
In the following, to avoid redundancy, the detailed comparison of the different models is only discussed for the case of the compressible MAP. However, the ability of the models is nevertheless demonstrated for both the compressible and quasi-incompressible dataset.

All NN-based models were trained by using the \emph{SLSQP optimizer} (Sequential Least Squares Programming). This has been shown to be favorable for the training of NNs with a small to moderate size \cite{Kalina2023,Linden2023}.\footnote{A comparison of the training with SLSQP and the more common Adam optimizer is given in \ref{app:adam}.} To exclude random effects from initialization, 100 trainings were carried out each if not explicitly stated otherwise. The best model is then used for the discussion of the results. The implementation of the models and the calibration workflow
was realized using \emph{Python}, \emph{Tensorflow} and \emph{SciPy}.

\subsubsection{Conventional model}

\begin{figure}
	\begin{center}
		\includegraphics{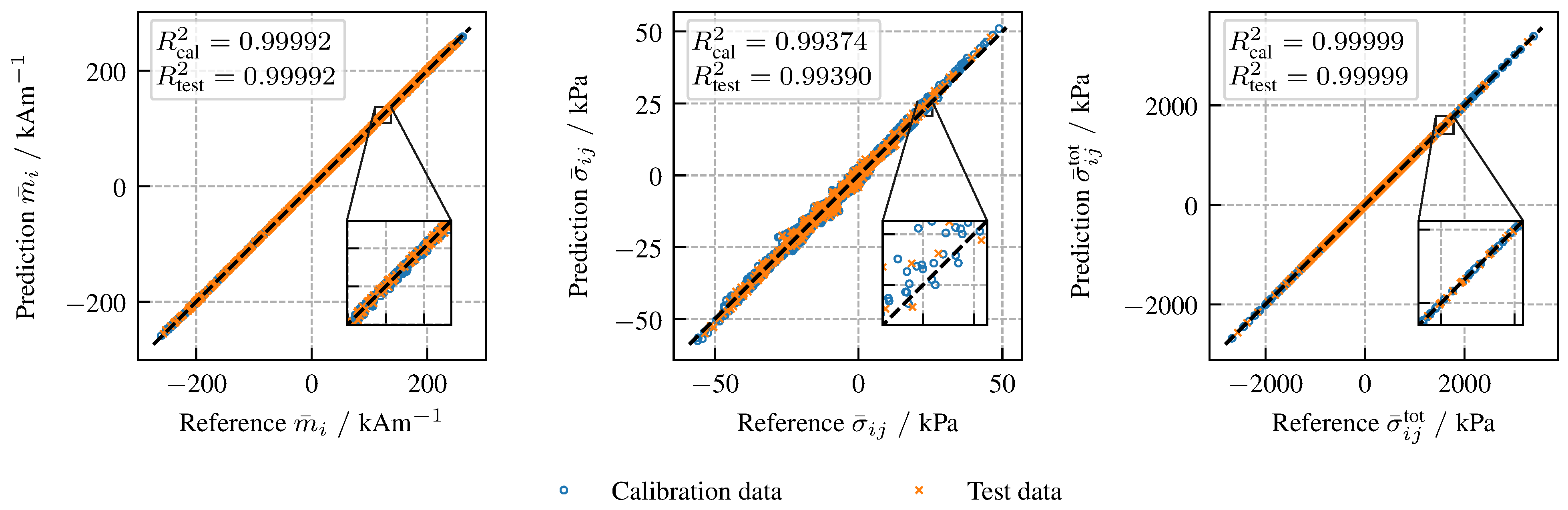}
	\end{center}
	\caption{Predictions of the conventional model~\eqref{eq:GebhartModel} for the compressible case compared to reference values obtained from 10 magneto-mechanical SVE simulations each: Shown are magnetization $\bve m$, mechanical stress $\bte \sigma = \bte \sigma^\text{tot} - \bte \sigma^\text{pon}$ and total stress $\bte \sigma^\text{tot}$. The ratio of calibration and test data  is $70/30$ and the parametrization has been done according to the procedure in Tab.~\ref{tab:fitting_Geb}.}
	\label{fig:corr_Geb_comp}
\end{figure}

To start with, the performance of the conventional macroscopic model \eqref{eq:GebhartModel} is analyzed. This model was calibrated by applying the tailored parametrization procedure given in Tab.~\ref{tab:fitting_Geb}, which is implemented in Matlab \cite{Gebhart2022,Gebhart2022a}.

First, the elastic part of the potential was fitted for the compressible MAP by using the set $\mathcal D^\text{mech}$ comprising only purely mechanical states. The calibrated purely elastic model part shows a very good prediction quality. This is assessed by the \emph{coefficient of determination} $R^2$, which is defined by 
\begin{align}
	R^2 := 1 - \frac{\alpha}{\beta} \text{ with } \alpha := \sum_{i=1}^{n} \left({}^iy^\text{ref} - {}^iy^\text{pred}\right)^2 \; ,
	\beta := \sum_{i=1}^{n} \left({}^iy^\text{ref} - \frac{1}{n}\sum_{j=1}^{n} {}^jy^\text{ref}\right)^2
\end{align}
for a set with $n$ states. A value of $R^2=1$ thus marks a perfect fit. If tensor-valued quantities are analyzed, a summation over the components is done in addition. For calibration and test data, values of $R^2_\text{cal}=0.99978$ and  $R^2_\text{test}=0.99976$ are achieved. Thus, the elastic ground stress can be predicted with high accuracy by the Mooney-Rivlin model \eqref{eq:Rivlin}. Note that the elastic model is also polyconvex, which is achieved by the restrictions for the parameters.

Afterwards, by fixing the parameters of the elastic potential part, the remaining parameters were determined.
The predictions for the coupled dataset $\mathcal D^\text{coup}$ are given in Fig.~\ref{fig:corr_Geb_comp}, where magnetization $\bve m$, stress $\bte \sigma$ and total stress $\bte \sigma^\text{tot}$ are shown. As can be seen, the predictions for $\bve m$ and $\bte \sigma^\text{tot}$ are close to perfect. Also, the predictions for $\bte \sigma$ are still quite good, but nevertheless significantly worse compared to $\bve m$ and $\bte \sigma^\text{tot}$. It can be further seen here that the total stress, which is dominated by the quadratic Maxwell stress contribution $\bte \sigma^\text{Max}$ given in Eq.~\eqref{eq:pon}, is about two orders of magnitude larger than $\bte \sigma$. However, again, for a macroscopic surrogate model for MAPs, it is absolutely important to predict $\bte \sigma$ with sufficient accuracy, which is due to the fact that this quantity is linked to macroscopic effects like the \emph{magnetostictive} or \emph{magnetorheological effect}, cf. Remark~\ref{rmk:total_stress}. 

The parametrized total energy density function $\bar \psi^\text{Geb}(\bte F, \bve B)$ is not polyconvex. This is due to the model structure, which does not allow the required positive semi-definite Hessian for all permissible states.
However, it fulfills the \emph{local polyconvexity} and thus also the \emph{local ellipticity} for all $\bte F$-$\bve B$ states included in $\mathcal D^\text{coup}$. Furthermore, the \emph{domain}  $\mathscr{P\!o\!l\!y}\subset\GL\times\Ln_1$, in which the local polyconvexity in the 2D version as given in \ref{app:poly_ell_2D} is fulfilled, was determined numerically\footnote{A number \num{1e7} states have been generated with LHS and the eigenvalues of the Hessian have been checked for these.} and is given by $\phi_3\in [0,2\pi]$, $\Phi_3\in [0,\pi]$, $\bar\lambda\in [0.3,3.0]$, $\bar J\in [0.3,3.0]$, $B\in [0,4]\,\text{T}$, $\theta\in [0,2\pi]$ with
\begin{align}
	\bte F = \bte R_{x_3}(\phi_3) \cdot \te U \text{ with } \te U := \bte R_{X_3}(\Phi_3) \cdot \diag(\bar \lambda, \bar \lambda^{-1} \bar J, 1) \cdot \bte R_{X_3}^T(\Phi_3) \text{ and } \bve B = \bar B (\cos \theta \ve \, \ve e_1 + \sin \theta \ve \, \ve e_2) \; . \label{eq:polyDomain}
\end{align}
From a practical point of view, the size of this domain is quite sufficient for the simulation of MAPs. In order to predict the complex magneto-mechanical material response with improved accuracy compared to the conventional model, the NN-based models will be calibrated in the following.

\subsubsection{Polyconvex NN model I}

\begin{figure}
	\begin{center}
		\includegraphics{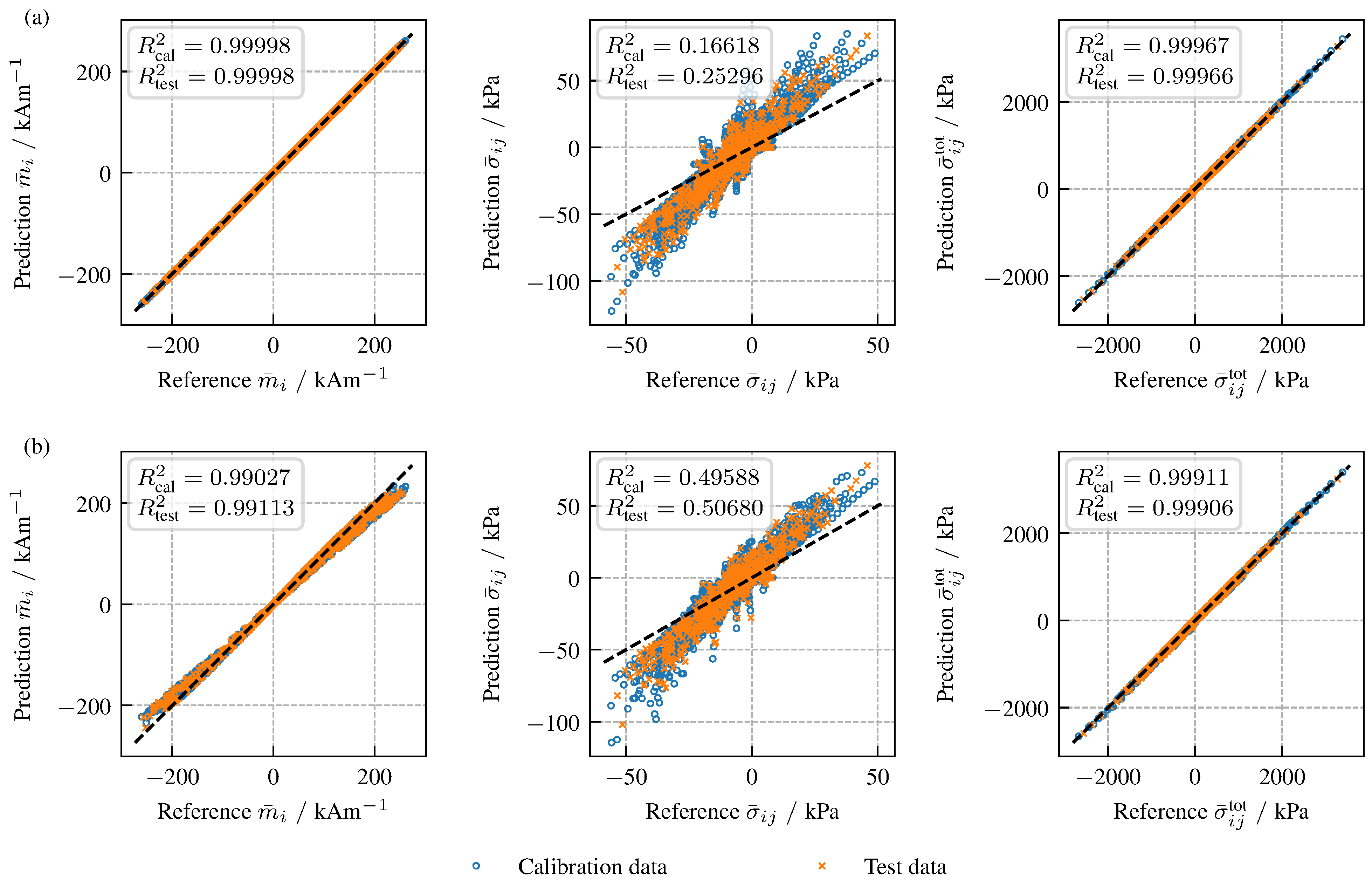}
	\end{center}
	\caption{Predictions of the polyconvex NN~model~I~\eqref{eq:modelI} for the compressible case compared to reference values obtained from 10 magneto-mechanical SVE simulations each: (a) training with the loss $\mathcal L = \mathcal L^{\sigma^\text{tot}} + \mathcal L^m$ and (b) with $\mathcal L = \mathcal L^{\sigma} + \mathcal L^m$. Shown are magnetization $\bve m$, mechanical stress $\bte \sigma = \bte \sigma^\text{tot} - \bte \sigma^\text{pon}$ and total stress $\bte \sigma^\text{tot}$. The ratio of calibration and test data  is $70/30$ and the training has been done with the SLSQP optimizer.}
	\label{fig:corr_polyModel}
\end{figure}

After the model developed by Gebhart~and~Wallmersperger~\cite{Gebhart2022a}, we apply the \emph{polyconvex NN~model~I} given in Eq.~\eqref{eq:modelI} to the data for the compressible MAP. One hidden layer with 6 neurons and two hidden layers with 10 neurons each have been chosen for the networks describing $\bar \psi_\text{ICNN}^\text{el}(\bar I_1,\bar I_2,\bar I_3,\bar I_1^*,\bar J)$ and $\bar \psi_\text{ICNN}^\text{cmv}(\bI^{\text{poly},*},\bar J)$, respectively.\footnote{These hyperparameters have been shown to be suitable for the NN models. An extensive hyperparameter study is not done in this work.} Furthermore, the \emph{softplus activation function} is used. The \emph{non-trainable constant} $\lambda_\text{gro}$ introduced in Eq.~\eqref{eq:growth} is pragmatically chosen to $\SI{0.1}{\kilo\pascal}$. Note that this choice is somewhat insignificant, since the growth term is taken into account in the training and the network adapts depending on the value of the constant.

First, the elastic potential was trained according to Eq.~\eqref{eq:cal_el} with data from $\mathcal D^\text{mech}$, where an almost perfect accuracy was achieved: $R^2_\text{cal} =0.999999$ and $R^2_\text{test} =0.999999$.
To check the predictive ability of the model for magneto-mechanical loadings step by step, a calibration with $\bte \sigma^\text{tot}$ and $\bve m$ was then performed, where the elastic part was fixed.
As shown in the correlation plots given in Fig.~\ref{fig:corr_polyModel}(a), these quantities can be predicted with very high precision. The model is also \emph{polyconvex} and thus \emph{global ellipticity} is also guaranteed automatically, i.e. by construction. The stress tensor $\bte \sigma$, however, is predicted very poorly which is not surprising for the chosen loss term $\mathcal L := \mathcal L^{\sigma^\text{tot}} + \mathcal L^m$. 

However, as depicted in Fig.~\ref{fig:corr_polyModel}(b), where the performance of the same polyconvex NN model calibrated with $\mathcal L := \mathcal L^{\sigma} + \mathcal L^m$ is shown, it is still not able to reproduce  $\bte \sigma$ in an adequate way. Obviously, then, the ICNN-based model is not able to accurately describe the behavior of MAPs. This is due to the claimed restrictions on the ICNN architecture, i.e., the convex and non-decreasing softplus activation function and non-negative weights. In the quasi-incompressible case, it is much better, but still far from the satisfactory prediction quality that can be achieved with NNs, cf. Tab.~\ref{tab:loss_inc}. Even increasing the hidden layers or the number of neurons in each layer did not bring any noticeable improvement. So, in summary, the model is too restrictive and should not be used as a macroscopic surrogate model for MAPs. 

\begin{rmk}
	To explain why the ICNN is not able to fit the data, we adapt the technique applied in  Kalina~et~al.~\cite{Kalina2022a} to determine the so-called stress coefficients. Accordingly, we assume a model without split of the total energy density which depends on the polyconvex invariants $\bI:=(\bar I_1,\bar I_2, \bar J, \bar I_4, \bar I_5)$, i.e., $\bar \psi(\bar\I)$. From the constitutive relations~\eqref{eq:constInv} and the definition of the mechanical stress tensor as $\bte \sigma = \bte \sigma^\text{tot} - \bte \sigma^\text{pon}$ we find
	\begin{align}
		\bte \sigma &= \bar{\mathcal F}_\alpha\left[\bte g_\alpha  + \left(\bve b \cdot \bve h_\alpha\right) \one 
		- \bve h_\alpha \otimes \bve b
		\right] - \frac{1}{2\mu_0} |\bve b|^2 \one \; , \; 
		\bve m = - \bar{\mathcal F}_\alpha \bve h_\alpha + \frac{1}{\mu_0} \bve b \text{ with } 
		\label{eq:fit_fs} \\
		\bte g_\alpha&:=\diffp{\bar{\mathcal I}_\alpha}{\bte F} \cdot \cof \left(\bte F^{-1}\right)
		\; , \bve h_\alpha := \diffp{\bar{\mathcal I}_\alpha}{\bve B} \cdot \bte F^{-1} \; .
	\end{align}
	Therein, $\bar{\mathcal F}_\alpha$ are the magneto-elastic equivalents to the stress coefficients, i.e., $\bar{\mathcal F}_\alpha:=\partial_{\bar{\mathcal I}_\alpha} \bar \psi$. Thus, it is assumed that there exists a potential and the data are isotropic.
	Relation~\eqref{eq:fit_fs} enables us to determine the introduced coefficients $\bar{\mathcal F}_\alpha$ for a single data tuple ${}^i\mathcal T:=\left({}^i\bte F, {}^i \bve B, {}^i \bte \sigma, {}^i\bve m\right)$ by a simple least squares fit, cf. \cite{Kalina2022a}. Consequently, no special form of energy is assumed here.
	However, since the derivatives of the ICNN with respect to the invariants can only be non-negative, which is due to the non-decreasing activation functions and the non-negative weights, we claim $\bar{\mathcal F}_\alpha\ge 0$ as a constraint in the least squares fit, except for $\bar{\mathcal F}_2 \in \R$ which corresponds to $\bar{\mathcal I}_2 = \bar J$. The least squares fit for a single state marked with ${}^i(\cdot)$ is thus given by
	\begin{align}
		{}^i \bar{\ve{\mathcal F}} = \underset{\bar{\ve{\mathcal F}} \in \mathcal C}{\arg\min}
		\left(\left\|\bte \sigma({}^i\bte F, {}^i \bve B, \bar{\ve{\mathcal F}}) - {}^i \bte \sigma \right\|^2
		+ \left|\bve m({}^i\bte F, {}^i \bve B, \bar{\ve{\mathcal F}}) - {}^i \bve m \right|^2 
		\right) \; \text{with }
		\mathcal C:= \R_{\ge 0} \times \R \times \R_{\ge 0} \times \R_{\ge 0} \times \R_{\ge 0}
		\; .
		\label{eq:leastSuares}
	\end{align} 
	
	The technique based on Eq.~\eqref{eq:leastSuares} has been exemplarily applied to the 25 states included in one loading path. Thereby, the fit leads to average relative errors of \SI{967.2}{\percent} and \SI{22.7}{\percent} in the norms of $\bte \sigma$ and $\bve m$, respectively. These huge errors illustrate that the behavior of MAPs cannot be described with an ICNN. Note that the applied method allows to determine the optimal values of the derivatives of the energy with respect to the invariants for a single state. Again, only the existence of a potential, the symmetry group and positive derivatives according to the invariants, except for $\bar J$, are assumed. No specific functional relation is postulated. Thus, the values resulting from Eq.~\eqref{eq:leastSuares} are the maximum of accuracy that can be achieved at all in a linear least squares sense, if one chooses an invariant-based approach in combination with ICNNs.
	It should also be noted that this test only shows that an ICNN cannot reproduce the data. If the errors were small, the inversion would not apply, since there must also be convexity for the potential with respect to the invariants.
	
	To strengthen the statement of the test, the least squares fit has been repeated without the non-negativity constraint and with $\bar I_6$ as an additional invariant, i.e., $\bar{\ve{\mathcal F}}\in \R^6$. This leads to significantly lower average errors of \SI{0.029}{\percent} and \SI{0.001}{\percent} for $\bte \sigma$ and $\bve m$, respectively. However, the fit results in negative coefficients $\bar{\mathcal F}_\alpha$ not only for $\bar{\mathcal F}_2$. Note that the errors are not zero, even in the numerical sense, since the data are not perfectly isotropic.
\end{rmk}

\begin{rmk}
	Note that \emph{ICNN-based models} have proven to be very suitable for the modeling of \emph{finite strain elasticity} even in the anisotropic case \cite{Klein2021,Linden2023}. The purely elastic portion can also be reproduced very well here with the ICNN-based model. Such models have also been successfully used for \emph{finite strain electro-elasticity} in \cite{Klein2022}, but, however, there only the electro-mechanical equivalent to the \emph{total stress} tensor is considered. Since for electro-active polymers the actuation fields are applied directly via electrodes, there is a clear difference here compared to MAPs especially for the relevance of the vacuum contribution, see also Remark~\ref{rmk:total_stress}.
\end{rmk}

\subsubsection{NN models II and III with weakly enforced local polyconvexity}

After the discussion on the polyconvex NN model I, we analyze the ability of \emph{NN model II}~\eqref{eq:modelII} and \emph{NN model III}~\eqref{eq:modelIII}, both with the relaxed requirement of the \emph{local polyconvexity condition}~\eqref{eq:localPoly}. 

\begin{figure}
	\begin{center}
		\includegraphics{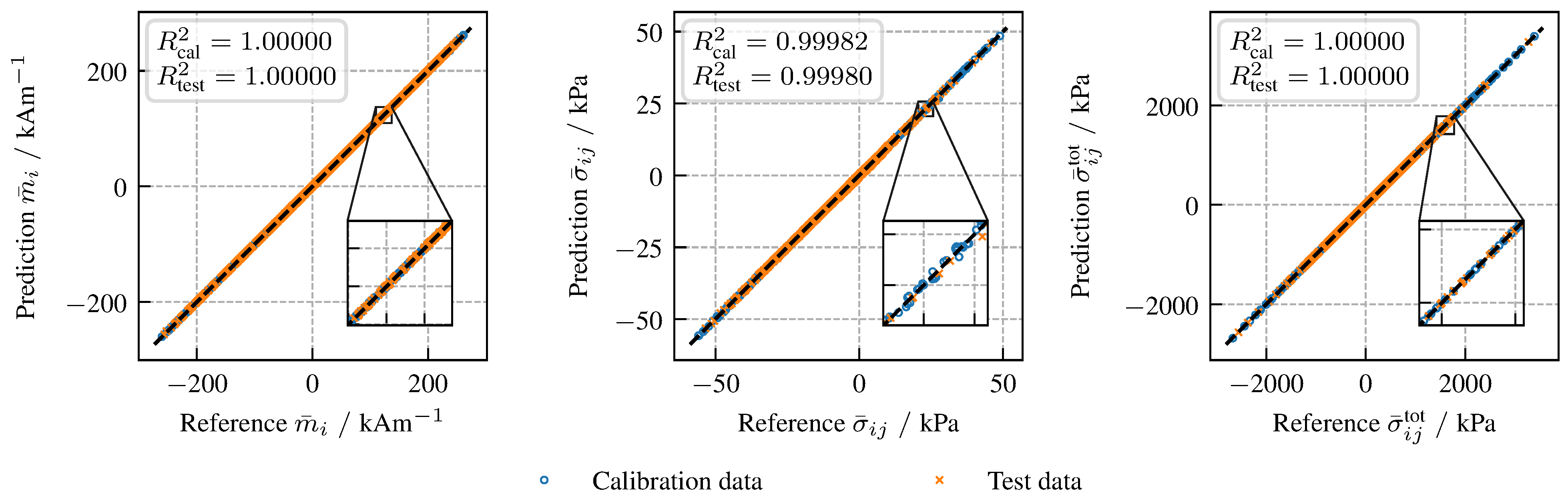}
	\end{center}
	\caption{Predictions of NN~model~II~\eqref{eq:modelII} for the compressible case compared to reference values obtained from 10 magneto-mechanical SVE simulations each: Shown are magnetization $\bve m$, mechanical stress $\bte \sigma = \bte \sigma^\text{tot} - \bte \sigma^\text{pon}$ and total stress $\bte \sigma^\text{tot}$. The ratio of calibration and test data  is $70/30$ and the training has been done with the loss $\mathcal L = \mathcal L^{\sigma} + \mathcal L^m$ and the SLSQP optimizer.}
	\label{fig:corr_nonPolyModel}
\end{figure}

\begin{figure}
	\begin{center}
		\includegraphics{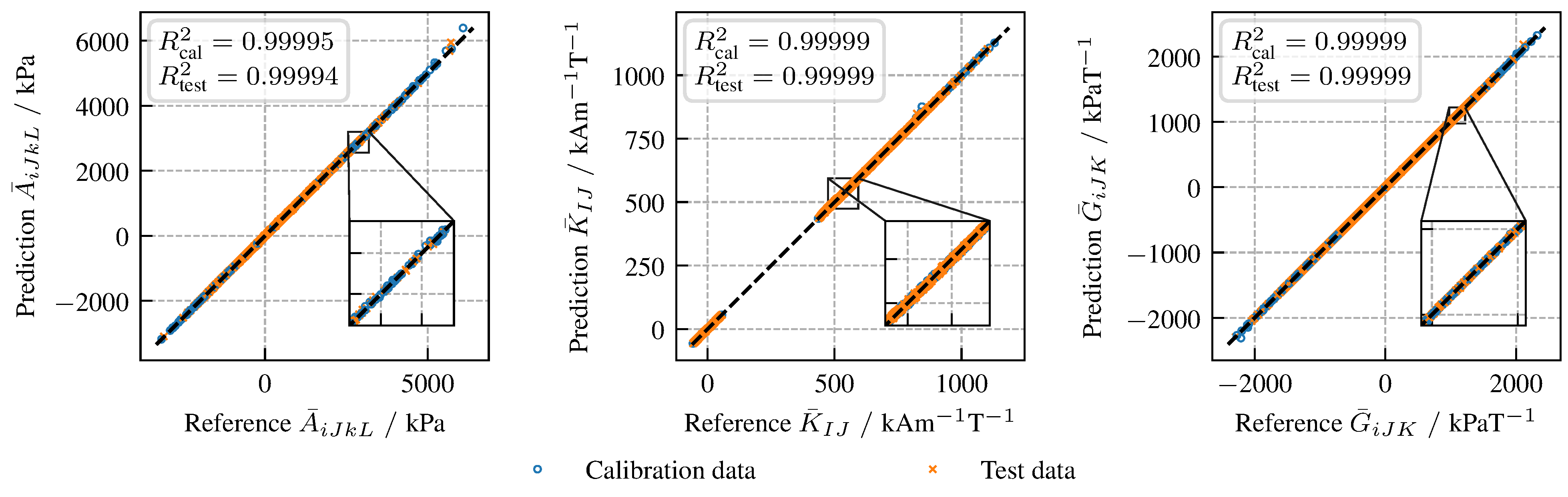}
	\end{center}
	\caption{Predictions of NN~model~II~\eqref{eq:modelII} for the compressible case compared to reference values obtained from 10 magneto-mechanical SVE simulations each: Shown are the coordinates of the tangents $\btttte A\in \Ln_4$, $\bte K \in \Ln_2$ and $\bttte G \in \Ln_3$. The ratio of calibration and test data  is $70/30$ and the training has been done with the loss $\mathcal L = \mathcal L^{\sigma} + \mathcal L^m$ and the SLSQP optimizer.}
	\label{fig:corr_tanNonPolyModel}
\end{figure}

\paragraph{Pre-calibration}
To begin with, NN~model~II is analyzed. The only change of this model to the polyconvex model is that for the description of the potential $\psi^\text{cmv}(\bI,\bar J)$ a PNN with the full invariant set $\bI\in\R^6$ is used instead of the ICNN. The \emph{elastic part} remains the same and is thus still \emph{polyconvex}.
One hidden layer with 6 neurons and two hidden layers with 10 neurons each have been chosen for the networks describing $\bar \psi_\text{ICNN}^\text{el}(\bar I_1,\bar I_2,\bar I_3,\bar I_1^*,\bar J)$ and $\bar \psi_\text{PNN}^\text{cmv}(\bI,\bar J)$, respectively. Again, the \emph{softplus activation function} is used, such that the output of the PNN can be ensured to be greater to zero.
The predictions of the model trained with $\mathcal L:=\mathcal L^\sigma + \mathcal L^m$ for $\bve m$, $\bte \sigma$ and $\bte \sigma^\text{tot}$ are shown in Fig.~\ref{fig:corr_nonPolyModel}. As one can see there, the accuracy of NN~model~II is very good for all three quantities and clearly better than the other two models tested before, i.e. the conventional model and NN~model~I. Thereby, the accurate prediction quality for the mechanical stress $\bte \sigma$ is to be particularly emphasized. As shown in the zoom plot in Fig.~\ref{fig:corr_nonPolyModel}, even this rather difficult to model quantity is predicted almost perfectly. As expected, however, even the \emph{local polyconvexity} is no longer fulfilled for both, $\mathcal D^\text{coup}_\text{cal}$ and $\mathcal D^\text{coup}_\text{test}$. Nonetheless, condition~\eqref{eq:ell_cond} for \emph{strict local ellipticity}  is not violated, although the additional loss term $\mathcal L^\text{ell}$ was not used for training. This is due to the fact that the prediction quality for the first derivatives of the potential, i.e., $\bte P^\text{tot}$ and $\bve H$, are so accurate that also the second derivatives, i.e., the \emph{tangent terms} $\btttte A$, $\bte K$ and $\bttte G$, are predicted in an adequate way, cf. Fig.~\ref{fig:corr_tanNonPolyModel}. Thus, since the data itself fulfill the ellipticity locally, also the calibrated NN model does not violate this condition. The simplified 2D condition~\eqref{eq:ell_2D} has been checked by sampling the unit vectors $\ve N(\phi), \ve V(\phi) \in\Vn$ in 180 steps for the angle $\phi\in[0,\pi]$ for each state.

\begin{figure}[b!]
	\begin{center}
		\includegraphics{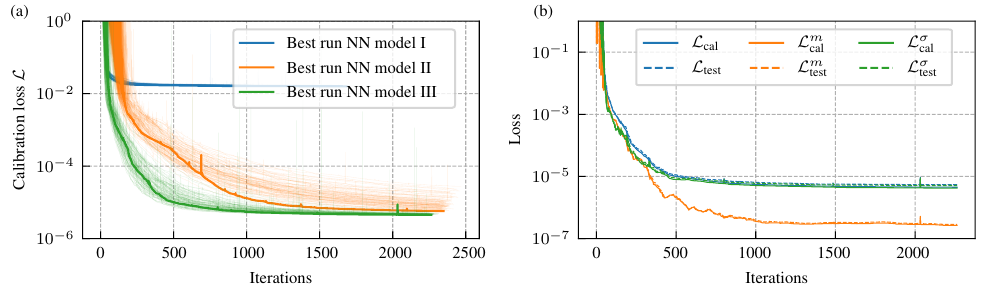}
	\end{center}
	\caption{Training process of NN models with the loss $\mathcal L = \mathcal L^{\sigma} + \mathcal L^m$ and the SLSQP optimizer: (a) overall loss courses of the polyconvex NN model I and the NN models II and III for 100 training runs each, and (b) calibration and test losses for the best training run of model III. The data from the compressible MAP were used and a split of calibration and test data with a ratio of $70/30$ has been done.}
	\label{fig:SLSQP}
\end{figure}

In the next step, before further investigating the enforcement of local polyconvexity, we compare \emph{NN model II} with \emph{NN model III}. In contrast to the other NN-based models, the latter one is closer to the structure of a conventional macroscopic model, i.e., a split into \emph{elastic, coupling, magnetic} and non-trainable \emph{vacuum} contribution is used. One hidden layer with 6 neurons, two hidden layers with 10 neurons each and one hidden layer with 5 neurons have been chosen for the networks describing $\bar \psi^\text{el}_\text{ICNN}(\bar I_1,\bar I_2,\bar I_3,\bar I_1^*,\bar J)$, $\bar \psi_\text{FNN}^\text{coup}(\bI,\bar J)$ and $\bar \psi_\text{FNN}^\text{mag}(\bar I_\text{mag})$, respectively. For all, the softplus activation function is used. The training of both NN models for $\mathcal L:=\mathcal L^\sigma + \mathcal L^m$ is compared in Fig.~\ref{fig:SLSQP}(a), where 100 training runs have been done. Also NN~model~I is included for completeness. For each model, the 100 runs are plotted as thin lines and the best run is marked with a thick line, respectively. Accordingly, the training of \emph{model~III} is more targeted, i.e., it converges faster, and the final loss is lower compared to \emph{model~II}. This can be explained in particular by the incorporation of the non-trainable vacuum part $\bar \psi^\text{vac}(\bar I_\text{vac})$. This way, the model does not have to learn it during the calibration process. Note again that the vacuum energy contribution is independent of the constitutive behavior. The loss curves for calibration and test data are shown in Fig.~\ref{fig:SLSQP}(b) for NN~model~III. As can be seen, the model \emph{generalizes very well}, which is evident from the fact that the loss terms for calibration and test are nearly the same. Furthermore, one can see that the magnetization $\bve m$ can be learned with higher precision compared to the stress $\bte \sigma$.

Thus, altogether, \emph{NN~model~III} shows the best performance of all three NN-based approaches. The results for the quasi-incompressible MAP are similar to that. The conventional model as well as all introduced NN models are compared by the number of trainable variables and the loss terms $\mathcal L^\sigma$, $\mathcal L^m$, $\mathcal L^\text{ell}$ and $\mathcal L^\text{poly}$ for both, the compressible and the quasi-incompressible MAP in Tabs.~\ref{tab:loss_comp} and \ref{tab:loss_inc}. It can be seen that the performance of the models is generally better for the quasi-incompressible case, since the MAP's effective material response is less complex here. 

\begin{figure}
	\begin{center}
		\includegraphics{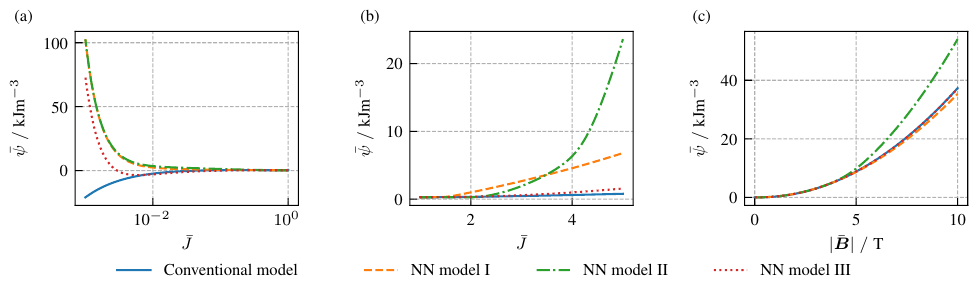}
	\end{center}
	\caption{Analysis of volumetric and magnetic growth conditions given in Eq.~\eqref{eq:growthConditions}: (a) compression loading by $\bte F=\bar J^{1/3} \one$, $\bve B = (1 \ve e_1) \, \text{T}$ with $\bar J\in[1,10^{-3}]$, (b) expansion loading by $\bte F=\bar J^{1/3} \one$, $\bve B = (1 \ve e_1) \, \text{T}$ with $\bar J\in[1,5]$, and (c) magnetic loading by $\bte F=\one$, $\bve B = |\bve B| \ve e_1$ with $|\bve B|\in[0,10]\,\text{T}$.}
	\label{fig:growth}
\end{figure}

After the comparison of the prediction quality, we now focus on the fulfillment of the growth conditions formulated in Eq.~\eqref{eq:growthConditions}, namely the \emph{volumetric growth condition} and the \emph{magnetic growth condition}. Both of these conditions are fulfilled by construction for the NN~models~I~and~II, which is not the case for NN~model~III and the conventional model. Thus, a numerical test has been done for specific loadcases. It is shown in Fig.~\ref{fig:growth}(a) and (b) for the volumetric growth condition, where compression and expansion are considered, i.e. $\bte F=\bar J^{1/3} \one$ and it is chosen $\bve B = (1 \ve e_1) \, \text{T}$. As can be seen, the conventional model does not fulfill the growth condition for compression. In contrast, for NN~model~III, a clear nonlinear growth of the energy for compression occurs for small values of $\bar J$. However, negative values for energy occur before this growth region.
Regarding volumetric expansion, positive energies occur for all models, whereby NN~model~II already shows a clearly progressive growth for $\bar J>2.5$. 
Regarding the magnetic growth condition, progressively increasing curves  can be seen for all models, see Fig.~\ref{fig:growth}(c).

\begin{table}
	\begin{center}
		\caption{Comparison of the macroscopic models by parameters and loss terms for the compressible MAP. All NN-based models were trained 100 times with the loss $\mathcal L = \mathcal L^{\sigma} + \mathcal L^m$, where the best training run was selected. The conventional model was parametrized according to Tab.~\ref{tab:fitting_Geb}. It should be noted that the calculation of individual loss terms is used here for model evaluation.}
		\label{tab:loss_comp}
		\begin{footnotesize}
			\begin{tabular}{llllll|ll}
				Model & Parameters & $\mathcal L^\sigma_\text{cal}$ & $\mathcal L^\sigma_\text{test}$ &  $\mathcal L^m_\text{cal}$ & $\mathcal L^m_\text{test}$ & $\mathcal L^\text{ell}$ & $\mathcal L^\text{poly}$\\
				\hline\hline
				Conventional model & 11 & \num{1.756e-04} & \num{1.941e-04} & \num{1.377e-05} & \num{1.473e-05} & \num{0.000e+00} & \num{0.000e+00}\\ 
				NN model I & 319 & \num{1.414e-02} & \num{1.571e-02} & \num{1.773e-03} & \num{1.687e-03} & \num{0.000e+00} & \num{0.000e+00}\\ 
				NN model II & 232 & \num{5.165e-06} & \num{6.278e-06} & \num{5.892e-07} & \num{6.382e-07} & \num{0.000e+00} & \num{2.801e-02}\\ 
				NN model III & 248 & \num{4.250e-06} & \num{5.084e-06} & \num{2.327e-07} & \num{2.400e-07} & \num{0.000e+00} & \num{1.279e-03}
			\end{tabular}
		\end{footnotesize}
	\end{center}
\end{table}

\begin{table}
	\begin{center}
		\caption{Comparison of the macroscopic models by parameters and loss terms for the quasi-incompressible MAP. All NN-based models were trained 100 times with the loss $\mathcal L = \mathcal L^{\sigma} + \mathcal L^m$, where the best training run was selected. The conventional model was parametrized according to Tab.~\ref{tab:fitting_Geb}. It should be noted that the calculation of individual loss terms is used here for model evaluation.}
		\label{tab:loss_inc}
		\begin{footnotesize}
			\begin{tabular}{llllll|ll}
				Model & Parameters & $\mathcal L^\sigma_\text{cal}$ & $\mathcal L^\sigma_\text{test}$ &  $\mathcal L^m_\text{cal}$ & $\mathcal L^m_\text{test}$ & $\mathcal L^\text{ell}$ & $\mathcal L^\text{poly}$\\
				\hline\hline
				Conventional model & 11 & \num{7.850e-05} & \num{8.547e-05} & \num{1.637e-05} & \num{1.675e-05} & \num{0.000e+00} & \num{0.000e+00}\\ 
				NN model I & 319 & \num{8.282e-04} & \num{8.779e-04} & \num{3.989e-06} & \num{4.111e-06} & \num{0.000e+00} & \num{0.000e+00}\\ 
				NN model II & 232 & \num{2.441e-06} & \num{2.590e-06} & \num{3.188e-07} & \num{3.199e-07} & \num{0.000e+00} & \num{1.059e-02}\\ 
				NN model III & 248 & \num{1.883e-06} & \num{1.947e-06} & \num{2.302e-07} & \num{2.320e-07} & \num{0.000e+00} & \num{3.089e-03}
			\end{tabular}
		\end{footnotesize}
	\end{center}
\end{table}

\begin{figure}
	\begin{center}
		\includegraphics{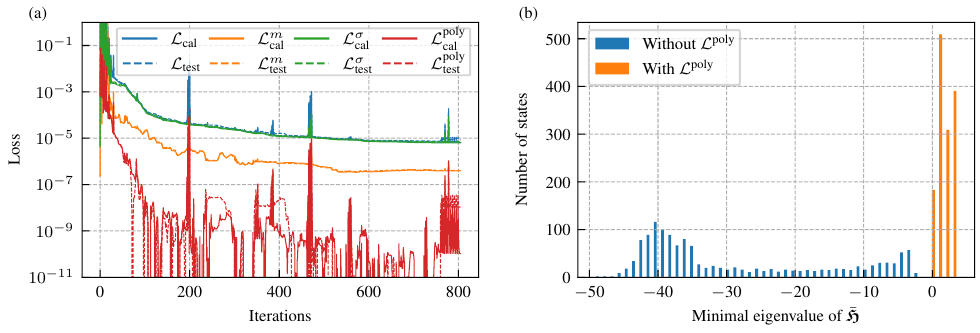}
	\end{center}
	\caption{Post-training of the NN~model~III~\eqref{eq:modelIII} with an additional penalty term for local polyconvexity: (a) calibration and test losses for training with $\mathcal L = \mathcal L^\sigma + \mathcal L^m + 100\, \mathcal L^\text{poly}$ and (b) minimal eigenvalues of the Hessian $\gt H$ for all states in the dataset $\mathcal D^\text{coup}$ for the compressible MAP with and without post-training with $\mathcal L^\text{poly}$. Data were split into $70/30$ calibration/test data.}
	\label{fig:poly}
\end{figure}

\paragraph{Post-calibration}
After the detailed comparison of all models, we now investigate the extent to which \emph{local polyconvexity} can be incorporated into the model in a \emph{weak sense}, i.e., by adding the loss term $\mathcal L^\text{poly}$ given in Eq.~\eqref{eq:loss_poly}. At this point we only investigate the behavior for \emph{NN~model~III}. The network architecture remains the same. After pre-calibration with $\mathcal L:=\mathcal L^\sigma + \mathcal L^m$, a post-training step with $\mathcal L:=\mathcal L^\sigma + \mathcal L^m + 100\,\mathcal L^\text{poly}$ has been done. It is exemplarily shown in Fig.~\ref{fig:poly}(a) for the compressible MAP. As depicted there, the loss terms $\mathcal L^\sigma$ and $\mathcal L^m$ of the \emph{pre-trained model} increase again at the beginning of the training, because the \emph{local polyconvexity} is violated and the corresponding penalty term has a great influence due to the chosen weight ratio of $w_\text{poly}/w_\sigma = w_\text{poly}/w_m = 100$. After a few hundred iterations, these loss terms then fall back to around their pre-training levels.
The loss for local polyconvexity is also on a low level of around \num{1e-9}, where the oscillations are due to the fact that this loss can rapidly change the value if a state changes from negative eigenvalues of $\bar{\gt H}$ to positive ones. In the end, the loss for local polyconvexity do not drop to zero. However, the post-training process is accompanied by the fact that the minimum \emph{eigenvalues} of the Hessian $\bar{\gt H}$ are shifted from the negative to the positive range, cf. Fig.~\ref{fig:poly}(b). Post-training was performed only once for each data set, using the best of the 100 pre-trained models for initialization in each case. Note that the post training process is more expensive compared to the pre-training, which is due to the fact the \emph{second derivatives} of the energy with respect to the set $\bar{\gt I}$ have to be calculated in each iteration step, cf. Eq.~\eqref{eq:compHessian}. However, although a higher order Sobolev training must be performed, the training can be performed within a few minutes on a conventional laptop.

\begin{figure}
	\begin{center}
		\includegraphics{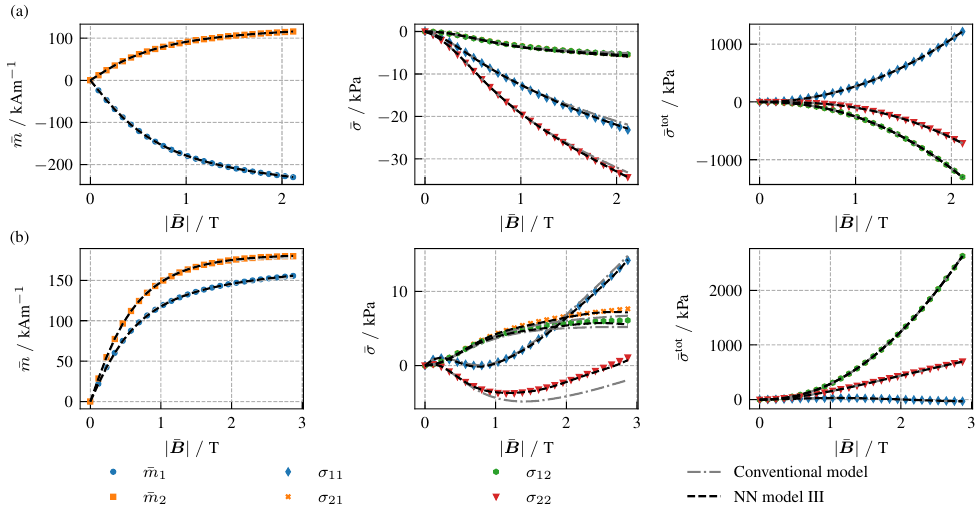}
	\end{center}
	\caption{Comparison of the predictions of the conventional model~\eqref{eq:GebhartModel} calibrated according to the procedure in Tab.~\ref{tab:fitting_Geb} and the NN~model~III trained with the loss  $\mathcal L = \mathcal L^\sigma + \mathcal L^m + 100\, \mathcal L^\text{poly}$ to the homogenized reference for two load cases. The final states are specified by: (a) $\bte F=0.989\, \ve e_1\otimes\ve e_1 + 0.965\, \ve e_2 \otimes \ve e_2 + \ve e_3 \otimes \ve e_3$ and $\bve B=(-1.88\,\ve e_1+0.97\,\ve e_2)\,\text{T}$, as well as (b) $\bte F=1.106\, \ve e_1\otimes\ve e_1 + 0.973\, \ve e_2 \otimes \ve e_2 + \ve e_3 \otimes \ve e_3$ and $\bve B=(1.74\,\ve e_1+2.28\,\ve e_2)\,\text{T}$. The case of the compressible MAP is shown.}
	\label{fig:interpolation}
\end{figure}

\begin{figure}
	\begin{center}
		\includegraphics{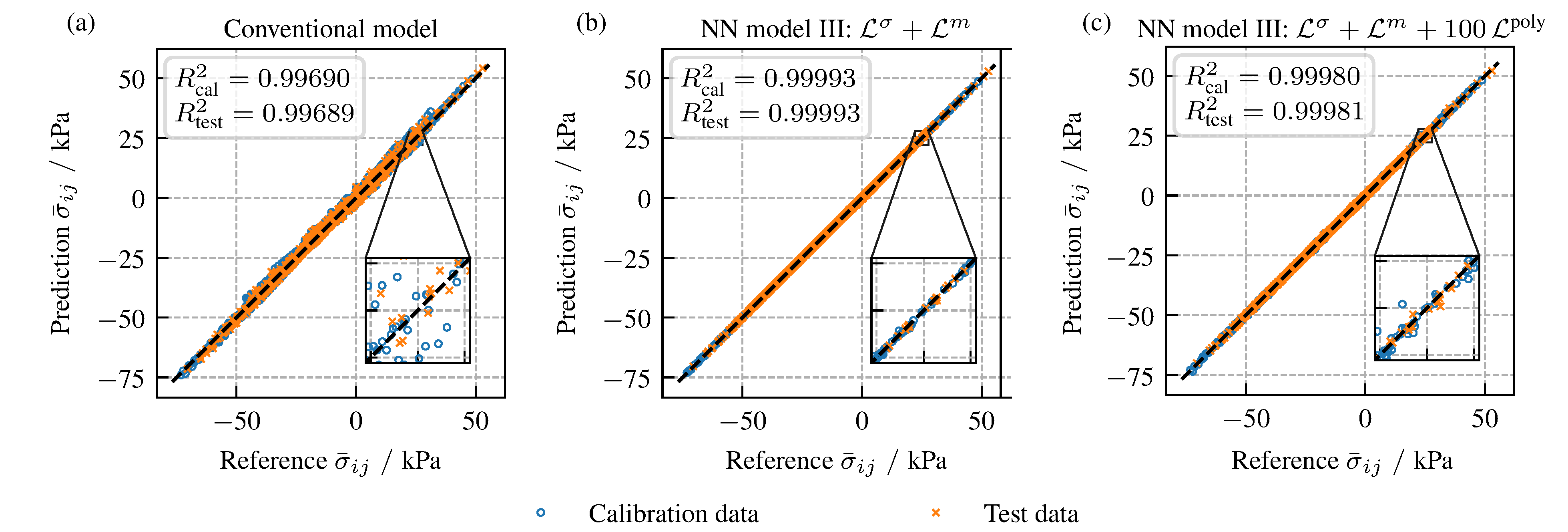}
	\end{center}
	\caption{Comparison of the model predictions for the mechanical stress $\bte \sigma$ for the quasi-incompressible MAP: (a) conventional model~\eqref{eq:GebhartModel}, (b) NN~model~III trained with $\mathcal L = \mathcal L^\sigma + \mathcal L^m$, and (c) NN~model~III post-trained with $\mathcal L = \mathcal L^\sigma + \mathcal L^m + 100\, \mathcal L^\text{poly}$. The reference data were generated by magneto-mechanical simulations of 10 SVEs each and a division into 70/30 calibration and test data has been done.}
	\label{fig:interpolationInc}
\end{figure}

All loss terms for the post-training of NN~model~III are given in Tab.~\ref{tab:post} for the compressible and the quasi-incompressible case. Interestingly, the loss for the local polyconvexity is equal to zero in the quasi-incompressible case for both calibration and test.
Furthermore, the predictions for two load cases are exemplarily shown in Fig.~\ref{fig:interpolation} for the compressible MAP, where also the conventional model is incorporated. As one can see, both models make \emph{close to perfect predictions} for magnetization $\bve m$ and total stress $\bte \sigma^\text{tot}$, whereas the performance of the NN-based model is significantly improved for $\bte \sigma$. For the quasi-incompressible MAP, correlation plots of $\bte \sigma$ are given in Fig.~\ref{fig:interpolationInc}, where the conventional model is compared to the NN model for training with $\mathcal L:=\mathcal L^\sigma + \mathcal L^m$ as well as $\mathcal L:=\mathcal L^\sigma + \mathcal L^m + 100\, \mathcal L^\text{poly}$. Here, too, it can be seen that the NN model is clearly superior over the conventional one. However, the improved prediction quality comes with the price of more model parameters. 

\begin{table}
	\begin{center}
		\caption{Loss terms for the NN~model~III after post-training with $\mathcal L = \mathcal L^{\sigma} + \mathcal L^m + 100\,\mathcal L^\text{poly}$. It should be noted that the calculation of individual loss terms is used here for model evaluation.}
		\label{tab:post}
		\begin{footnotesize}
			\begin{tabular}{lllllll|l}
				Dataset & $\mathcal L^\sigma_\text{cal}$ & $\mathcal L^\sigma_\text{test}$ &  $\mathcal L^m_\text{cal}$ & $\mathcal L^m_\text{test}$ & $\mathcal L^\text{poly}_\text{cal}$ & $\mathcal L^\text{poly}_\text{test}$ & $\mathcal L^\text{ell}$\\
				\hline\hline
				Compressible & \num{6.272e-06} & \num{7.035e-06} & \num{3.925e-07} & \num{4.117e-07} & \num{1.052e-10} & \num{1.055e-08} & \num{0.000e+00}\\ 
				Quasi-incompressible & \num{5.168e-06} & \num{5.323e-06} & \num{5.091e-07} & \num{4.907e-07} & \num{0.000e+00} & \num{0.000e+00} & \num{0.000e+00}
			\end{tabular}
		\end{footnotesize}
	\end{center}
\end{table}

\begin{table}
	\begin{center}
		\caption{Local polyconvexity domains $\mathscr{P\!o\!l\!y}\subset\GL\times\Ln_1$ for the conventional model and NN~model~III both calibrated with data from the quasi-incompressible MAP. 
			The domains were determined numerically by testing \num{1e7} states generated with LHS. The specification of $\bte F$-$\bve B$ states is given by $\bte F = \bte R_{x_3}(\phi_3) \cdot \bte U \text{ with } \bte U := \bte R_{X_3}(\Phi_3) \cdot \diag(\bar \lambda, \bar \lambda^{-1} \bar J, 1) \cdot \bte R_{X_3}^T(\Phi_3)$  and  $\bve B = \bar B (\cos \theta \ve \, \ve e_1 + \sin \theta \ve \, \ve e_2)$.}
		\label{tab:polyDomain}
		\begin{footnotesize}
			\begin{tabular}{lllllll}
				Model & $\phi_3$ & $\Phi_3$ & $\bar\lambda$ & $\bar J$ & $\bar B \, /\, \text{T}$ & $\theta$\\
				\hline\hline
				Conventional model & $[0,2\pi]$ & $[0,\pi]$ & $[0.3,4.0]$ & $[0.95,1.05]$ & $[0,4]$ & $[0,2\pi]$ \\ 
				NN~model~III & $[0,2\pi]$ & $[0,\pi]$ & $[0.8,1.3]$ & $[0.99,1.01]$ & $[0,3]$ & $[0,2\pi]$
			\end{tabular}
		\end{footnotesize}
	\end{center}
\end{table}

The \emph{local polyconvexity domain} $\mathscr{P\!o\!l\!y}\subset\GL\times\Ln_1$ was not determined for NN~model~III in the compressible case. This is due to the fact the corresponding loss is not equal to zero after training and the condition is thus not fulfilled even for the states in the calibration and test datasets. However, for the quasi-incompressible case, where the loss term $\mathcal L^\text{poly}$ is equal to zero after post training, the domain was  numerically determined by testing \num{1e7} states generated with LHS. The domains are given in Tab.~\ref{tab:polyDomain} for the conventional model and NN~model~III. As one can see, the domain is significantly larger for the conventional model.

\begin{rmk}
	Within the interpolation study, the alternative loss term $\mathcal L^\text{ell}$ is not applied for post-calibration. This is due to the fact that the \emph{local ellipticity} condition was already fulfilled for the considered datasets after the pre-training step with $\mathcal L^\sigma + \mathcal L^m$. However, the proposed loss term can be useful if NN-based models have to calibrated with experimental data or data from MD simulations if even \emph{local polyconvexity} is too restrictive to achieve reasonable compatibility with data.
\end{rmk}

\subsection{Extrapolation behavior of the macroscopic models}

\begin{figure}[b!]
	\begin{center}
		\includegraphics{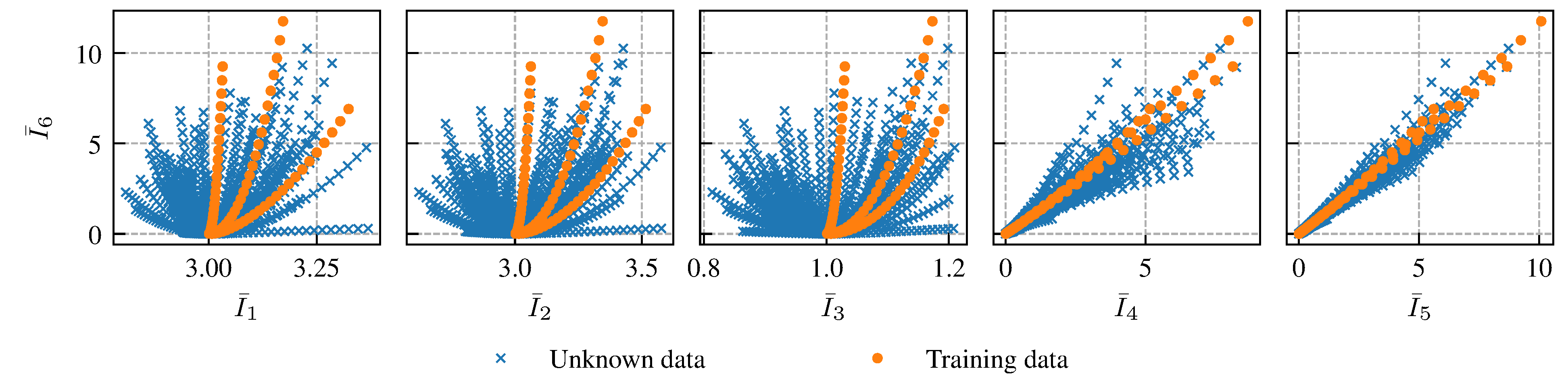}
	\end{center}
	\caption{Dataset $\mathcal D^\text{coup}$ for the extrapolation study with the compressible MAP: sampled invariant space visualized in five sectional planes. Only 3 load cases out of $\mathcal D^\text{coup}$ comprising a total number of 75 tuples are used for calibration.}
	\label{fig:extInv}
\end{figure}

After the detailed analysis of the model's ability to learn the constitutive behavior of MAPs from a large dataset, the \emph{extrapolation behavior} is now studied. Thereby, only the \emph{conventional model}~\eqref{eq:GebhartModel} and \emph{NN~Model~III}~\eqref{eq:modelIII} are considered and we restrict the investigation to the compressible case.
For the extrapolation study, only three load cases of $\mathcal D^\text{mech}$ and $\mathcal D^\text{coup}$ each are used. Thus a total number of 75 purely mechanical and 75 coupled magneto-mechanical states are available for calibration. The projection of the selected $\bte F$-$\bve B$ states from $\mathcal D^\text{coup}$ into the \emph{invariant space} is shown in Fig.~\ref{fig:extInv} within five sectional planes. Accordingly, large parts of the invariant space are not covered by the calibration data for the extrapolation study.

\begin{figure}
	\begin{center}
		\includegraphics{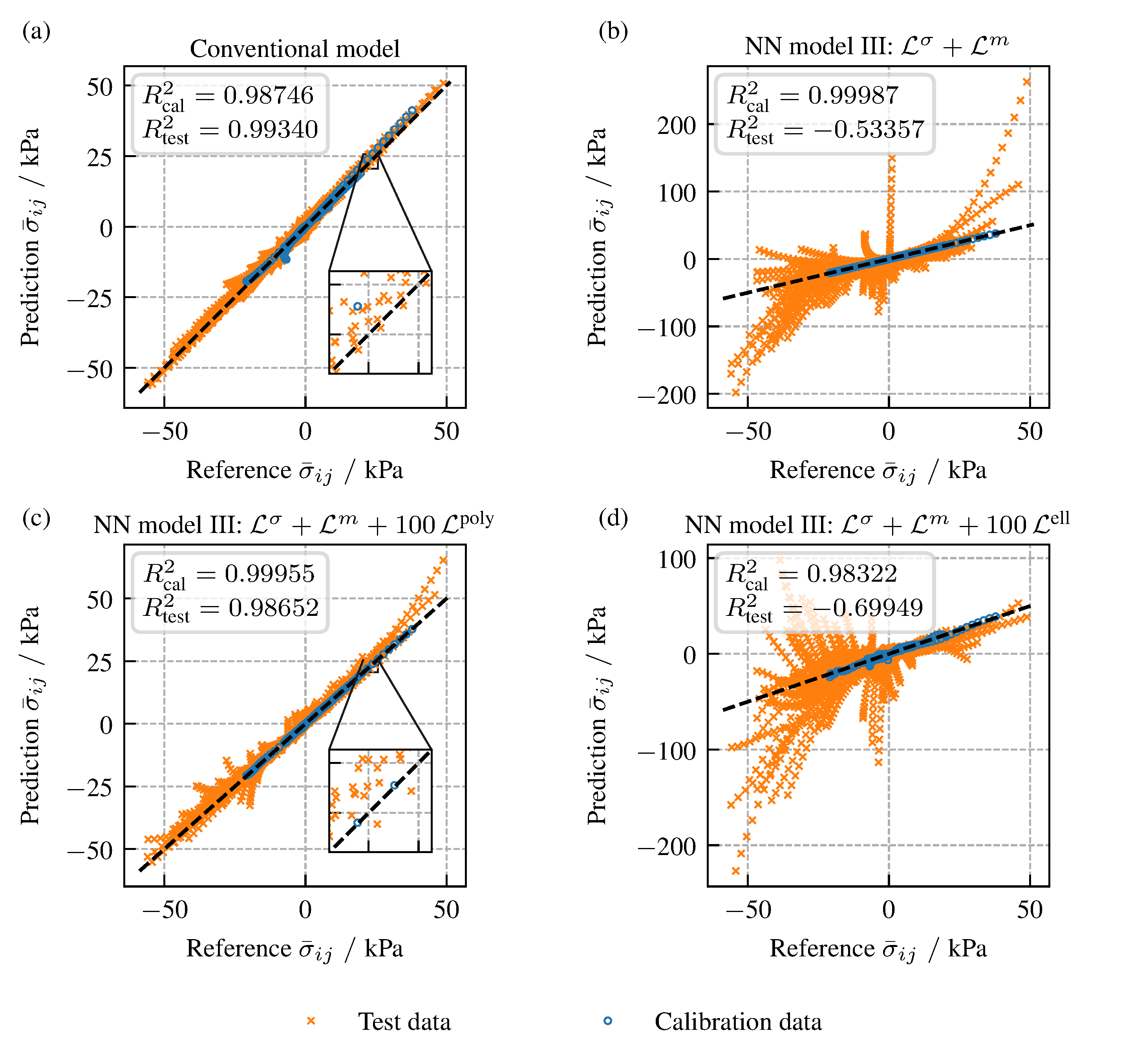}
	\end{center}
	\caption{Comparison of model predictions for the mechanical stress $\bte \sigma$ for extrapolation: (a) conventional model~\eqref{eq:GebhartModel}, (b) NN~model~III trained with $\mathcal L = \mathcal L^\sigma + \mathcal L^m$, (c) NN~model~III post-trained with $\mathcal L = \mathcal L^\sigma + \mathcal L^m + 100\, \mathcal L^\text{poly}$, and (d) NN~model~III post-trained with $\mathcal L = \mathcal L^\sigma + \mathcal L^m + 100\, \mathcal L^\text{ell}$. The reference data were generated by magneto-mechanical simulations of 10 SVEs each. Only 75 tuples of $\mathcal D^\text{mech}$ and $\mathcal D^\text{coup}$ each have been used for the calibration, cf. Fig.~\ref{fig:extInv}.}
	\label{fig:extrapolation}
\end{figure}

\begin{figure}
	\begin{center}
		\includegraphics{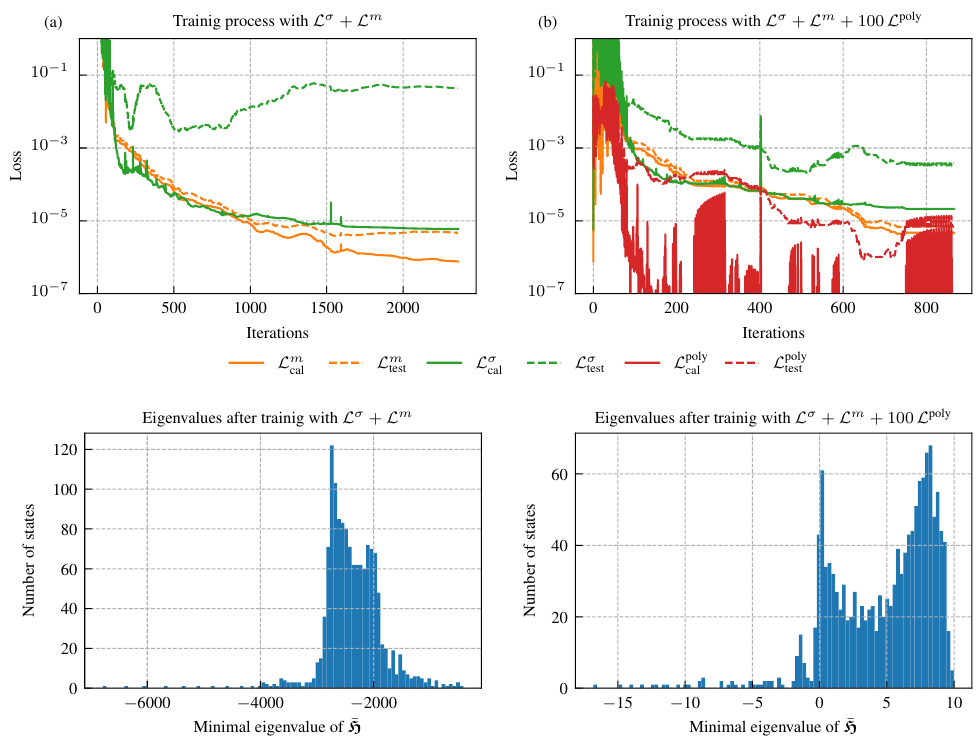}
	\end{center}
	\caption{Behavior of the NN~model~III trained with only 75 tuples from $\mathcal D^\text{mech}$ and $\mathcal D^\text{coup}$ each: (a) training with the loss $\mathcal L = \mathcal L^\sigma + \mathcal L^m$, and (b) post-training with the loss $\mathcal L = \mathcal L^\sigma + \mathcal L^m + 100\, \mathcal L^\text{poly}$. The first line shows the training course and the second line shows the minimum eigenvalues of the Hessian $\gt H$ after training.}
	\label{fig:polyExt}
\end{figure}

With the reduced datasets with $|\mathcal D_\text{cal}^\text{mech}|=75$ and $|\mathcal D_\text{cal}^\text{coup}|=75$, the conventional model has been calibrated by using the procedure introduced in Tab.~\ref{tab:fitting_Geb}. Since the mechanical stress has proven to be the most complicated variable to predict, we will now focus only on it in the discussion. Regarding the correlation plot for the mechanical stress $\bte \sigma$ in Fig.~\ref{fig:extrapolation}(a), one can see that the predictions of the conventional model are still good. Surprisingly, they are in the same range as for the model trained with the larger dataset, cf. Fig.~\ref{fig:corr_Geb_comp}. It is thus sufficient to calibrate this model with only a few load cases. This favorable behavior results from the predefined model structure compared to more general NN approaches.

The results for \emph{NN~model~III} trained with the reduced data and $\mathcal L:= \mathcal L^\sigma + \mathcal L^m$ are given in Fig.~\ref{fig:extrapolation}(b). As shown there, the calibration data have been learned quite well. However, the stress predictions for the test data, which require the model to extrapolate from the calibration domain, are very poor. The coefficient $R^2_\text{test} = -0.5337$ is even negative here. Thus, the model is not able to make good predictions for unknown data if the sparse calibration data set is used.

Finally, a \emph{post-training} of the \emph{NN~model~III} with the loss $\mathcal L:=\mathcal L^\sigma + \mathcal L^m + 100\, \mathcal L^\text{poly}$ has been done. To reinforce the influence of the polyconvex penalty term, the weight $w_\text{poly}$ has been chosen to 100 here again. The predictions of the \emph{post-trained model} are depicted in Fig.~\ref{fig:extrapolation}(c). As shown there, the performance for the calibration data remains approximately the same. However, the prediction quality for the test data is now improved drastically. Although such behavior has been observed for ICNNs when applied to finite strain elasticity \cite{Linden2023}, it is quite surprising here since the introduced penalty loss for \emph{local polyconvexity} can enforce the condition \eqref{eq:localPoly} only in a weak sense and only for the calibration data. 
Training course and and minimal eigenvalues after the calibration are shown in Fig.~\ref{fig:polyExt}(a) and (b) for pre-training and post-training, respectively. The plot shows again clearly how the test loss for sigma improves in post-training. Furthermore, it can be seen that the entire smallest eigenvalues of the Hessian $\bar{\gt H}$ are shifted to the right and this despite the fact that only 75 data tuples of $\mathcal D^\text{coup}$ are used in the post-training. Furthermore, although the model is not locally polyconvex for all states of the data set, it is \emph{locally elliptic} for all these states after post-training, which was not the case before.

\begin{figure}
	\begin{center}
		\includegraphics{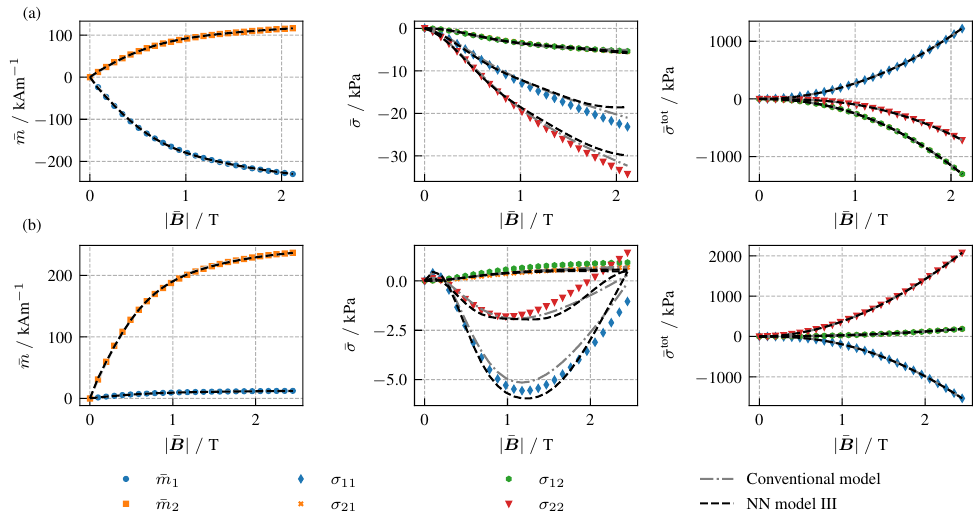}
	\end{center}
	\caption{Comparison of the extrapolation behavior of the conventional model~\eqref{eq:GebhartModel} and the NN~model~III~\eqref{eq:modelIII} post-trained with $\mathcal L = \mathcal L^\sigma + \mathcal L^m + 100\, \mathcal L^\text{poly}$. The final states of the unseen load cases are specified by: (a) $\bte F=0.989\, \ve e_1\otimes\ve e_1 + 0.965\, \ve e_2 \otimes \ve e_2 + \ve e_3 \otimes \ve e_3$ and $\bve B=(-1.88\,\ve e_1+0.97\,\ve e_2)\,\text{T}$, as well as (b) $\bte F=1.069\, \ve e_1\otimes\ve e_1 + 0.980\, \ve e_2 \otimes \ve e_2 + \ve e_3 \otimes \ve e_3$ and $\bve B=(0.115\,\ve e_1+2.445\,\ve e_2)\,\text{T}$. Both models were calibrated with only 75 tuples from $\mathcal D^\text{mech}$ and $\mathcal D^\text{coup}$ each, respectively.}
	\label{fig:LCsExtrapolation}
\end{figure}

The model predictions of the conventional model and NN~model~III post-trained with the loss for local polyconvexity are given in Fig.~\ref{fig:LCsExtrapolation} for two selected load cases. As shown there, the predictions are well for small induction fields, i.e., in the linear magnetic regime with $\bve m \propto \bve b$, and get then worse for higher inductions. This is not surprising, since the magneto-mechanical behavior of MAPs typically gets much more complicated when the particles' magnetic saturation behavior begins.

Additionally, for comparative reasons, the predictions of NN~model~III after post-training with $\mathcal L:=\mathcal L^\sigma + \mathcal L^m + 100\, \mathcal L^\text{ell}$ are given in Fig.~\ref{fig:extrapolation}(d). As can be seen, the predictions for the unknown data are still poor here. Apparently, this post-training has no special influence on the extrapolation ability. It should be noted here, that local polyconvexity is a more rigorous requirement than local ellipticity. The former is sufficient for the latter but not necessary.

\section{Conclusions}\label{sec:conc}

In the present work, an \emph{NN-based} approach  for the modeling of \emph{isotropic finite strain magneto-elasticity} is proposed. Three models are formulated in such a way, that they account for a reasonable number of \emph{physical conditions} a priori, i.e., by construction, whereby these models are denoted as PANN \cite{Linden2023,Klein2022}. Their prediction ability is investigated and compared to conventional models by a calibration to data generated with a \emph{computational homogenization} approach, where compressible as well quasi-incompressible MAPs are considered.

Starting with basic finite strain kinematics and a summary of the Maxwell and balance equations, an introduction to magneto-elasticity is given. This also includes a discussion on \emph{global and local ellipticity} and \emph{polyconvextiy}. Furthermore, a \emph{relaxed local polyconvexity}, which implies local ellipticity, is introduced. In addition, a Hill-type computational homogenization framework is summarized. 
Appropriate microscale constitutive models for the magnetizable particles and the elastic matrix are formulated afterwards. In the following section, based on
the given theoretical basis, a conventional macroscopic model for MAPs from the literature \cite{Gebhart2022a} is introduced and a total of three NN-based macroscopic models are developed. This also includes appropriate loss terms for training. 
Afterwards, the models are applied as macroscopic surrogates for \emph{computational homogenizations} in an examples part. For the data generation, an \emph{SVE} approach, which is applied to keep the size of the cells for the homogenization in a reasonable range, is used. In addition, an invariant-based pre-sampling leaned to \cite{Kalina2022a,Kalina2023} is applied. With the generated datasets, a detailed analysis of the models' interpolation and extrapolation capability has been done. It turns out that \emph{ICNNs}, which allow to construct polyconvex models and have been shown to be very suitable for modeling anisotropic finite strain elasticity \cite{Klein2021,Linden2023}, are too restrictive for modeling MAPs. Instead, a model which is not polyconvex but is able to fulfill the less restrictive \emph{local polyconvexity} has shown to be favorable. Thereby, this condition can be enforced in \emph{weak sense} via an additional loss term. This has been shown to have no negative effect on the approximation quality. The investigation of the model's extrapolation behavior even shows that this leads to a remarkable improvement. Compared to the conventional model, the proposed NN-based approach is able to provide clearly better predictions for interpolation and similar quality for extrapolation. As already mentioned, this comes with the cost of a significantly increased number of model parameters. However, these parameters, namely weights and biases, can be determined in an automated manner by applying efficient training approaches.

Summarizing, the presented NN-based models have shown to be highly accurate \emph{surrogate models} for computationally expensive SVE simulations for finite strain magneto-elasticity. The incorporation of physical principles into the NN models in an a priori manner ensures that even for extrapolation the underlying physics are not violated and thus a good generalization is ensured. This also enables comparatively small network architectures, see also \cite{Klein2021,Kalina2023}. However, it has been shown that polyconvexity in particular, which is enforced via ICNNs, is a condition which is to strong.
A weakened requirement, the introduced \emph{local polyconvexity}, is favorable in order to not restrict the approximation quality too much.

Several applications and extensions of our approach are planned in the future. For example, the models are to be tested using 3D data, whereby the data generation is a non-trivial task for MAPs. In addition, an extension to \emph{anisotropic magneto-elasticity}, which enables the modeling of pre-structured MAPs with chain-like particle arrangements \cite{Danas2012,Hiptmair2015}, is planned. Furthermore, in order to exploit the advantages of NN-based models, an integration into \emph{multiscale schemes} as \emph{FE}${}^\textit{ANN}$ \cite{Kalina2023} is planned, so that macroscopic specimens and components can be simulated automatically and with acceptable computational effort and at the same time with high precision.
Finally, an extension to \emph{dissipative} constitutive behavior \cite{Rosenkranz2023,Vlassis2021,Masi2021,Fuhg2023} has to be done.

\section*{Acknowledgement}
All presented computations were performed on a PC-Cluster at the Center for Information Services and High Performance Computing (ZIH) at TU Dresden. The authors thus thank the ZIH for generous allocations of computer time. This work was supported by a postdoc
fellowship of the German Academic Exchange Service (DAAD).

\section*{CRediT authorship contribution statement}
\textbf{Karl A. Kalina:} Conceptualization, Formal analysis, Investigation, Methodology, Visualization, Software, Validation, Visualization, Writing -- original draft, Writing -- review \& editing, Funding acquisition. 
\textbf{Philipp Gebhart:} Conceptualization, Writing – review \& editing.
\textbf{J\"{o}rg Brummund:} Formal analysis, Writing – review \& editing.
\textbf{Lennart Linden:} Writing – review \& editing.
\textbf{WaiChing Sun:} Conceptualization, Resources, Writing -- review \& editing.
\textbf{Markus Kästner:} Resources, Writing -- review \& editing.

\appendix

\section{Neural network architectures}\label{app:NNs}
To model the free energy expressions, NNs combined with normalization layers for in- and output are used in this work. This allows to restrict the weights to a range favorable for optimization without having to normalize the training data. In particular, this offers some advantages when using invariants, since each invariant is typically in a different range and is computed from tensor quantities during the prediction process. In addition, it simplifies to incorporate trained models into FE codes. More precisely, the normalization layers can be integrated into the original architecture after training simply by multiplying the weights.  

A model which have to be trained by a dataset $\mathcal D$ consisting of tuples ${}^i\mathcal T := ({}^i \ve{\mathscr X}, {}^i Y) \in \R^n \times \R$, with the generalized vector ${}^i \ve{\mathscr X} :=({}^i X_1,{}^i X_2,\ldots,{}^i X_n)$, is given by
\begin{align}
	f^\text{NN}: \R^n \to \R\,,\; \ve{\mathscr X} \mapsto f^\text{NN}(\ve{\mathscr X}) := (n^\text{out} \circ g^\text{NN} \circ \ve{\mathscr n}^\text{in})(\ve{\mathscr X}) \; .
\end{align}
Therein, we define the normalization layers without summation over multiple indices as
\begin{align}
	n_\alpha^\text{in}&: \R \to [x_\alpha^\text{min},x_\alpha^\text{max}] \subset \R \,,\; X_\alpha\mapsto n_\alpha^\text{in}(X_\alpha)
	:= X_\alpha \frac{x_\alpha^\text{max}-x_\alpha^\text{min}}{X^\text{max}_\alpha-X^\text{min}_\alpha} + \frac{x_\alpha^\text{max}X_\alpha^\text{min}-x_\alpha^\text{min}X_\alpha^\text{max}}{X_\alpha^\text{min}-X_\alpha^\text{max}} \text{ and}\\
	n^\text{out}&: [y^\text{min},y^\text{max}] \to \R  \,,\; y\mapsto n^\text{out}(y)
	:= \frac{Y^\text{max}-Y^\text{min}}{y^\text{max}-y^\text{min}}\left( y - \frac{y^\text{max}Y^\text{min}-y^\text{min}Y^\text{max}}{Y^\text{min}-Y^\text{max}} \right)\; ,
\end{align}
where $X_\alpha^\text{min}$, $X_\alpha^\text{max}$, $Y^\text{min}$ and $Y^\text{max}$ have to be determined from the data set, whereas $x^\text{min}_\alpha$, $x^\text{max}_\alpha$, $y^\text{min}$ and $y^\text{max}$ have to be prescribed. The function $g^\text{NN}(\ve{\mathscr x})$ can be represented by a standard FNN, a PNN\footnote{Note that in the case of a PNN the output normalization is given by
	\begin{align*}
		n^\text{out}: [y^\text{min},y^\text{max}] \to \R  \,,\; y\mapsto n^\text{out}(y)
		:= \frac{Y^\text{max}-Y^\text{min}}{y^\text{max}-y^\text{min}} y \; ,
	\end{align*}
	which is to enforce $f^\text{NN}(X) \ge 0\,\forall X$}
or an ICNN. Thereby, an FNN with $H$ hidden layers is given by
\begin{align}
	\label{eq:psi_NN_ML}
	o^{[1]}_\alpha &=  
	\mathcal{A}\Big(\sum_{\beta=1}^n w_{\alpha\beta}^{[1]} x_\beta + b_\alpha^{[1]}\Big)\;, \; \alpha \in\{1,2,\ldots,N^\text{nn,1}\}\; , \\
	o^{[h]}_\alpha &=  
	\mathcal{A}\Big(\sum_{\beta=1}^{N^{\text{NN},h-1}}w_{\alpha\beta}^{[h]} o_\beta^{[h-1]}+b_\alpha^{[h]} \Big)\;, \; \alpha \in\{1,2,\ldots,N^\text{nn,h}\}\;, h\in\{2,\dotsc,H\}\;, \\
	g^\text{FNN}(\ve{\mathscr x})&= \sum_{\alpha=1}^{N^{\text{NN},H}} W_{\alpha}\, o^{[H]}_\alpha + B \in\R\; ,
\end{align}
whereby there are no restrictions for the weights, i.e., $w_{\alpha\beta}^{[1]}, b_{\alpha}^{[1]}, w_{\alpha\beta}^{[h]}, b_{\alpha}^{[h]}, W_\alpha, B \in \R$. The weights and bias values are summarized in the $k$-dimensional vector
\begin{align}
	\w \in \FNN := \left\{ 
	w_{\alpha\beta}^{[h]}, b_{\alpha}^{[h]}, W_\alpha, B \in \R \; | \; h\in\{1,\ldots,H\}
	\right\} \; .
	\label{eq:setFNN}
\end{align}
Note that for clarity, a somewhat simplified mathematical notation is used here for the set $\FNN$.

The PNN architecture is almost the same. To satisfy that $g^\text{PNN}(\ve{\mathscr x}) \ge 0 \; \forall \ve{\mathscr x} \in \R^n$, the activation function in the last hidden layer has to be greater equal to zero for all outputs of the former layer. A possible choice is thus the softplus activation function $\softplus(z) := \ln(1+\exp z)$. Furthermore, weights and bias of the output layer have to be non-negative, i.e., $W_\alpha, B \in \R_{\ge 0}$. The restrictions are summarized as
\begin{align}
	\w \in \PNN := \left\{
	w_{\alpha\beta}^{[h]}, b_{\alpha}^{[h]} \in \R ; W_\alpha, B \in \R_{\ge 0} \; | \; h\in\{1,\ldots,H\}
	\right\} \; .
	\label{eq:setPNN}
\end{align}
Note that the chosen normalization approach does not disturb the required positivity. 

Finally, the ICNN architecture with skip connections \cite{Amos2017} is given by
\begin{align}
	\label{eq:ICNN}
	o^{[1]}_\alpha &=  
	\mathcal{A}(\sum_{\beta=1}^n w_{\alpha\beta}^{[1]} x_\beta + b_\alpha^{[1]})\;, \; \alpha \in\{1,2,\ldots,N^\text{nn,1}\}\; , \\
	o^{[h]}_\alpha &=  
	\mathcal{A}\Big(\sum_{\beta=1}^{N^{\text{NN},h-1}}w_{\alpha\beta}^{[h]} o_\beta^{[h-1]}+
	\sum_{\beta=1}^n s_{\alpha\beta}^{[h-1]} x_\beta + b_\alpha^{[h]} \Big) \;, \; \alpha \in\{1,2,\ldots,N^\text{nn,h}\}\;,\; h\in\{2,\dotsc,H\}\;, \\
	g^\text{ICNN}(\ve{\mathscr x})&= \sum_{\alpha=1}^{N^{\text{NN},H}} W_{\alpha}\, o^{[H]}_\alpha + 
	\sum_{\beta=1}^n S_{\beta} x_\beta + B \in\R\; .
\end{align}
To ensure convexity with respect to $\ve{\mathscr x}$, a convex and non-decreasing activation function, e.g., the softplus function, and non-negative weights beyond the first layer are chosen. Note that this is a sufficient, albeit not necessary condition for convexity.
If the model have to be convex with respect to another quantity from which the input follows, also the weights in the first layer and thus also the weights of the skip connections have to be non-negative \cite{Klein2021,Linden2023}. This is the case for models formulated in invariants, e.g., $I_\alpha(\te F, \ve B)$.
Thus, the restrictions $w_{\alpha\beta}^{[1]}, w_{\alpha\beta}^{[h]}, s_{\alpha\beta}^{[h]}, W_\alpha \in \R_{\ge 0}$ and $b_{\alpha}^{[1]}, b_{\alpha}^{[h]}, B \in \R$ hold for the ICNN, which is summarized in a generalized vector
\begin{align}
	\w \in \ICNN := \left\{w_{\alpha\beta}^{[1]}, w_{\alpha\beta}^{[h]}, s_{\alpha\beta}^{[h-1]}, W_\alpha, S_\beta \in \R_{\ge 0}; \, b_{\alpha}^{[1]}, b_{\alpha}^{[h]}, B \in \R \; | \; 
	h\in\{2,\ldots,H\}\right\} \; . \label{eq:setICNN}
\end{align}

\section{Modeling in the F-H setting}\label{app:FH}
As an alternative to the presented $\te F$-$\ve B$ setting, an $\te F$-$\ve H$ formulation can be used for the modeling of finite strain magneto-elasticity. It follows from the Legendre-Fenchel transformation
\begin{equation}
	\psi^*(\te F, \ve H) := \underset{\ve B \in \Ln_1}{\inf} [\psi(\te F, \ve B)- \ve B \cdot \ve H]  
\end{equation}
and is characterized by the constitutive relations
\begin{equation}
	\te P^\text{tot} = \frac{\partial \psi^*}{\partial \te F} \text{ and }
	\ve B = -\frac{\partial \psi^*}{\partial \ve H} \; .
\end{equation}
Note that the listed requirements on the potential $\psi(\te F,\ve B)$ given in Sect.~\ref{subsec:magnetoElasticity} similarly hold for $\psi^*(\te F, \ve H)$, expect for the non-negativity of the potential, the polyconvexity and the local polyconvexity. These conditions cannot be used here. The (strict) ellipticity condition for this setting is equivalent to Eq.~\eqref{eq:ellipticity}, but the generalized acoustic tensor is given by 
\begin{align}
	\te{\mathit\Gamma^*}(\ve N) := \te Q^*(\ve N) - \frac{[\ttte G^* :(\ve N \otimes \ve N)]\otimes [\ttte G^* :(\ve N \otimes \ve N)]}{\te K^* :(\ve N \otimes \ve N)} \; , \ve N \in \Vn \; ,  \label{eq:acousticTensorFH}
\end{align}
with $\te Q^*(\ve N) = A^*_{iJkL} N_J N_L \ve e_i \otimes \ve e_k$ and the moduli
\begin{equation}
	\tttte A^* := \diffp{{}^2\psi^*}{\te F \partial \te F} \in \Ln_4 \; , \te K^* := \diffp{{}^2 \psi^*}{\ve H \partial \ve H} \in \Sym \text{ and }
	\ttte G^* := \diffp{{}^2 \psi^*}{\te F \partial \ve H} \in \Ln_3 \; .
\end{equation}

The $\te F$-$\ve H$ setting is a mixed energy-enthalpy formulation. Thus, the solution of variational principles leads to saddle point problems. This formulation is favorable if the magnetic scalar potential $\ve H=: -\nablaX \varphi$ is used.

\section{Ellipticity and relaxed local polyconvexity}\label{app:poly_ell}

Within this appended section, we show that the \emph{relaxed local polyconvexity} defined in Eq.~\eqref{eq:localPoly} implies \emph{local ellipticity} in a state and its neighborhood. For reasons of clarity, we switch to index notation with Einstein summation convention for the most computations here. First, for reasons of completeness, the ellipticity condition is derived shortly.

\subsection{Ellipticity in magneto-elasticity}\label{app:ell}

Following \cite{Destrade2011,Rudykh2013,Polukhov2020}, we consider \emph{infinitesimal homogeneous plane waves} superimposed to an equilibrium state defined by $\ve \varphi^0$, $\te F^0 = (\nablaX \ve \varphi^0)^T$ and $\ve B^0$. The equilibrium state is homogeneous in $\te F^0$ and $\ve B^0$, i.e. $F_{lK,M}^0 = B_{K,M}^0 = 0$. The plane waves are given by
\begin{align}
	\delta \varphi_l = a_l f(\gamma) \; \text{and} \;
	\delta B_K = V_K g(\gamma) \; \text{with } \gamma = X_Q N_Q \pm ct \; , \label{eq:planeWaves}
\end{align}
where $\ve N \in \Vn$ is the propagation direction and $\ve a, \ve V \in \Vn$ are amplitude directions of the plane waves, respectively. The parameter $c$ denotes the \emph{speed of the waves}, which has to be \emph{real valued} \cite{Ebbing2010,Destrade2011}, and the functions $f(\gamma)$ as well as $g(\gamma)$ are assumed to be sufficiently continuously differentiable. By inserting the solutions \eqref{eq:planeWaves} in the incremental balance of linear momentum,  Amp\`{e}re's law and Gauss's law, i.e. $\delta P_{lK,K}^\text{tot} = \varrho_0 \delta \ddot \varphi_l$, $e_{KLM} \delta H_{M,L}=0$ and $\delta B_{K,K}=0$, we get 
\begin{align}
	A_{lKmN} a_l a_m N_K N_N + G_{lKM} a_l N_K V_M \frac{g'(\gamma)}{f''(\gamma)} &= \varrho_0 c^2 (>) \ge 0 \; \text{,}\label{eq:locInert} \\
	e_{KLM} \left[G_{iJM} a_i N_J N_L f''(\gamma) + K_{MN} V_N N_L g'(\gamma)\right] &= 0 \; \text{and} \label{eq:locAmpere} \\
	N_K V_K &= 0 \; , \label{eq:orthoNV}
\end{align}
where the tangent terms, which follow from the second derivatives of the potential $\psi(\te F,\ve B)$ at $\te F^0$ and $\ve B^0$ are given by
\begin{align}
	A_{lKmN} = \diffp{{}^2 \psi}{F_{lK}\partial F_{mN}} \bigg|_{\te F^0, \ve B^0} \; ,
	G_{lKM} = \diffp{{}^2 \psi}{F_{lK}\partial B_M} \bigg|_{\te F^0, \ve B^0} \;\text{and} \; 
	K_{MN} = \diffp{{}^2 \psi}{B_M\partial B_N} \bigg|_{\te F^0, \ve B^0} \; .
\end{align}
As mentioned, Eq.~\eqref{eq:locInert} states that the wave speeds $c$ have to be real valued.
From Eq.~\eqref{eq:locAmpere} we can find that 
\begin{align}
	Z_M N_L - Z_L N_M = 0 \text{ with }
	Z_M = G_{iJM} a_i N_J  f''(\gamma) + K_{MN} V_N  g'(\gamma) \; , \label{eq:anti}
\end{align}
which is a result of the antisymmetry.
By multiplying the above equation with $V_M N_L$ we get 
\begin{align}
	V_M Z_M  = 0 \; \to \; g'(\gamma) = - \frac{G_{iJM} a_i N_J V_M f''(\gamma)}{V_P V_Q K_{PQ}} \; . \label{eq:gs}
\end{align}
Finally, after inserting Eq.~\eqref{eq:gs} into Eq.~\eqref{eq:locInert}, one finds 
\begin{align}
	a_l a_m \Gamma_{lm} = \varrho_0 c^2 (>) \ge 0 \text{ with }
	\Gamma_{lm} := A_{lKmN} N_K N_M  - 
	\frac{G_{lKP}N_K V_P G_{mRS} N_R V_S}{V_U V_V K_{UV}} \; , (\ve N, \ve V)\in\mathcal E \; ,\label{eq:acoustic1}
\end{align}
wherein $\te{\mathit\Gamma} \in \Sym$ is the \emph{generalized magneto-elastic acoustic tensor} which has been also derived in \cite{Polukhov2020,Rudykh2013} and $\mathcal E:=\{\ve N, \ve V \in \Vn \,|\, \ve V \cdot \ve N = 0\}$. The equation above is the \emph{(strict) strong ellipticity condition} which is fulfilled for all $\ve a\in \Vn$ if $\te{\mathit\Gamma}$ is positive semi-definite. For strict strong ellipticity $\te{\mathit\Gamma}$ has to be positive definite.

By multiplying Eq.~\eqref{eq:anti} with $K_{PL}^{-1} K_{QM}^{-1}$, inserting it again into Eq.~\eqref{eq:locInert} and after some tensor manipulations, we derive an alternative generalized magneto-elastic acoustic tensor and the strong ellipticity condition is given by
\begin{align}
	a_l a_m \Xi_{lm} = \varrho_0 c^2 (>) \ge 0 \text{ with }
	\Xi_{lm} := A_{lKmN} N_K N_M  - 
	G_{lKP} N_K K_{PQ}^{-1}\left(\delta_{QM} - \frac{N_Q K_{MS}^{-1} N_S}{N_U N_V K_{UV}^{-1}} \right) G_{mNM} N_N \; .
\end{align}
The tensor $\te{\mathit\Xi} \in \Sym$ is also introduced in \cite{Ortigosa2016a,Destrade2011}. If the ellipticity has to be proven numerically in the 3D case, this version is favorable over the one given in Eq.~\eqref{eq:acoustic1}, since only one unit vector has to be sampled here.

Finally, we derive a third version of the strong ellipticity condition by just summing together Eq.~\eqref{eq:locInert} and Eq.~\eqref{eq:gs}${}_1$ multiplied with $g'(\gamma)/(f''(\gamma))^2$. This gives 
\begin{align}
	A_{lKmN} a_l a_m N_K N_N + 2 G_{lKM} a_l N_K V_M \frac{g'(\gamma)}{f''(\gamma)} 
	+ K_{KL} V_K V_L \left(\frac{g'(\gamma)}{f''(\gamma)}\right)^2 = \varrho_0 c^2 (>)\ge 0 \; . \label{eq:ell3}
\end{align}

\subsection{Implication of the local polyconvexity condition}\label{app:poly}

Now we consider a magneto-elastic potential $\psi(\te F,\ve B)$ which is \emph{locally polyconvex} in the neighborhood of a state $(\te F,\ve B) \in \GL \times \Ln_1$, see Eq.~\eqref{eq:localPoly} for the definition of this concept. The potential can thus be expressed by $\mathcal P(\gt I)$ which is locally polyconvex with respect to the argument set $ \gt I := \left(\te F, \cof \te F, \det \te F, \te F \cdot \ve B, \ve B\right)$.

Leaned on the methodology in Bonet~et~al.~\cite{Bonet2015} and Ortigosa~and~Gil~\cite{Ortigosa2016a}, we express the second derivatives of $\psi(\te F,\ve B)$ by $\mathcal P(\gt I)$ in the following, which allows us to demonstrate that the introduced local polyconvexity implies strong ellipticity at a state. Thus, we use the \emph{chain rule} to get
\begin{align}
	A_{lKmN} &= \diffp{{}^2 \psi}{F_{lK}\partial F_{mN}} =
	\diffp{{}^2 \mathcal P}{\mathfrak I_\alpha\partial \mathfrak I_\beta}
	\diffp{\mathfrak I_\alpha}{F_{lK}} \diffp{\mathfrak I_\beta}{F_{mN}}
	+ \diffp{\mathcal P}{\mathfrak{I}_\alpha} 
	\diffp{{}^2 \mathfrak I_\alpha}{F_{lK}\partial F_{mN}} \; , \label{eq:FF}\\
	G_{lKM} &= \diffp{{}^2 \psi}{F_{lK}\partial B_{M}} =
	\diffp{{}^2 \mathcal P}{\mathfrak I_\alpha\partial \mathfrak I_\beta}
	\diffp{\mathfrak I_\alpha}{F_{lK}} \diffp{\mathfrak I_\beta}{B_M}
	+ \diffp{\mathcal P}{\mathfrak{I}_\alpha} 
	\diffp{{}^2 \mathfrak I_\alpha}{F_{lK}\partial B_M} \; \text{and} \label{eq:FB}\\
	K_{KL} &= \diffp{{}^2 \psi}{B_K\partial B_{L}} =
	\diffp{{}^2 \mathcal P}{\mathfrak I_\alpha\partial \mathfrak I_\beta}
	\diffp{\mathfrak I_\alpha}{B_K} \diffp{\mathfrak I_\beta}{B_L}
	+ \diffp{\mathcal P}{\mathfrak{I}_\alpha} 
	\diffp{{}^2 \mathfrak I_\alpha}{B_K\partial B_L} \; . \label{eq:BB}
\end{align}
Next, we will show that the second summands in the above equations always vanish when these expressions are substituted into Eq.~\eqref{eq:ell3}. To prove for that, the second derivatives of the arguments $\mathfrak I_\alpha$ are calculated. They are mostly equal to zero except for\begin{align}
	\diffp{{}^2 (J F_{Ji}^{-1})}{F_{lK} \partial F_{mN}} = e_{ilm} e_{JKN} \; ,
	\diffp{{}^2 J}{F_{lK} \partial F_{mN}} = e_{lmp} e_{KNQ} F_{pQ} \; , 
	\diffp{{}^2 (F_{iJ} B_J)}{F_{lK} \partial B_M} = \delta_{il} \delta_{KM} \; . \label{eq:Rterms}
\end{align}
By using these expressions and inserting Eqs.~\eqref{eq:FF}--\eqref{eq:BB} into Eq.~\eqref{eq:ell3}, we get
\begin{align}
	\diffp{{}^2 \mathcal P}{\mathfrak I_\alpha\partial \mathfrak I_\beta} \left[
	\diffp{\mathfrak I_\alpha}{F_{lK}} \diffp{\mathfrak I_\beta}{F_{mN}} a_l a_m N_K N_N
	+ 2 \diffp{\mathfrak I_\alpha}{F_{lK}} \diffp{\mathfrak I_\beta}{B_M} a_l N_K V_M \frac{g'(\gamma)}{f''(\gamma)} 
	+ \diffp{\mathfrak I_\alpha}{B_K} \diffp{\mathfrak I_\beta}{B_L} V_K V_L \left(\frac{g'(\gamma)}{f''(\gamma)}\right)^2 \right] + R = \varrho_0 c^2 \; ,
\end{align}
where the term $R$ results from the expressions given in Eq.~\eqref{eq:Rterms} and follows to zero:
\begin{align}
	R = \diffp{\mathcal P}{\left(J F_{Ji}^{-1}\right)} e_{ilm} e_{JKN} a_l a_m N_K N_N   + 
	\diffp{\mathcal P}{J} e_{lmp} e_{KNQ} F_{pQ} a_l a_m N_K N_N + 
	\diffp{\mathcal P}{\left(F_{iJ}B_J\right)} a_i N_K V_K = 0 \; .
\end{align}
This is due to the summation of antisymmetric and symmetric quantities for the first two summands and the orthogonality of $\ve N$ and $\ve V$ for the last, cf. Eq.~\eqref{eq:orthoNV}. Thus, we get
\begin{align}
	\diffp{{}^2 \mathcal P}{\mathfrak I_\alpha\partial \mathfrak I_\beta} \left[
	\mathfrak a_\alpha \mathfrak a_\beta + 2 \mathfrak a_\alpha \mathfrak b_\beta
	+ \mathfrak b_\alpha \mathfrak b_\beta
	\right] = 
	\diffp{{}^2 \mathcal P}{\mathfrak I_\alpha\partial \mathfrak I_\beta} \underbrace{\left[
		\mathfrak a_\alpha + \mathfrak b_\alpha\right]}_{\mathfrak c_\alpha}
	\left[
	\mathfrak a_\beta + \mathfrak b_\beta\right] = 
	\diffp{{}^2 \mathcal P}{\mathfrak I_\alpha\partial \mathfrak I_\beta} \mathfrak c_\alpha \mathfrak c_\beta= \varrho_0 c^2  \ge 0 \; ,
\end{align}
which is fulfilled at a state $(\te F,\ve B) \in \GL \times \Ln_1$ in which the Hessian $\mathfrak H_{\alpha\beta}$ of $\mathcal P(\gt I)$ is \emph{positive semi-definite}. In the equation above, $\mathfrak a_\alpha$ and $\mathfrak b_\alpha$ are given by
\begin{align}
	\mathfrak a_\alpha := \diffp{\mathfrak I_\alpha}{F_{lK}} a_l N_K \text{ and }
	\mathfrak b_\alpha := \diffp{\mathfrak I_\alpha}{B_M} V_M \frac{g'(\gamma)}{f''(\gamma)} \; .
\end{align}
Thus, it is verified that \emph{local polyconvexity} as introduced in Eq.~\eqref{eq:localPoly} implies \emph{local strong ellipticity}.

\section{Simplifications for 2D plane strain states}\label{app:2D}

\subsection{Restrictions for tensor quantities}
To reduce from a 3D to a 2D problem, only field distributions satisfying the plane strain condition are considered. The position of the Cartesian basis vectors $\ve e_k$ and $\ve e_K$ are chosen in such a way that for the coordinates of $\te F$, $\te P^\text{tot}$, $\ve B$ and $\ve H$ the relations
\begin{align}
	[F_{kL}] = \begin{bmatrix}
		F_{11} & F_{12} & 0 \\
		F_{21} & F_{22} & 0 \\
		0 & 0 & 1
	\end{bmatrix} \; , \;
	[P^\text{tot}_{kL}] = \begin{bmatrix}
		P^\text{tot}_{11} & P^\text{tot}_{12} & 0 \\
		P^\text{tot}_{21} & P^\text{tot}_{22} & 0 \\
		0 & 0 & P^\text{tot}_{33}
	\end{bmatrix} \; ,\;
	[B_{K}] = \begin{bmatrix}
		B_1\\
		B_2\\
		0
	\end{bmatrix}
	\text{ and }
	[H_{K}] = \begin{bmatrix}
		H_1\\
		H_2\\
		0
	\end{bmatrix}
\end{align}
hold. From the definition of the vector potential and the assumptions made, finally the
reduction to only one nonzero coordinate pointing in the $X_3$-direction results, i.e., $\ve A = A_3 \ve e_3$. Due to that, the Coulomb gauge is automatically fulfilled in the 2D case.

The reduction to plane strain problems also affects the microscopic BCs~\eqref{eq:peri1}--\eqref{eq:peri4}: Here, the periodicity is now also to be restricted for the considered fluctuation components to the $x_1$-$x_2$-plane so that positive and negative surfaces are reduced to two boundary edges, respectively. For the vector potential one must also make the following change:
\begin{align}
	\ve A \in \mathcal A(\bte B) &:= \left\{\ve A \in \Ln_1 \, | \, A_3 = e_{3KL} \bar B_K X_L + \tilde A_3\; \text{with } \tilde A_3^+ = \tilde A_3^-,\; A_1 = A_2 = 0\right\} \; , \label{eq:periA2D}\\
	\ve J \in \mathcal J &:= \left\{\ve J \in \Ln_1 \, | \, J_3^+ = - J_3^-,\; J_1=J_2=0\right\} \; ,
	\label{eq:periK2D} 
\end{align}
see also \cite{Danas2017,Kalina2021a}. 

\subsection{Simplifications for ellipticity and local polyconvexity}\label{app:poly_ell_2D}
\paragraph{Ellipticity}
Regarding the ellipticity, only the tensor coordinates belonging to the $x_1$-$x_2$ plane have to be taken into account in the \emph{plane strain} case \cite{Polukhov2020}. This is due to the fact that the considered plane waves, cf. \ref{app:ell}, are restricted to the $x_1$-$x_2$ plane in this case and consequently it holds $a_3=N_3=V_3=0$ for Ineq.~\eqref{eq:ellipticityLocal}. To account for this, we define $(\ve N, \ve V) \in \mathcal E:= \{\ve N, \ve V\in\Vn \, |\, \ve N\cdot \ve V = 0,\; N_3=V_3=0\}$ for the orthogonal directions. This leads to the fact that only two antiparallel vectors $\ve V$ are possible for a given $\ve N$, where both are equivalent. We can thus describe both vectors by only one angle as
\begin{align}
	[N_K] = \begin{bmatrix}
		\cos(\phi)\\
		\sin(\phi)\\
		0
	\end{bmatrix} \text{ and }
	[V_K] = \begin{bmatrix}
		\cos(\phi+\pi/2)\\
		\sin(\phi+\pi/2)\\
		0
	\end{bmatrix} \; .
\end{align}
The generalized acoustic tensor for a given $\te F$-$\ve B$ state consequently only depends on $\phi$, i.e., $\te{\mathit\Gamma}(\phi)$.
The requirement that a state $(\te F,\ve B)\in\GL\times\Ln_1$ fulfills the strict local ellipticity condition is thus reformulated to
\begin{align}
	\left(\gamma_1 = \Gamma_{11} > 0 \; \wedge \; \gamma_2 = \Gamma_{11}\Gamma_{22} - \Gamma_{12}\Gamma_{21} > 0\right) \; \forall 
	\phi \in [0,\pi] \; .
	\label{eq:ell_2D}
\end{align}
The calculation of the loss term $\mathcal L^\text{ell}$ is thus also simplified as follows:
\begin{align}
	\mathcal L^\text{ell} &:= \frac{1}{|\mathcal D^\text{coup}_\text{cal}|}\sum_{i=1}^{|\mathcal D^\text{coup}_\text{cal}|} \Bigg[ \frac{1}{n_{\gamma_1}} \relu\left(-{}^i\gamma_1^\text{min} \right)
	+ \frac{1}{n_{\gamma_2}} \relu\left(-\gamma_2^\text{min} \right) 
	\Bigg]\; \text{with} \; \label{eq:ell2D}\\
	{}^i\gamma_\alpha^\text{min}&:=\underset{\phi\in[0,\pi]}{\inf}  \gamma_\alpha({}^i\te F, {}^i\ve B, \w, \phi) 
	, \; \alpha \in\{1,2\} \; .	\label{eq:opt_ell2D}
\end{align}

\paragraph{Local polyconvexity}
Finally, also for the local polyconvexity defined by Eq.~\eqref{eq:localPoly}, a simplified 2D \emph{plane strain} version has to be formulated. This should still imply local ellipticity in the simplified 2D version of Ineq.~\eqref{eq:ell3}, i.e., 
\begin{align}
	A_{lKmN} a_l a_m N_K N_N + 2 G_{lKM} a_l N_K V_M \frac{g'(\gamma)}{f''(\gamma)} 
	+ K_{KL} V_K V_L \left(\frac{g'(\gamma)}{f''(\gamma)}\right)^2 = \varrho_0 c^2 \ge 0 \label{eq:ell3_2D}
\end{align} 
with $a_3=0$ and $(\ve N, \ve V) \in \mathcal E:= \{\ve N, \ve V\in\Vn \, |\, \ve N\cdot \ve V = 0,\; N_3=V_3=0\}$. Thus, by considering
\begin{align}
	&\diffp{{}^2 \mathcal P}{\mathfrak I_\alpha\partial \mathfrak I_\beta} \left[
	\mathfrak a_\alpha \mathfrak a_\beta + 2 \mathfrak a_\alpha \mathfrak b_\beta
	+ \mathfrak b_\alpha \mathfrak b_\beta
	\right] = 
	\diffp{{}^2 \mathcal P}{\mathfrak I_\alpha\partial \mathfrak I_\beta} \underbrace{\left[
		\mathfrak a_\alpha + \mathfrak b_\alpha\right]}_{\mathfrak c_\alpha}
	\left[
	\mathfrak a_\beta + \mathfrak b_\beta\right] = 
	\diffp{{}^2 \mathcal P}{\mathfrak I_\alpha\partial \mathfrak I_\beta} \mathfrak c_\alpha \mathfrak c_\beta= \varrho_0 c^2  \ge 0 \; \text{with} \\
	&\mathfrak a_\alpha = \diffp{\mathfrak I_\alpha}{F_{lK}} a_l N_K \text{ and }
	\mathfrak b_\alpha = \diffp{\mathfrak I_\alpha}{B_M} V_M \frac{g'(\gamma)}{f''(\gamma)} \; ,
\end{align}
we find that a reduction to the $x_1$-$x_2$ plane coordinates of the tensors included in the set $\gt I=(\te F, \cof \te F, J, \te F \cdot \ve B, \ve B)$ can be done for $\te F$, $\te F\cdot \ve B$ and $\ve B$, i.e., one only needs $F_{iJ}, F_{iJ} B_J, B_I$ with $i,I,J\in\{1,2\}$. This is due to the fact that
\begin{align}
	\diffp{F_{iJ}}{F_{lK}} a_l N_K &= a_i N_J = 0 \text{ for } i,J=3 \; ,\\
	\diffp{(F_{iJ}B_J)}{F_{lK}} a_l N_K &= a_i N_K B_K = 0 \text{ for } i=3 \; ,\\
	\diffp{(F_{iJ}B_J)}{B_{K}} V_K &= F_{iK} V_K = 0 \text{ for } i=3 \; \text{and}\\
	\diffp{B_I}{B_{K}} V_K &= V_I = 0 \text{ for } I=3 \; ,
\end{align} 
so the corresponding second derivatives of $\mathcal P(\gt I)$ with respect to $\gt I$ do not give  contributions in the 2D plane strain case.
In contrast, for $\cof \te F$, we get
\begin{align}
	\diffp{(J F_{Ui}^{-1})}{F_{lK}} a_l N_K &= J (F_{Kl}^{-1}F_{Ui}^{-1} - F_{Ul}^{-1}F_{Ki}^{-1}) a_l N_K = 0 \text{ for } i \in\{1,2\}, U=3 \text{ and } i=3, U\in\{1,2\} \; ,
\end{align} 
which is not zero for $(i,U) = (3,3)$. Therefore, in addition to the in-plane coordinates of $\cof \te F$, the component $J F_{33}^{-1}$ must also be considered.
The relevant part of the Hessian $\gt H$ represented as a matrix is thus $14\times 14$, which results in 14 eigenvalues that must be considered.

\section{Comparison of Adam and SLSQP}\label{app:adam}

\begin{figure}
	\begin{center}
		\includegraphics{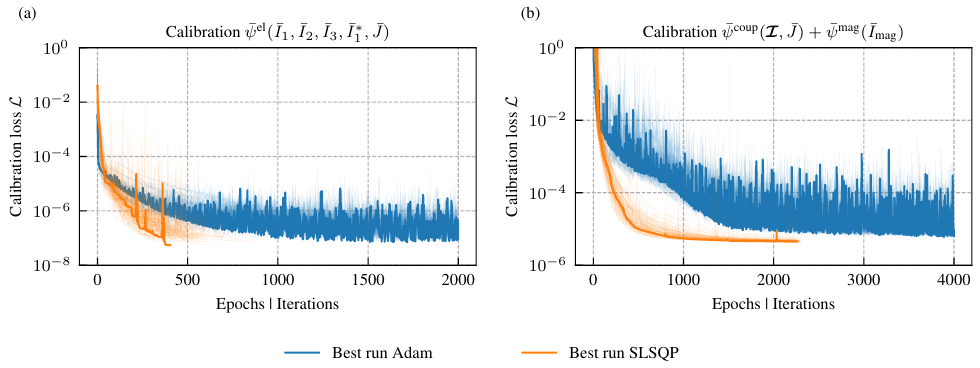}
	\end{center}
	\caption{Comparison of the optimizers Adam and SLSQP for the calibration of NN~model~III~\eqref{eq:modelIII}: (a) elastic part with mini batch size 32 for Adam and (b) coupling part and magnetic part with fixed parameters of the elastic part and mini batch size 64 for Adam. The learning rate for the Adam optimizer was 0.01.}
	\label{fig:Adam}
\end{figure}

In this appended section, the trainings of the proposed NN~model~III~\eqref{eq:modelIII} with the optimizer \emph{SLSQP (Sequential Least Squares Programming)} and \emph{Adam (Adaptive Moment Estimation)} are exemplarily compared to each other.

SLSQP belongs to the class of quasi-Newton methods, whereas Adam is a stochastic gradient-based optimizer. Following the common terms from machine learning, a training step with SLSQP is denoted as iteration whereas it is called epoch for Adam. For training with SLSQP, the full dataset is processed in each iteration. In case of Adam, mini batches are used, which introduces a regularization effect \cite{Kollmannsberger2021}. These are recomposed in a random manner after each epoch.

The training of NN~model~III with data $\mathcal D^\text{mech}$ and $\mathcal D^\text{coup}$ for the compressible MAP is analyzed now. A division of the overall datasets into calibration and test sets with a ratio of $70/30$ was chosen. One hidden layer with 6 neurons, two hidden layers with 10 neurons each and one hidden layer with 5 neurons have been chosen for the networks describing $\bar \psi^\text{el}_\text{ICNN}(\bar I_1,\bar I_2,\bar I_3,\bar I_1^*,\bar J)$, $\bar \psi_\text{FNN}^\text{coup}(\bI,\bar J)$ and $\bar \psi_\text{FNN}^\text{mag}(\bar I_\text{mag})$, respectively. For all, the softplus activation function is used. To exclude random effects in the comparison of Adam and SLSQP, 100 training runs each have been done for both the training of the elastic part by $\mathcal D^\text{mech}$ and the remaining energy contributions by $\mathcal D^\text{coup}$. The mini batch size was chosen to 32 and 64 for training by $\mathcal D^\text{mech}$ and $\mathcal D^\text{coup}$, respectively. The learning rate was chosen to 0.01.
As shown in Fig.~\ref{fig:Adam}, the optimization with SLSQP is clearly superior to Adam. It converges faster and a lower loss is reached. It should be noted, however, that SLSQP is unsuitable for significantly larger networks, since the computational cost increases nonlinearly with the number of trainable variables.

\bibliographystyle{unsrtnat} 
\bibliography{PANN_MRE.bib}

\end{document}

%% file: StyleSetup.tex
\usepackage{amssymb}
\usepackage{amsthm}
\usepackage{amsmath}
\usepackage{siunitx}
\usepackage{mathtools}		
\mathtoolsset{centercolon}	
\usepackage{newtxmath}      
\usepackage{lineno}
\usepackage{pifont}
\usepackage[dvipsnames]{xcolor}

\newcommand{\R}{\mathbb R} 
\newcommand{\N}{\mathbb N} 
\newcommand{\Sym}{\mathscr{S\! y \! m}} 
\newcommand{\Ln}{\mathcal L} 
\newcommand{\SO}{\mathscr{S\!O}(3)} 
\newcommand{\Othree}{\mathscr{O}(3)} 
\newcommand{\GL}{\mathcal{G\!L}^+(3)} 
\newcommand{\Vn}{\mathcal{N}} 
\newcommand{\G}{\mathcal{G}} 

\newcommand{\B}{\mathcal{B}} 
\newcommand{\Sj}{\mathcal{S}} 
\newcommand{\I}{\boldsymbol{\mathcal I}}
\newcommand{\bI}{\,\bar{\boldsymbol{\!\I}}}
\newcommand{\bfI}{\bar{\boldsymbol{\mathfrak i}}}
\newcommand{\w}{\boldsymbol{\mathscr w}}
\newcommand{\p}{\boldsymbol{\mathscr p}}
\newcommand{\FNN}{\mathcal{F\!N\!N}} 
\newcommand{\PNN}{\mathcal{P\!N\!N}} 
\newcommand{\ICNN}{\mathcal{I\!C\!N\!N}} 

\DeclareMathOperator{\relu}{ReLU}
\DeclareMathOperator{\softplus}{SP}
\DeclareMathOperator{\tr}{tr}
\DeclareMathOperator{\cof}{cof}
\DeclareMathOperator{\sym}{sym}
\DeclareMathOperator{\skw}{skw}
\DeclareMathOperator{\diag}{diag}
\DeclareSIUnit[number-unit-product = \,]{\promille}{\textperthousand}

\newcommand{\dx}{\mathrm d} 
\newcommand{\ve}[1]{\boldsymbol{#1}} 
\newcommand{\te}[1]{\boldsymbol {#1}} 
\newcommand{\tg}[1]{\boldsymbol #1} 
\newcommand{\ttte}[1]{\mathbf #1} 
\newcommand{\tttte}[1]{\mathbb #1} 
\newcommand{\one}{\textit{\textbf{1}}}
\newcommand{\zero}{\textit{\textbf{0}}}
\newcommand{\gt}[1]{\boldsymbol{\mathfrak #1}}

\newcommand{\bve}[1]{\bar{\boldsymbol{#1}}} 
\newcommand{\bte}[1]{\bar{\boldsymbol #1}} 
\newcommand{\bttte}[1]{\bar{\mathbf #1}} 
\newcommand{\btttte}[1]{\bar{\mathbb #1}} 

\newcommand{\diffp}[2]{\frac{\partial #1}{\partial #2}} 

\newcommand{\jump}[1]{\left\llbracket #1\right\rrbracket}

\newcommand{\nablaX}{\nabla_{\!\!{\ve X}}}

\newcommand{\cmark}{\textcolor{ForestGreen}{\ding{51}}}%
\newcommand{\cbmark}{\textcolor{ForestGreen}{(\ding{51})}}%
\newcommand{\xmark}{\textcolor{BrickRed}{\ding{55}}}%

\makeatletter
\newcommand*\bigcdot{\mathpalette\bigcdot@{.5}}
\newcommand*\bigcdot@[2]{\mathbin{\vcenter{\hbox{\scalebox{#2}{$\m@th#1\bullet$}}}}}
\makeatother